\documentclass[12pt,a4paper]{article}
\usepackage[utf8]{inputenc}
\usepackage[english]{babel}
\usepackage{amsmath}
\usepackage{amsfonts}
\usepackage{amssymb}
\usepackage{graphicx}
\usepackage{cancel}
\usepackage{color}
\usepackage{slashed, epsf, latexsym}
\usepackage{epsfig}
\usepackage{graphics}
\usepackage{jheppub}

\begin{document}

    \preprint{TIFR/TH/18-11}
	\title{Black holes in presence of cosmological constant: Second  order in ${1\over D}$}
	\author[a]{Sayantani Bhattacharyya} 
	\author[a]{, Parthajit Biswas}
	\author[b]{, Yogesh Dandekar}
	\affiliation[a]{National Institute of Science Education and Research, HBNI, Bhubaneshwar 752050, Odisha, India}
	\affiliation[b]{Tata Institute of Fundamental Research, Mumbai 400005, India}
	\emailAdd{sayanta@niser.ac.in}
	\emailAdd{parthajit.biswas@niser.ac.in}
	\emailAdd{yogesh@theory.tifr.res.in}

	\abstract{We have extended the results of \cite{arbBack} upto second subleading order in an expansion around large dimension $D$.  Unlike the previous case,  there are non-trivial metric corrections at this order. Due to  our `background-covariant' formalism, the dependence on  Ricci and  the Riemann curvature tensor of the background is manifest here. The gravity system is dual to a dynamical membrane coupled with a  velocity field. The dual membrane is embedded in some smooth  background geometry that also satisfies the Einstein equation in presence of cosmological constant. We explicitly computed the corrections to the equation governing the membrane-dynamics. Our results match with earlier derivations in appropriate limits. We calculated the spectrum of QNM from our membrane equations and matched them against similar results derived from gravity.}

\maketitle


\section{Introduction}\label{intro}
Recently it has been shown that in large number of dimensions, black hole solutions simplify a lot. \footnote{See \cite{arbBack,membrane,Chmembrane,yogesh1,yogesh2,radiation,Dandekar:2017aiv} for the work related to the formulation of Membrane paradigm at large D. See \cite{Emparan:2013moa,Emparan:2013xia, Emparan:2014cia,EmparanCoupling} and also \cite{Giribet:2013wia,Prester:2013gxa,Emparan:2013oza} for initial work introducing the large D limit in General Relativity. See \cite{Emparan:2014jca,QNM:Emparan,Romero-Bermudez:2015bma,Suzuki:2015iha,Suzuki:2015axa,Emparan:2015gva,Tanabe:2015hda,Tanabe:2015isb,Chen:2015fuf,EmparanHydro,Sadhu:2016ynd,Herzog:2016hob,Tanabe:2016pjr,Tanabe:2016opw,Rozali:2016yhw,Chen:2016fuy,Chen:2017wpf,Chen:2017hwm,Rozali:2017bll,Chen:2017rxa,Emparan:2018bmi,Herzog:2017my,BinChen:2018apr} for other parallel work which uses the technique of large D expansion.} The effect of the black hole is essentially confined around its event horizon in a  parametrically thin region whose thickness is proportional to the inverse of the number of dimension. Further, the spectrum of the linearized fluctuation (Quasi Normal Modes or QNMs) develops a large gap proportional to the number of dimension. In  \cite{membrane} authors have shown how one can formulate the autonomous nonlinear theory of the low lying modes. They combine to form a dynamical black hole solution  to Einstein equation which could be determined in an expansion in inverse powers of $D$. 

In \cite{arbBack} the authors have extended the calculation of \cite{membrane} (which was for pure Einstein Gravity) to solutions in presence of cosmological constant and in general for any asymptotic background provided it is a solution of the gravity equation. The method used in \cite{arbBack} has manifest background covariance but the calculation were done only upto the first subleading order in $\left(\frac{1}{D}\right)$. 

In this note,  we would like to extend the calculation of \cite{arbBack} to the second subleading order. The key motivation is two-fold. Firstly from the result of \cite{arbBack} we know that at the first subleading order the background curvature does not appear explicitly in any of the equation or the solution. However it should appear explicitly at second subleading order (which, very roughly speaking, captures the effect of two derivatives on the background). Secondly  from the experience of the `flat space computation', it is expected that at this order we should see the entropy production from a dynamical black hole.

However, in this note we shall confine ourselves only to the computation of the membrane equation of motion and the metric correction upto the second subleading order in $\left(\frac{1}{D}\right)$ expansion. We leave  the `study of entropy production' for future.\\

As a consistency check of our results we shall linearize our membrane equation and compare the spectrum with that of the low lying QNMs (already determined in \cite{QNM:Emparan}). We shall find a perfect match  upto the relevant order.\\

The organization of this note is as follows.\\
In section (\ref{setup}) we  have described the basic set-up of our problem in terms of equations and also the final result for the corrections to metric and the membrane equations. Next in section (\ref{sec:compute}) we gave a sketch of the computation, which turns out to be quite tedious in this case. Many of the details we collected in the appendices. In section (\ref{sec:check}) we have performed several checks. Some of them are about the internal consistency of our set of equations (see subsection - \ref{check1}) and the rest are about the calculation of the  linearized spectrum of our membrane around different static backgrounds. We have also matched them against the known results of QNMs (see subsection - \ref{check2}). Finally in section (\ref{sec:future}) we discuss about the future directions.

\section{Set up and final result}\label{setup}
In this section we shall briefly define the basic set-up of our problem in terms of equations. It is essentially an extension of section-2 of \cite{arbBack}. So we shall be very brief here.\\
 We are dealing with pure gravity in presence of cosmological constant. The Action and the equation of motion are given by the following.
 \begin{equation}\label{eq:LagEom}
 {\cal{S}}=\int\sqrt{-G}[{\cal R}-\Lambda]
 \end{equation}
Where the  dimension (denoted as $D$) dependence of $\Lambda$ is parametrized as follows 
\begin{equation}
\Lambda=[(D-1)(D-2)]\lambda,\qquad \lambda\sim{\cal{O}}(1)
\end{equation}
Varying \eqref{eq:LagEom} with respect to the metric we get the equation of motion
\begin{equation}\label{eq:ein}
E_{AB}\equiv R_{AB}-\left(\frac{{\cal R}-\Lambda}{2}\right)G_{AB}=0
\end{equation}
Our aim is to solve these equations perturbatively as a series in inverse power of $D$. Schematically our solution will take the form
\begin{equation}\label{schemesol}
G_{AB} = G^{(0)}_{AB} +\left(\frac{1}{D}\right) G^{(1)}_{AB} +\left(\frac{1}{D}\right)^2 G^{(2)}_{AB}  + \cdots
\end{equation}
We take our starting ansatz $G^{(0)}_{AB}$ to be the following\footnote{See \cite{arbBack}, \cite{membrane}, \cite{Chmembrane} for detailed explanation for this choice}
\begin{equation}\label{ansatz}
\begin{split}
G^{(0)}_{AB} &= g_{AB} + \psi^{-D}O_A O_B
\end{split}
\end{equation}
Here $g_{AB}$ is the background metric which could be any smooth solution of the starting equation \eqref{eq:LagEom}.\\
 $O\equiv O_A dX^A$ is a one-form that is null with respect to the background metric $g_{AB}$.\\
It turns out that this starting solution has an event horizon, given by the null hypersurface - ${\cal S}:\psi =1$. We define the function $\psi$ in a way  so that $\psi =1$ is the horizon to all order in $\left(\frac{1}{D}\right)$ expansion. Further it satisfies the following equation (which we shall refer to as `subsidiary condition-1')
\begin{equation}\label{eq:sub2}
\nabla^2\psi^{-D} =0
\end{equation}
\footnote{Throughout this note `$\cdot$' denotes the contraction with respect to the background $g_{AB}$ and $\nabla$ is a covariant derivative with respect to the background. All raising and lowering of indices will also be with respect to the background. Otherwise it would be written explicitly.}We can always determine $\psi$ explicitly in an expansion in $\left(\frac{1}{D}\right)$ solving equation \eqref{eq:sub2} with the initial condition that $\psi =1$ coincides with the horizon \cite{radiation}.
We fix the normalization of $O^A$ by demanding that the inner product between $O_A$ and the unit normal to the $\psi = \text{constant}$ surface (viewed as a hypersurface embedded in the background $g_{AB}$) is always one. In terms of equation this implies
\begin{equation}\label{eq:normalO}
\begin{split}
O\cdot n\equiv\left[\frac{O\cdot \partial\psi}{\sqrt{(\partial\psi)\cdot(\partial\psi)} }\right]=1
\end{split}
\end{equation}
Note that using the above normalization we can define a unit normalized velocity field $u^A$.
\begin{equation}\label{eq:defua}
u^A \equiv -\left(O^A - n^A\right),~~~u\cdot u =-1
\end{equation}
It turns out that $u^A$ is the null generator of the $\psi=1$ hypersurface (viewed as a null hypersurface embedded in $G^{(0)}_{AB}$).\\
 However, just the normalization cannot fix all components of the null one-form $O_A$  everywhere. We fix this ambiguity demanding that $O_A$ satisfies the following geodesic constraint ( which we shall refer to as `subsidiary condition-2')
 \begin{equation}\label{eq:sub1}
\begin{split}
(O\cdot \nabla)O_A \propto O_A,~~O\cdot O =0
\end{split}
\end{equation}
\eqref{eq:sub1} again could be solved in an expansion in $\left(\frac{1}{D}\right)$ provided we have an unambiguous initial condition to all order. We fix this condition by demanding that $u^A$ as  defined in \eqref{eq:defua}  is the null generator of the horizon to all order \cite{radiation}.\\
We shall determine the metric corrections in terms of the well defined $\psi$ and $O_A$ fields and their derivatives.

\subsection{Solution at the first subleading order}\label{results}
As mentioned in the introduction,  $G^{(1)}_{AB}$ - the metric correction at first subleading order has already been determined \cite{arbBack}. For convenience, here we shall quote the first order solution.\\
It turns out that Einstein equations could be solved provided the extrinsic  curvature of the $\psi=1$ hypersurface  (viewed as a hypersurface embedded in the background) and the velocity field $u^A$ together satisfy the following constraint equations on the horizon. The constraint equation can be written as an intrinsic equation to the membrane.
\begin{equation}\label{eq:constraint}
\begin{split}
&{\cal{P}}^\nu_\mu\left[ \frac{\hat\nabla^2 u_\nu}{\cal K} -\frac{\hat\nabla_\nu{\cal K}}{\cal K}+  u_\alpha {\cal K}^\alpha_\nu - (u\cdot\hat\nabla)u_\nu \right] = {\cal O}\left(\frac{1}{D}\right),~~\hat{\nabla}\cdot u={\cal{O}}\left(\frac{1}{D}\right)\\
\text{where}~~&~ {\cal{P}}_{\mu\nu}=\hat{g}_{\mu\nu}+u_\mu u_\nu
\end{split}
\end{equation}
Here $\hat g_{\mu\nu}$ denotes the induced metric on the membrane ($\psi=1$ hypersurface)
and $\hat{\nabla}$ is the covariant derivative with respect to $\hat g_{\mu\nu}$.  The velocity field $u_\mu$ is the pull back of the bulk velocity field $u_A$  and
${\cal{K}}_{\mu\nu}$ is the pull back of the extrinsic curvature of the membrane onto the hypersurface \footnote {In terms of equations what we mean is the following. \\
The space time form of the extrinsic curvature is given by
$$K_{AB}=\Pi^C_A\nabla_A n_B, \text{ where   } \Pi_{AB}=g_{AB}-n_A n_B$$  $u_\mu$ and ${\cal K}_{\mu\nu} $ is defined as 
\begin{equation}\label{extrinsic}
\begin{split}
&u_\mu =\left( \frac{\partial X^A}{\partial y^\mu} \right)u_A,~~~
{\cal K}_{\mu\nu} = \left( \frac{\partial X^M}{\partial y^\mu} \right) \left( \frac{\partial X^N}{\partial y^\nu} \right) K_{MN}
\end{split}
\end{equation}
where $X^M$ denotes the coordinates of the full space time and $y^\mu$ denotes coordinates on the membrane. }. ${\cal{K}}$ is the trace of the extrinsic curvature.

For every solution of the above constraint equations we could determine $G^{(1)}_{AB}$. It turns out $G^{(1)}_{AB}$ simply vanishes given our choice of subsidiary conditions.\\
In this note our goal is to find corrections to equation \eqref{eq:constraint} to the next order in $\left(\frac{1}{D}\right)$ expansion and also  $G^{(2)}_{AB}$. \\

But before getting into any details of the computation, we shall first present our final result .

\subsection{Final Result:\\ Metric and membrane equation at second subleading order}
In this subsection we shall present the subleading correction to the membrane equation \eqref{eq:constraint} and the solution to $G^{(2)}_{AB}$\\
The metric correction would take the following form.
\begin{equation}\label{eq:parameter1}
\begin{split}
G^{(2)}_{AB}& =\bigg[O_A O_B\left(\sum_{n=1}^2 f_n(R)~\mathfrak{s}_n \right) +  t(R)~\mathfrak{t}_{AB} 
+  v(R)~\big(~ \mathfrak{v}_A O_B + \mathfrak{v}_B O_A\big)\bigg]\\
&\text{where}~~ R\equiv D(\psi-1),~~P_{AB} = g_{AB} - n_A n_B +u_A u_B\\
&\text{and,}~~n^A [\mathfrak{v}_n]_A = u^A[\mathfrak{v}_n]_A=0,~~n^A [\mathfrak{t}_n]_{AB} =u^A  [\mathfrak{t}_n]_{AB}=0,~~g^{AB}[\mathfrak{t}_n]_{AB}=0
\end{split}
\end{equation}\\
where
\begin{equation}\label{structure}
\begin{split}
&\mathfrak{t}_{AB}=P^C_A P^{D}_B\bigg[ \bar{R}_{FCDE}O^E O^F+\frac{K}{D}\bigg(K_{CD}-\frac{\nabla_C u_{D}+\nabla_{D}u_C}{2}\bigg)\\
&~~~~~~~~~~~~~~~~~~- P^{EF}(K_{EC}-\nabla_E u_C)(K_{FD}-\nabla_F u_{D})\bigg]\\
&\mathfrak{v}_{A}={P}^B_{A}\Bigg[\frac{K}{D}\left(n^D u^E O^F \bar{R}_{FBDE}\right)+\frac{K^2}{2D^2}\left(\frac{{\nabla}_B {K}}{K}+(u\cdot{\nabla})u_B-2~u^D {K}_{D B}\right)\\
&~~~~~~~~~~~~~-{P}^{F D} \left(\frac{{\nabla}_F {K}}{D}-\frac{K}{D} (u^E {K}_{E F})\right)\left({K}_{D B}-\nabla_D u_B\right)\Bigg]\\
&\mathfrak{s}_1=u^E u^F n^D n^C\bar{R}_{CEFD}+\left(\frac{u\cdot{\nabla}K}{K}\right)^2+\frac{\tilde{\nabla}_A {K}}{K}\left[4~u^B {K}^A_B-2\left[(u\cdot{\nabla})u^A\right]-\frac{\tilde{\nabla}^A {K}}{K}\right]\\
&-(\tilde{\nabla}_A u_B)(\tilde{\nabla}^A u^B)-(u\cdot {K}\cdot u)^2-\left[(u\cdot\tilde{\nabla})u_A\right][(u\cdot\tilde{\nabla})u^A]+2\left[(u\cdot{\nabla})u^A\right](u^B {K}_{BA})\\
&~~~~~~~~~~~~~~-3~(u\cdot {K}\cdot {K}\cdot u)-\frac{K}{D}\left(\frac{u\cdot{\nabla}{K}}{K}-u\cdot {K}\cdot u\right)\\
&\mathfrak{s}_2=\frac{K^2}{D^2}\Bigg[-\frac{{K}}{D}\left(\frac{u\cdot{\nabla}{K}}{K}-u\cdot {K}\cdot u\right)- 2~\lambda- (u\cdot {K}\cdot {K}\cdot u)+2 \left(\frac{{\nabla}_A{K}}{K}\right)u^B {K}^A_B-\left(\frac{u\cdot{\nabla}{K}}{K}\right)^2\\
&+2\left(\frac{u\cdot{\nabla}{K}}{K}\right)(u\cdot {K}\cdot u)-\left(\frac{\tilde{\nabla}^D K}{K}\right)\left(\frac{\tilde{\nabla}_D K}{K}\right)-(u\cdot {K}\cdot u)^2+n^B n^D u^E u^F\bar{R}_{FBDE}\Bigg]
\end{split}
\end{equation}
Where, $\bar{R}_{ABCD}$ is the Riemann tensor\footnote{ Riemann tensor is defined by the relation $$[\nabla_A,\nabla_B]\omega_C=R_{ABC}^{~~~~~~D}\omega_D$$} of the background metric $g_{AB}$ and $\tilde{\nabla}$ is defined as follows:
 for any general  tensor with $n$ indices $W_{A_1A_2\cdots A_n}$
\begin{equation}\label{tildedef}
\tilde{\nabla}_A W_{A_1A_2\cdots A_n}=\Pi_A^C~\Pi_{A_1}^{C_1}\Pi_{A_2}^{C_2}\cdots \Pi_{A_n}^{C_n}\left(\nabla_C W_{C_1C_2\cdots C_n}\right), \quad \text{with}\quad \Pi_{AB}=g_{AB}-n_A n_B
\end{equation}
\begin{equation}\label{funcn}
\begin{split}
&t(R)=-~2\left(\frac{D}{K}\right)^2\int_R^{\infty}\frac{y~dy}{e^y-1}\\
&v(R)=2\left(\frac{D}{K}\right)^3\bigg[\int_R^{\infty}e^{-x}dx\int_0^x\frac{y~e^y}{e^y-1}dy~-~e^{-R}\int_0^{\infty}e^{-x}dx\int_0^x\frac{y~e^y}{e^y-1}dy\bigg]\\
&f_1(R)=-2\left(\frac{D}{K}\right)^2\int_R^{\infty}x~e^{-x}dx+2~e^{-R}\left(\frac{D}{K}\right)^2\int_0^{\infty}x~e^{-x}dx\\
\end{split}
\end{equation}
\begin{equation}
\begin{split}
&f_2(R)=\left(\frac{D}{K}\right)\Bigg[\int_R^{\infty}e^{-x}dx\int_0^x\frac{v(y)}{1-e^{-y}}dy-e^{-R}\int_0^{\infty}e^{-x}dx\int_0^x\frac{v(y)}{1-e^{-y}}dy\Bigg]\\
&~~~~~~~~-\left(\frac{D}{K}\right)^4\Bigg[\int_R^{\infty}e^{-x} dx\int_0^x\frac{y^2~ e^{-y}}{1-e^{-y}}dy-e^{-R}\int_0^{\infty}e^{-x} dx\int_0^x\frac{y^2 ~e^{-y}}{1-e^{-y}}dy\Bigg]
\end{split}
\end{equation}
As we can see that our solution is parametrized by the shape of the constant $\psi$ hypersurfaces (encoded in its extrinsic curvature $K_{AB}$)  along with the velocity field $u^A$. However, because of our subsidiary conditions if we know $K_{AB}$ and $u^A$ along one constant $\psi$ hypersurface, they are determined everywhere else. In this sense the real data our class of solutions are to be provided only along one simple surface; the most natural choice of which is the horizon or the hypersurface $\psi=1$.

As we have mentioned before, we cannot choose any arbitrary shape of the membrane  and velocity field as our initial data. The metric, presented above, would solve  Einstein equation \eqref{eq:ein} only if the data satisfy some constraint - the equation \eqref{eq:constraint} with subleading corrections. This will lead to the following  corrected membrane equation at this order.
\begin{equation}\label{eq:constraint22}
\begin{split}
&\left[\frac{\hat{\nabla}^2 u_\alpha}{\cal{K}}-\frac{\hat{\nabla}_\alpha{\cal{K}}}{\cal{K}}+u^\beta {\cal{K}}_{\beta\alpha}-u\cdot\hat{\nabla} u_\alpha\right]{\cal P}^\alpha_\gamma + \Bigg[-\frac{u^\beta {\cal{K}}_{\beta \delta} {\cal{K}}^\delta_\alpha}{\cal{K}}+\frac{\hat{\nabla}^2\hat{\nabla}^2 u_\alpha}{{\cal{K}}^3}-\frac{(\hat{\nabla}_\alpha{\cal{K}})(u\cdot\hat{\nabla}{\cal{K}})}{{\cal{K}}^3}\\
&-\frac{(\hat{\nabla}_\beta{\cal{K}})(\hat{\nabla}^\beta u_\alpha)}{{\cal{K}}^2}-\frac{2{\cal{K}}^{\delta \sigma}\hat{\nabla}_\delta\hat{\nabla}_\sigma u_\alpha}{K^2} -\frac{\hat{\nabla}_\alpha\hat{\nabla}^2{\cal{K}}}{{\cal{K}}^3}+\frac{\hat{\nabla}_\alpha({\cal{K}}_{\beta\delta} {\cal{K}}^{\beta\delta} {\cal{K}})}{K^3}+3\frac{(u\cdot {\cal{K}}\cdot u)(u\cdot\hat{\nabla} u_\alpha)}{{\cal{K}}}\\
&-3\frac{(u\cdot {\cal{K}}\cdot u)(u^\beta {\cal{K}}_{\beta\alpha})}{{\cal{K}}}
-6\frac{(u\cdot\hat{\nabla}{\cal{K}})(u\cdot\hat{\nabla} u_\alpha)}{{\cal{K}}^2}+6\frac{(u\cdot\hat{\nabla}{\cal{K}})(u^\beta {\cal{K}}_{\beta\alpha})}{{\cal{K}}^2}+3\frac{u\cdot\hat{\nabla} u_\alpha}{D-3}\\
&-3\frac{u^\beta {\cal{K}}_{\beta\alpha}}{D-3}-\frac{(D-1)\lambda}{{\cal{K}}^2}\left(\frac{\hat{\nabla}_\alpha {\cal{K}}}{\cal{K}}-2u^\sigma {\cal{K}}_{\sigma\alpha}+2(u\cdot\hat{\nabla})u_\alpha\right)\Bigg]{\cal P}^\alpha_\gamma = {\cal{O}}\left(\frac{1}{D}\right)^2
\\
\\
&\hat{\nabla}\cdot u - \frac{1}{2{\mathcal K}}\left( \hat{\nabla}_{(\alpha}u_{\beta)}\hat{\nabla}_{(\gamma}u_{\delta)}{\cal P}^{\beta\gamma}{\cal P}^{\alpha\delta} \right)= {\cal{O}}\left(\frac{1}{D}\right)^2 
\end{split}
\end{equation}
Where $\hat{\nabla}$ is the covariant derivative with respect to $\hat{g}_{\mu\nu}$, the induced metric on $\psi=1$ hypersurface. $\cal{K}_{\mu\nu}$  and $u_\mu$ are defined in \eqref{extrinsic}. $\hat{\nabla}_{(\alpha}u_{\beta)}$ is defined as
$$ \hat{\nabla}_{(\alpha}u_{\beta)}\equiv\hat{\nabla}_{\alpha}u_{\beta}+\hat{\nabla}_{\beta}u_{\alpha}$$

\section{Sketch of the  computation}\label{sec:compute}
It turns out that though the computation to determine the second order metric correction is tedious,  conceptually it is a straightforward extension of what has been done in \cite{arbBack}.
Therefore in this section, we shall omit most of the derivations and mention only those where there are some differences from \cite{arbBack}.

We shall follow the same convention as in \cite{arbBack}.  In particular our choice of gauge is also the same, namely
$$O^BG^{(2)}_{AB} =0$$
With this gauge choice the second order correction could be parametrized as
\begin{equation}\label{eq:parameter}
\begin{split}
G^{(2)}_{AB}& =\bigg(O_A O_B\sum_n f_n(R)~\mathfrak{s}_n +\frac{1}{D} P_{AB} \sum_n h_n(R)~\mathfrak{s}_n+ \sum_n t_n(R)~[\mathfrak{t}_n]_{AB} \\
&~~~~~~~~~~~~~~~~~~~~~~~~~+ \sum_n v_n(R)~\big(~ [\mathfrak{v}_n]_A O_B + [\mathfrak{v}_n]_B O_A\big)\bigg)\\
&\text{where}~~ R\equiv D(\psi-1),~~P_{AB} = g_{AB} - n_A n_B +u_A u_B\\
&\text{and,}~~n^A [\mathfrak{v}_n]_A = u^A[\mathfrak{v}_n]_A=0,~~n^A [\mathfrak{t}_n]_{AB} =u^A  [\mathfrak{t}_n]_{AB}=0,~~g^{AB}[\mathfrak{t}_n]_{AB}=0
\end{split}
\end{equation}
Here $\mathfrak{s}_n $,  $[\mathfrak{v}_n]_A$,  $[\mathfrak{t}_n]_{AB} $ are different independent scalar, vector and tensor structures, constructed out of the membrane data. \\
Evaluating  \eqref{eq:ein} on 
$\left[G_{AB} = G^{(0)}_{AB} +\left(\frac{1}{D}\right)G^{(1)}_{AB} +\left(\frac{1}{D}\right)^2G^{(2)}_{AB} + {\cal O}\left(\frac{1}{D}\right)^3 \right]$  upto order ${\cal O}(1)$, we got a set of coupled, ordinary but inhomogeneous differential equation for  the unknown functions in equation \eqref{eq:parameter}.   Boundary conditions for these differential equations are set by the following physical  conditions.
\begin{enumerate}
\item The surface $(\psi=1)$ or $(R=0)$ is the event horizon  and therefore a null hypersurface to all orders.
\item $u^A$ is the null generator of this event horizon to all orders.
\item Bulk metric $G_{AB}$  to all orders approaches $g_{AB}$ as $R\rightarrow\infty$.
\end{enumerate}
These conditions translate to the following constraints on the unknown functions.
\begin{equation}\label{eq:bc}
\begin{split}
&f_n(R=0) = v_n(R=0) =0,~~~~~h_n(R=0) = t_n(R=0) = \text{finite},\\
&\lim_{R\rightarrow\infty}f_n(R)=\lim_{R\rightarrow\infty}h_n(R)=\lim_{R\rightarrow\infty}v_n(R) =\lim_{R\rightarrow\infty}t_n(R)=0
\end{split}
\end{equation}

The homogeneous part $H_{AB}$ (i.e., the part  that acts like a differential operator on the space of unknown functions appearing in $G^{(2)}_{AB}$) is universal. It will have the same form as in the `first order' calculation and we do not need to recalculate it. For convenience, here we shall quote the results for the homogeneous part as derived in \cite{arbBack}.
\begin{equation}\label{eq:decomphomo}
\begin{split}
H_{AB}& \equiv  H^{(1)} O_A O_B + H^{(2)} (n_A O_B + n_B O_A) + H^{(3)} n_A n_B + H^{(tr)} P_{AB}\\
&~~+\left(O_A P^C_B + O_B P^C_A\right) H^{(V_1)}_C  +\left(n_A P^C_B + n_B P^C_A\right) H^{(V_2)}_C+  H^{(T)}_{AB}\\
\end{split}
\end{equation}
where, 
\begin{equation}\label{eq: comphomo}
\begin{split}
&H^{(1)}=-\frac{N^2}{2}(1-e^{-R})\sum_{n}\mathfrak{s}_n \left( f_n^{\prime\prime}+f_n^{\prime} \right)-\frac{N}{2}e^{-R}\sum_{n}\frac{(\nabla\cdot\mathfrak{v}_n)}{D}v_n+\frac{N^2}{4}e^{-R}(1-e^{-R})\sum_n \mathfrak{s}_n h_n^{\prime}\\
&~~~~~~~~H^{(2)}=\frac{N^2}{2}\sum_{n}\mathfrak{s}_n\left(f_n^{\prime\prime}+f_n^{\prime} \right)+\frac{N}{2}\sum_{n}\frac{(\nabla\cdot\mathfrak{v}_n)}{D}v_n^{\prime}-\frac{N^2}{4}e^{-R}\sum_n\mathfrak{s}_n h_n^{\prime}\\
&~~~~~~~~~~~~~~~~~~~~~~~~~~~~~~~~~~H^{(3)}=-\frac{N^2}{2}\sum_n \mathfrak{s}_n h_n^{\prime\prime}\\
&~~~~~~~~~~~~~~~~~~~~~~~~~~~~~~~~~~~~~~~H^{(tr)}=0\\
&~~~~~~~~~~~~~~~~~~~~H^{(V_1)}_C=-\frac{N^2}{2}(1-e^{-R})\sum_n\left(v_n^{\prime\prime}+v_n^{\prime}\right)[\mathfrak{v}_n]_C\\
&~~~~~~~~~~~~~~H^{(V_2)}_C=\frac{N^2}{2}\sum_{n}\left(v_n^{\prime\prime}+v_n^{\prime}\right) [\mathfrak{v}_n]_C+\frac{N}{2D}\sum_{n} t_n^{\prime}\left(\nabla_D[\mathfrak{t}_n]^D_C\right)\\
&~~~~~~~~~~~~~~~~~~~~~H^{(T)}_{AB}=-\frac{N^2}{2}\sum_{n}\left[t_n^{\prime\prime}(1-e^{-R})+t_n^{\prime}\right][\mathfrak{t}_n]_{AB}\\
\end{split}
\end{equation}
Here for any  $R$ dependent function, $X'(R)$ denotes ${dX(R)\over dR}$.

The `source' parts of these equations are determined by evaluating the Einstein equation on the first order corrected metric. By construction the order ${\cal O}(D^2)$ and order ${\cal O}(D)$ pieces of these equations will vanish and first non-zero contribution, relevant for the computation of this note , will be of ${\cal O}(1)$.

From the above discussion it follows that the key  part of the computation  is to determine the source term, which we denote here by $S_{AB}$.
Since $G^{(1)}_{AB}$ vanishes,  just like in \cite{arbBack} here also the source will be given by $E_{AB}$ calculated on $\left(G^{(0)}_{AB}\right)$, however the complication lies in the fact that the calculation has to  be carried out upto order ${\cal O}(1)$.\\
Here we are presenting the final result for the source. See appendix \ref{app:calsource} for the details.
For convenience, we shall decompose $S_{AB}$ into its different components.
\begin{equation}\label{eq: decompsource}
\begin{split}
S_{AB}& \equiv  S^{(1)} O_A O_B + S^{(2)} (n_A O_B + n_B O_A)+S^{(3)}n_A n_B +S^{(tr)} P_{AB}\\
&~~~~~+\left(O_A P^C_B + O_B P^C_A\right) S^{(V_1)}_C+ \left(n_A P^C_B + n_B P^C_A\right) S^{(V_2)}_C +  S^{(T)}_{AB}\\
\text{where}&~~~
O^AS^{(T)}_{AB} = n^A S^{(T)}_{AB} =0,~~~S^{(T)}_{AB}P^{AB}=0~~~\text{and}~~~P_{AB} \equiv g_{AB} +u_A u_B - n_A n_B
\end{split}
\end{equation}
The explicit expression for the different components are the following.
\begin{equation}\label{eq: compsource}
\begin{split}
&S^{(1)}=e^{-2R}\left(\frac{K}{2}\right)E^\text{scalar}
+\left(e^{-R}-e^{-2R}\right)\mathfrak{s}_1+e^{-2R}\left(\frac{R^2}{2}\right)\left(\frac{D}{K}\right)^2\mathfrak{s}_2-R\left(\frac{e^{-2R}}{2}\right)\left({\tilde{\nabla}\cdot E^\text{vector}}\right)_{R=0}\\ \\
&S^{(2)}=e^{-R}\left[-\mathfrak{s}_1+\left(\frac{K}{2}\right)E^{\text{scalar}}\right]_{R=0}-R\left(\frac{e^{-R}}{2}\right)\left({\tilde{\nabla}\cdot E}\right)_{R=0}+e^{-R}\left(\frac{R^2}{2}\right)\bigg[\left(\frac{D^2}{K^2}\right)\mathfrak{s}_2\bigg]_{R=0}\\ \\
&S^{(V_1)}_C=\frac{e^{-R}}{2}\left[KE_C^{\text{vector}}-2~R\left(\frac{D}{K}\right)\mathfrak{v}_C\right],~~~~~S^{(T)}_{AB}=e^{-R}~\mathfrak{t}_{AB}\\
&S^{(3)} = S^{tr} =0,~~~~S_C^{(V_2)} =0
\end{split}
\end{equation}
Where 
\begin{equation}\label{conteq}
\begin{split}
&E^\text{scalar}=\left[\left(\tilde{\nabla}\cdot u\right)\bigg{|}_{\psi=1}-\frac{1}{2K}\left[\nabla_{(A}u_{B)}\nabla_{(C}u_{D)}P^{BC} P^{AD}\right]\right]\\
\end{split}
\end{equation}
\begin{equation}\label{eq:constraint2}
\begin{split}
E_{C}^{\text{vector}}&=\bigg[\frac{\tilde{\nabla}^2 u_A}{{K}}-\frac{\tilde{\nabla}_A{{K}}}{{K}}+u^B {{K}}_{B A}-u\cdot\tilde{\nabla} u_A\bigg]{ P}^A_C\\
&+\bigg[-\frac{u^B {{K}}_{B D} {{K}}^D_A}{{K}}+\frac{\tilde{\nabla}^2\tilde{\nabla}^2 u_A}{{{K}}^3}-\frac{(\tilde{\nabla}_A{{K}})(u\cdot\tilde{\nabla}{{K}})}{{{K}}^3}-\frac{(\tilde{\nabla}_B{{K}})(\tilde{\nabla}^B u_A)}{{{K}}^2}\\
&~~~~-\frac{2{{K}}^{D E}\tilde{\nabla}_D\tilde{\nabla}_E u_A}{K^2}-\frac{\tilde{\nabla}_A\tilde{\nabla}^2{{K}}}{{{K}}^3}+\frac{\tilde{\nabla}_A({{K}}_{BD} {{K}}^{BD} {{K}})}{K^3}+3\frac{(u\cdot {{K}}\cdot u)(u\cdot\tilde{\nabla} u_A)}{{{K}}}
\\
&~~~~-3\frac{(u\cdot {{K}}\cdot u)(u^B {{K}}_{B A})}{{{K}}}-6\frac{(u\cdot\tilde{\nabla}{{K}})(u\cdot\tilde{\nabla} u_A)}{{{K}}^2}+6\frac{(u\cdot\tilde{\nabla}{{K}})(u^B {{K}}_{B A})}{{{K}}^2}\\
&~~~~+3\frac{u\cdot\tilde{\nabla} u_A}{D-3}-3\frac{u^B {{K}}_{B A}}{D-3}-\frac{(D-1)\lambda}{{{K}}^2}\bigg(\frac{\tilde{\nabla}_A {{K}}}{{K}}-2u^D {{K}}_{D A}+2(u\cdot\tilde{\nabla})u_A\bigg)\bigg]{ P}^A_C
\end{split}
\end{equation}
See equation  \eqref{structure} for the definitions of ${\mathfrak{s}_1}$, $\mathfrak{s}_2$, $\mathfrak{v}_C$, $\mathfrak{t}_{AB}$.\\
$\tilde{\nabla}$ is defined as follows:\\
for any general  tensor with $n$ indices $W_{A_1A_2\cdots A_n}$
\begin{equation}\label{tildedef2}
\tilde{\nabla}_A W_{A_1A_2\cdots A_n}=\Pi_A^C~\Pi_{A_1}^{C_1}\Pi_{A_2}^{C_2}\cdots \Pi_{A_n}^{C_n}\left(\nabla_C W_{C_1C_2\cdots C_n}\right)
\end{equation}
The final set of coupled differential equations that we have to solve is simply
 \begin{equation}\label{fineq}
 H_{AB} + S_{AB} = 0
 \end{equation}
As explained in \cite{arbBack}, the homogeneous part $H_{AB}$ could be decoupled after taking  its appropriate projection on different directions. Similar projections applied on $S_{AB}$ will generate the sources for the scalar, vector, tensor and the trace sectors. \\
However, just as in the first order calculation, there is an `integrability' condition. Note that $H^{(1)}$ and $H_C^{(V_1)}$ vanish at $R=0$\footnote{To see the vanishing of $H^{(1)} $ at $R=0$ we have to use the fact   that $v_n(R)$ vanishes at $R=0$ as a consequence of our boundary condition. See equation \eqref{eq:bc}} . Hence consistency demands that $S^{(1)}$ and $S_C^{(V_1)}$ should also vanish on $R=0$.
In other words, these set of equations could be consistently solved only if on the horizon the velocity field $u_A$ and the extrinsic curvature of the $\psi =1$ membrane (viewed as a hypersurface embedded in the background) together satisfy the following equations. 
\begin{equation}\label{constq}
\begin{split}
&S^{(1)}\vert_{R=0} = \left(K\over 2\right) E^\text{scalar}\vert_{R=0}=0\\
&S_C^{(V_1)}\vert_{R=0} =\left(K\over 2\right) E_C^\text{vector}\vert_{R=0} = 0
\end{split}
\end{equation}
By appropriate pull-back these equations could be recast as an intrinsic equation on the hypersurface and they generate the  next order correction to the constraint equation \eqref{eq:constraint}.
We have described them in equations \eqref{eq:constraint22}.

Once the constraint equations are satisfied, we could see that in the source $S_{AB}$ only two scalar structures (${\mathfrak s}_1$ and ${\mathfrak s}_2$), one vector structure (${\mathfrak v}_C$) and one tensor structure (${\mathfrak t}_{AB}$) appear. So altogether we have 6 unknown functions  (2 functions for the scalar sector, 2 in the trace sector, 1 in the vector sector and 1 in the tensor sector)  into solve for. \\
The decoupled ODEs for different unknown metric functions:
\begin{itemize}
\item Scalar sector: \\
 For $h_n(R)$: ~~$H^{(3)} + S^{(3)}=0$
~~~~~~   for $f_n(R)$: ~~$H^{(1)} + S^{(1)}=0$,~~~$n=1,2$
\item  Vector sector:\\
For $v(R)$:~~$H_C^{(V_1)} + S_C^{(V_1)}=0$
\item Tensor sector:\\
For $t(R)$:~~$H^{(T)}_{AB} + S^{(T)}_{AB} =0$
\end{itemize}
Now we shall give the explicit form of the equations  sector by sector.\\

{\textbf{Tensor sector:}}\\
Here the explicit form of the equation is as follows
\begin{equation}
t^{\prime\prime}(1-e^{-R})+t^{\prime}=\frac{2}{N^2}~e^{-R}= 2\left(D\over K\right)^2e^{-R}
\end{equation} 
We can integrate this equation. After imposing 
$$t(R=0) = \text{finite} ~~\text{and }~~\lim_{R\rightarrow\infty}t(R) =0$$
 we find the result as presented in the first equation of \eqref{funcn}.\\

{\textbf{Vector sector:}}\\
Here the explicit form of the equation is as follows
\begin{equation}
(1-e^{-R})\frac{d}{dR}(e^R v^{\prime})+2\left(\frac{D}{K}\right)^3 R=0
\end{equation}
After imposing 
$$v(R=0) = 0 ~~\text{and}~~ \lim_{R\rightarrow\infty}v(R) =0$$
 we find the result as presented in the second  equation of \eqref{funcn}.\\

{\textbf{Trace sector:}}\\
The equations for $h_n(R)$ is simply given by
\begin{equation}
\begin{split}
-\frac{N^2}{2}\sum_{n}h_n^{\prime\prime}\mathfrak{s}_n=0
\end{split}
\end{equation}
Integrating this differential equation with the boundary condition \eqref{eq:bc}, we found correction in the trace sector vanishes i.e., $h_n(R)=0$\\

{\textbf{Scalar sector:}}\\
The equations for $f_1(R)$ and $f_2(R)$ are given by
\begin{equation}
\begin{split}
&e^{-R}(1-e^{-R})~ \frac{d}{dR}\left[e^R f_1^{\prime}\right]=2\left(\frac{D}{K}\right)^2e^{-R}(1-e^{-R})\\
&e^{-R}(1-e^{-R})~ \frac{d}{dR}\left[e^R f_2^{\prime}\right]=-\left(\frac{D}{K}\right)e^{-R} ~v(R)+\left(\frac{D}{K}\right)^4R^2~e^{-2R}
\end{split}
\end{equation}
To derive the second equation we have used the fact (see appendix \ref{divderivation2} for derivation)
\begin{equation}\label{vecdiv2}
\begin{split}
{\nabla\cdot\mathfrak{v}\over D}&=
\mathfrak{s}_2
\end{split}
\end{equation}
After imposing 
$$f_n(R=0) = 0 ~~\text{and}~~ \lim_{R\rightarrow\infty}f_n(R) =0,~~~~n=1,2$$
 we find the result as presented in the third and the fourth  equation of \eqref{funcn}.\\

\section{Checks}\label{sec:check}
In this section we shall perform several checks on our calculation. Roughly the checks could be of two types. The first is the internal consistency of our solutions and the systems of equations, i.e,  to verify that  if we simply substitute our solution in the system of equations \eqref{fineq}, each and every component of it vanishes upto corrections of order ${\cal O}\left(1\over  D\right)$. The details of it would be presented in subsection \ref{check1}. 

The second type of checks are the ones where we shall take several limits and match our results with some answers, known previously.
One trivial check in this category that we have tried on every stage of our computation is to match with the known results in asymptotically flat case \cite{yogesh1}, by setting the cosmological constant  $\Lambda$ to zero. The corrected constraint equation \eqref{eq:constraint22} manifestly matches with equation no (4.5) and (4.12) respectively of \cite{yogesh1},  if we set $\Lambda$ to zero. At this stage it is difficult to match the two metrics even after setting $\Lambda$ to zero, since our subsidiary conditions are different from that of \cite{yogesh1} and we leave it for future.

The other significant check that we have performed is the matching of the spectrum of linearized fluctuation derived from our constraint equations to that of the Quasi-Normal modes already calculated in \cite{QNM:Emparan}. In subsection \ref{check2}  we shall give the details of this computation.

\subsection{Check for internal consistency}\label{check1}
In this subsection we shall explicitly verify that our solution for the metric along with the membrane equations constraining the membrane data, does satisfy equation \eqref{fineq} i.e., each of its components vanishes upto corrections of order ${\cal O}\left(1\over D\right)$.

Let ${\cal E}_{AB}$ denote the LHS of equation \eqref{fineq}.
$${\cal E}_{AB} \equiv H_{AB} + S_{AB}$$
From the list of the decoupled ODEs (see the discussion below equation \eqref{fineq}) it is clear that the 4 of the 7 independent components of ${\cal E}_{AB}$ must be satisfied since we have solved for the metric functions by integrating them. These components are 
$$u^A u^B {\cal E}_{AB},~~O^A O^B {\cal E}_{AB}, ~~u^A P^{C} _B{\cal E}_{AC},~~P^C_AP^{C'}_B\left[{\cal E}_{CC'} - \left(\frac{{\cal E}}{D-2}\right)P_{CC'}\right]$$
where ${\cal E}$ denotes the projected trace of ${\cal E}_{AB} $  i.e.,  ${\cal E} = P^{AB}{\cal E}_{AB}$ \\
From the explicit expressions of $H_{AB}$ it is clear that $u^A H_{AB} u^B = H^{(1)}$ and $u^A H_{AC}P^C_B = H^{(V_1)}_B$ vanish at $\psi =1$ and membrane equations ensure that the same is true for the source.\\
As explained in \cite{arbBack}, if we consider `the variation of the metric as we go away from the horizon' as `dynamics', 
then the membrane equations play the role of  `constraint equations', whereas the equations we solved to determine the metric corrections are like the `dynamical' ones.
Now in any theory of gravity, it is enough to solve the `dynamical equations' everywhere and the constraint equation only along one constant `time slice' (in our case which would be a constant $\psi$ slice);   gauge invariance will ensure that the full set of equations are solved everywhere \cite{WaldBook}. This theorem guarantees that the rest of the three independent components of ${\cal E}_{AB}$ must vanish provided we have solved the equations correctly.  These components
\begin{equation*}
\begin{split}
&u^A O^B {\cal E}_{AB} = H^{(2)} + S^{(2)}\equiv{\cal E}^{(2)}\\
&\frac{1}{D}P^{AB} {\cal E}_{AB} = H^{(tr)} + S^{(tr)}\equiv {\cal E}^{(tr)}\\
&O^A P^C_B {\cal E}_{AC} = H^{(V_2)}_B + S^{(V_2)}_B\equiv {\cal E}^{(V_2)}_B
\end{split}
\end{equation*}
 Therefore the fact that these components do vanish on our solution is an important consistency check of our whole procedure and the final answer. Computationally it  turns out to be quite non-trivial.  In fact we have to take help from Mathematica to prove them.
 
 \subsubsection{Vanishing of ${\cal E}^{(2)}$}
From eq \eqref{eq: comphomo} it follows that 
 \begin{equation}
 \begin{split}
 H^{(2)}&=\frac{1}{2}\left(K\over D\right)^2\sum_{n=1}^2\mathfrak{s}_n\left(f_n^{\prime\prime}+f_n^{\prime}  -{e^{-R}\over 2}h_n^{\prime}\right)+\left(K\over D\right)\left(\frac{\nabla\cdot\mathfrak{v}}{2D}\right)v^{\prime}\\
 &=\frac{1}{2}\left(K\over D\right)^2\mathfrak{s}_1(f_1^{\prime\prime}+f_1^{\prime})+\frac{1}{2}\left(K\over D\right)\mathfrak{s}_2\left[N\left(f_2^{\prime\prime}+f_2^{\prime}\right)+v^{\prime}\right] \\
 &=e^{-R}\mathfrak{s}_1+\frac{1}{2}\left(K\over D\right)\mathfrak{s}_2\left[-\frac{e^{-R}}{1-e^{-R}}~v+\left(D\over K\right)^3\frac{R^2~ e^{-R}}{1-e^{-R}}+v^{\prime}\right]\\
&=e^{-R} \mathfrak{s}_1-\frac{1}{2}\left(D\over K\right)^2~R^2e^{-R} ~\mathfrak{s}_2
 \end{split}
 \end{equation}
 Here we have used the fact that metric correction in the trace sector  (i.e., $h_n(R)$) vanishes. Also we have used equation \eqref{vecdiv2}   for the divergence of  ${\mathfrak v}_C$ and the last three equations from \eqref{funcn} for the expressions of $f_n(R)$ and $v(R)$.\\
From equation \eqref{eq: compsource} we could  see that $H^{(2)}$ is exactly the minus of $S^{(2)}$ as required.

 \subsubsection{Vanishing of ${\cal E}^{(tr)}$}
 This follows trivially from \eqref{eq: compsource} and \eqref{eq: comphomo}, as both $S^{(tr)}$ and $H^{(tr)}$ vanish at this order.
  \subsubsection{Vanishing of ${\cal E}^{(V_2)}_B$}
 From equation \eqref{eq: compsource} we see that  $S_C^{(V_2)}=0$, therefore $H_C^{(V_2)}$ should also vanish on our solution. The equation below checks that this is true.
\begin{equation}
\begin{split}
H_C^{(V_2)}&\equiv\frac{1}{2}\left(K\over D\right)^2\left(v^{\prime\prime}+v^{\prime}\right) \mathfrak{v}_C+\frac{1}{2}\left(K\over D\right) t^{\prime}\frac{\left(\nabla_D\mathfrak{t}^D_C\right)}{D}\\
&=\frac{1}{2}\left(K\over D\right)\left[\left(K\over D\right)(v^{\prime\prime}+v^{\prime} )+t^{\prime}\right]\mathfrak{v}_C\\
&=0
\end{split}
\end{equation}
In the second line we have used the identity (see Appendix \ref{divderivation} for the derivation),
\begin{equation}\label{eq:divtensor2}
\nabla_D\left(\mathfrak{t}^D_C\right)=D~ \mathfrak{v}_C
\end{equation}
In the last line we have used the first and the second equation of \eqref{funcn} for the expressions of $v(R)$ and $t(R)$.

\subsection{Quasinormal Modes for Schwarzschild black hole in background AdS/dS spacetime}\label{check2}

Now as a check for our membrane equations, we will calculate the light quasinormal mode frequencies for Schwarzschild black hole in AdS/dS background. As expected, we find that the answers for the frequencies of light quasinormal modes match exactly with those derived in \cite{QNM:Emparan} from gravitational analysis. 

 As before, we shall follow \cite{arbBack} for the computation. Many steps and arguments  are exactly same as in \cite{arbBack}. For such steps we shall simply refer to \cite{arbBack} or  quote them in the appendix. And here we shall present only those parts of computation where we have to do some extension of what has been done in \cite{arbBack}.
 
We shall write the  background AdS/dS in global coordinates as
\begin{equation}\label{globmet}
 ds^2_{(bgd)} = g_{AB} dX^A dX^B= -\left( 1-\sigma\frac{r^2}{L^2} \right)dt^2 + \frac
 {dr^2}{\left( 1-\sigma\frac{r^2}{L^2} \right)} + r^2 d\Omega^2_{D-2}.
\end{equation}
Where
\begin{equation}
\begin{split}
\Lambda &= \frac{\sigma}{L^2} (D-1)(D-2)\\
L&=\text{AdS/dS radius}\\
 \sigma &= 0 \quad \text{for Flat} \\
 &= 1 \quad \text{for dS} \\
 &= -1 \quad \text{for AdS}
 \end{split}
\end{equation}
And the Schwarzschild black hole in this coordinate system is
\begin{equation}\label{eq:blackstatic}
ds^2_{(BH)} =  -\left( 1-\sigma\frac{r^2}{L^2} - \left( \frac{r_0}{r}\right)^{D-3}  \right)dt^2 + \frac
 {dr^2}{\left( 1-\sigma\frac{r^2}{L^2}- \left( \frac{r_0}{r}\right)^{D-3} \right)} + r^2 d\Omega^2_{D-2}.
\end{equation}
Where, $r_0$ is an arbitrary constant. Note that the position of  horizon is $r =r_H$  where $r_H$ is the zero of the function $f(r) =\left( 1-\sigma\frac{r^2}{L^2} - \left( \frac{r_0}{r}\right)^{D-3}  \right)$.
\begin{equation}\label{euc}
r_H = r_0 \left( 1 - \frac{1}{D}\ln\left(1-\frac{\sigma r_0^2}{L^2}\right) + {\cal O}(D^{-2}) \right)
\end{equation}
From now on we choose $r_H=1$  or in other words $r_0$ will be set to
$$r_0 =  \left( 1 + \frac{1}{D}\ln\left(1-\frac{\sigma }{L^2}\right) + {\cal O}(D^{-2}) \right) $$ for convenience. We will later reinstate the factors of $r_0$ from dimensional analysis.

A small fluctuation around a static black hole corresponds to a small fluctuation around a spherical membrane along with a small fluctuation in the velocity field, which is purely in the time direction at zeroth order. We will work upto linear order in the amplitude of fluctuations, which we denote by $\epsilon$.
\begin{equation}\label{bhflu}
 \begin{split}
  r &= 1 + \epsilon~ \delta r(t,a) \\
  u &= u_0 ~dt + \epsilon ~\delta u_\mu (t,a) dx^\mu
 \end{split}
\end{equation}
Here, we denote the angle coordinates along $(D-2)$ dimensional sphere by $a$ and the coordinates $\mu$ on the membrane worldvolume contain time $t$ and angles $a$.
The induced metric on the membrane worldvolume (viewed as a hypersurface embedded in the background metric \eqref{globmet}) upto linear order in $\epsilon$ is (where we denote the metric components by $g_{\mu\nu}^{(ind)}$)
\begin{equation}\label{lind}
 ds^2_{(ind)} =g_{\mu\nu}^{(ind)}dy^{\mu}dy^{\nu}= -\left( 1-\sigma\frac{1+2\epsilon \delta r}{L^2} \right)dt^2 + (1+2\epsilon \delta r) d\Omega^2_{D-2}
\end{equation}
Also, $u_\mu g^{\mu\nu}_{(ind)} u_\nu=-1$ implies
\begin{equation}
 \begin{split}
& u_0 = -\left( 1-\frac{\sigma}{L^2} \right)^{\frac{1}{2}} ~~\text{and}~~~
  \delta u_t(t,a) = \left( 1-\frac{\sigma}{L^2} \right)^{-\frac{1}{2}}\left(\frac{\sigma}{L^2}\right)\delta r(t,a)
 \end{split}
\end{equation}
The membrane equations are
\begin{equation}\label{membrane}
\hat{\nabla}\cdot u = \frac{1}{2{\mathcal K}}\left( \hat{\nabla}_{(\alpha}u_{\beta)}\hat{\nabla}_{(\gamma}u_{\delta)}{\cal P}^{\beta\gamma}{\cal P}^{\alpha\delta} \right)
\end{equation}
\begin{equation}\label{2ndorderfinal}
\begin{split}
& E^{tot}_\mu \equiv\left[\frac{\hat{\nabla}^2 u_\alpha}{\cal{K}}-\frac{\hat{\nabla}_\alpha{\cal{K}}}{\cal{K}}+u^\beta {\cal{K}}_{\beta\alpha}-u\cdot\hat{\nabla} u_\alpha\right]{\cal P}^\alpha_\gamma + \Bigg[-\frac{u^\beta {\cal{K}}_{\beta \delta} {\cal{K}}^\delta_\alpha}{\cal{K}}+\frac{\hat{\nabla}^2\hat{\nabla}^2 u_\alpha}{{\cal{K}}^3}\\
&-\frac{(\hat{\nabla}_\alpha{\cal{K}})(u\cdot\hat{\nabla}{\cal{K}})}{{\cal{K}}^3}-\frac{(\hat{\nabla}_\beta{\cal{K}})(\hat{\nabla}^\beta u_\alpha)}{{\cal{K}}^2}-\frac{2{\cal{K}}^{\delta \sigma}\hat{\nabla}_\delta\hat{\nabla}_\sigma u_\alpha}{K^2} -\frac{\hat{\nabla}_\alpha\hat{\nabla}^2{\cal{K}}}{{\cal{K}}^3}+\frac{\hat{\nabla}_\alpha({\cal{K}}_{\beta\delta} {\cal{K}}^{\beta\delta} {\cal{K}})}{K^3}\\
&+3\frac{(u\cdot {\cal{K}}\cdot u)(u\cdot\hat{\nabla} u_\alpha)}{{\cal{K}}}-3\frac{(u\cdot {\cal{K}}\cdot u)(u^\beta {\cal{K}}_{\beta\alpha})}{{\cal{K}}}
-6\frac{(u\cdot\hat{\nabla}{\cal{K}})(u\cdot\hat{\nabla} u_\alpha)}{{\cal{K}}^2}+6\frac{(u\cdot\hat{\nabla}{\cal{K}})(u^\beta {\cal{K}}_{\beta\alpha})}{{\cal{K}}^2}\\
&+3\frac{u\cdot\hat{\nabla} u_\alpha}{D-3}-3\frac{u^\beta {\cal{K}}_{\beta\alpha}}{D-3}-\frac{(D-1)\lambda}{{\cal{K}}^2}\left(\frac{\hat{\nabla}_\alpha {\cal{K}}}{\cal{K}}-2u^\sigma {\cal{K}}_{\sigma\alpha}+2(u\cdot\hat{\nabla})u_\alpha\right)\Bigg]{\cal P}^\alpha_\gamma = 0 
\end{split}
\end{equation}
In \eqref{membrane} and \eqref{2ndorderfinal}, the covariant derivative with respect to metric \eqref{lind} is denoted by $\hat\nabla$. The extrinsic curvature of membrane is denoted by ${\cal K}_{\mu\nu}$ and its trace by ${\cal K}$. The projector orthogonal to $u_\mu$ is denoted by ${\cal{P}}_{\mu\nu}$.

It turns out that $E_t^{tot}$ vanishes at linear order in $\epsilon$.
Using \eqref{eq:list1} and \eqref{eq:list2}, we evaluate the vector membrane equation in the angular directions 
\begin{equation}\label{lineq}
\begin{split}
& E_a^{tot} \equiv  \left[ (D-2)-\left( 1-\frac{\sigma}{L^2} \right)^{-1}\frac{\sigma}{L^2} \right]^{-1}
\bigg[-\left( 1-\frac{\sigma}{L^2} \right)^{-\frac{1}{2}}(\epsilon \partial_t^2 \delta u_a) + \left( 1-\frac{\sigma}{L^2} \right)^{-1}\frac{\sigma}{L^2} (\epsilon \partial_t \bar\nabla_a \delta r) \\&+ \left( 1-\frac{\sigma}{L^2} \right)^{\frac{1}{2}} \epsilon \bar\nabla^2 \delta u_a + \epsilon \partial_t \bar\nabla_a \delta r\bigg] 
- \left[ (D-2)-\left( 1-\frac{\sigma}{L^2} \right)^{-1}\frac{\sigma}{L^2} \right]^{-1} \times 
\\ & \bigg[ \left( 1-\frac{\sigma}{L^2} \right)^{-1}(\epsilon \partial^2_t \bar\nabla_a \delta r)-\left( 1-\frac{\sigma}{L^2} \right)^{-1}\frac{\sigma}{L^2}(\epsilon \bar\nabla_a \delta r) -\epsilon \bar\nabla_a \bar\nabla^2 \delta r - (D-2)(\epsilon \bar\nabla_a \delta r) \bigg] 
\\ & +\left[ \left( 1-\frac{\sigma}{L^2} \right)^{-1} (-\epsilon \partial_t \bar\nabla_a \delta r) + \left( 1-\frac{\sigma}{L^2} \right)^{\frac{1}{2}}(\epsilon \delta u_a) +\left( 1-\frac{\sigma}{L^2} \right)^{-\frac{1}{2}}\frac{\sigma}{L^2} \delta u_a \right] 
\\ & -\left[ \left( 1-\frac{\sigma}{L^2} \right)^{-\frac{1}{2}}(\epsilon \partial_t \delta u_a)-\left( 1-\frac{\sigma}{L^2} \right)^{-1}\frac{\sigma}{L^2}(\epsilon \bar\nabla_a \delta r) \right]
 \\ &-\frac{1}{D-2}  \bigg[\left( 1-\frac{\sigma}{L^2} \right)^{-2}\frac{\sigma}{L^2}\epsilon \partial_t \bar\nabla_a \delta r - \left( 1-\frac{\sigma}{L^2} \right)^{-1}\epsilon \partial_t \bar\nabla_a \delta r + \left( 1-\frac{\sigma}{L^2} \right)^{\frac{1}{2}} \epsilon \delta u_a \\&- \left(\frac{\sigma}{L^2}\right)^2\left( 1-\frac{\sigma}{L^2} \right)^{-\frac{3}{2}}\delta u_a \bigg] 
+ \frac{1}{(D-2)^3}\left( 1-\frac{\sigma}{L^2} \right)^{-\frac{3}{2}}\left[\bar\nabla^2\bar\nabla^2 \delta u_a\right] \\ &-\frac{2}{(D-2)^2} \left( 1-\frac{\sigma}{L^2} \right)^{-\frac{1}{2}}\epsilon\hat\nabla^2 \delta u_a  
 - \frac{1}{(D-2)^3} \left( 1-\frac{\sigma}{L^2} \right)^{-2}\left[ - \hat{\nabla}_a\hat{\nabla}^2\hat{\nabla}^2 \delta r - (D-2) \hat{\nabla}_a\hat{\nabla}^2 \delta r \right] 
\\ &+ \frac{1}{(D-2)^3}\left( 1-\frac{\sigma}{L^2} \right)^{-1} \left[ -3(D-2) \epsilon\left( \hat{\nabla}_a\hat{\nabla}^2 \delta r+(D-2)\hat{\nabla}_a\delta r \right) \right] 
\\ &+\frac{3}{D-2} \left[ \left( 1-\frac{\sigma}{L^2} \right)^{-\frac{3}{2}} \frac{\sigma}{L^2} \epsilon \partial_t \delta u_a - \left( 1-\frac{\sigma}{L^2} \right)^{-2}\left(\frac{\sigma}{L^2}\right)^2 \epsilon \hat{\nabla}_a \delta r \right] 
\\ &- \frac{3}{D-2}\left( 1-\frac{\sigma}{L^2} \right)^{-\frac{1}{2}} \left[ -\left( 1-\frac{\sigma}{L^2} \right)^{-\frac{3}{2}}\left(\frac{\sigma}{L^2}\right) \epsilon \partial_t \hat{\nabla}_a \delta r + \frac{\sigma}{L^2} \epsilon \delta u_a + \left(\frac{\sigma}{L^2}\right)^2\left( 1-\frac{\sigma}{L^2} \right)^{-1}\delta u_a  \right] 
\\ &+ \frac{3}{D-2} \left[ \left( 1-\frac{\sigma}{L^2} \right)^{-\frac{1}{2}}(\epsilon \partial_t \delta u_a)-\left( 1-\frac{\sigma}{L^2} \right)^{-1}\frac{\sigma}{L^2}(\epsilon \bar\nabla_a \delta r) \right] 
\\ &-\frac{3}{D-2} \left[ \left( 1-\frac{\sigma}{L^2} \right)^{-1} (-\epsilon \partial_t \bar\nabla_a \delta r) + \left( 1-\frac{\sigma}{L^2} \right)^{\frac{1}{2}}(\epsilon \delta u_a) + \frac{\sigma}{L^2}\left( 1-\frac{\sigma}{L^2} \right)^{-\frac{1}{2}}\delta u_a \right] 
\\ &-\frac{2}{D-2}\frac{\sigma}{L^2} \left[ \left( 1-\frac{\sigma}{L^2} \right)^{-\frac{3}{2}}(\epsilon \partial_t \delta u_a)-\left( 1-\frac{\sigma}{L^2} \right)^{-2}\frac{\sigma}{L^2}(\epsilon \bar\nabla_a \delta r) \right] 
\\ & +\frac{2}{D-2} \frac{\sigma}{L^2} \left[ \left( 1-\frac{\sigma}{L^2} \right)^{-2} (-\epsilon \partial_t \bar\nabla_a \delta r) + \left( 1-\frac{\sigma}{L^2} \right)^{-\frac{1}{2}}(\epsilon \delta u_a)+ \frac{\sigma}{L^2}\left( 1-\frac{\sigma}{L^2} \right)^{-\frac{3}{2}}\delta u_a \right] 
\\ & -\frac{1}{D-2} \frac{\sigma}{L^2}\left( 1-\frac{\sigma}{L^2} \right)^{-2} \bigg[ \frac{1}{D-2} \bigg(-\epsilon \bar\nabla_a \bar\nabla^2 \delta r - (D-2) \epsilon \bar\nabla_a \delta r\bigg)  \bigg]
\end{split}
\end{equation}
\newpage
Where, in \eqref{lineq} we have neglected the terms which are order ${\cal O}\left(1/D^2\right)$ or higher. We denote the covariant derivative with respect to a unit sphere metric in $D-2$ dimensions by $\bar\nabla_a$.
Similarly we evaluate the membrane equation \eqref{membrane}
\begin{equation}\label{veldiv}
 \hat{\nabla}.u  = \epsilon \bar\nabla^a \delta u_a + \epsilon \left( 1-\frac{\sigma}{L^2} \right)^{-\frac{1}{2}} (\partial_t \delta r)(D-2) = 0
\end{equation}
 We choose to divide the fluctuation $\delta u_a$ in two parts (see Section (5) of \cite{Chmembrane})
\begin{equation}\label{velcomp}
 \delta u_a = \delta v_a + \bar\nabla_a \Phi~,\quad \text{with}\quad\bar\nabla^a \delta v_a =0 
\end{equation}
Substituting \eqref{velcomp} into \eqref{veldiv} we find 
\begin{equation}\label{paru}
 \bar\nabla^2\Phi = -(D-2)\left( 1-\frac{\sigma}{L^2} \right)^{-\frac{1}{2}} (\partial_t \delta r) 
\end{equation}
Now we evaluate $\bar\nabla^aE_a^{tot}$. We use the identity $\bar\nabla^a\bar\nabla^2V_a = ((D-3)+\bar\nabla^2)\bar\nabla^aV_a$ for simplification. We find 

\begin{equation*}
\begin{split}
& \bar\nabla^a E_a^{tot} \equiv  \left[ (D-2)-\left( 1-\frac{\sigma}{L^2} \right)^{-1}\frac{\sigma}{L^2} \right]^{-1}
\bigg[(D-2)\left( 1-\frac{\sigma}{L^2} \right)^{-1}(\epsilon \partial_t^3 \delta r) 
\\ & + \left( 1-\frac{\sigma}{L^2} \right)^{-1}\frac{\sigma}{L^2} (\epsilon \partial_t \bar\nabla^2 \delta r)
 -(D-2) \epsilon \partial_t(D-3+\bar \nabla^2)\delta r + (\epsilon \partial_t \bar\nabla^2 \delta r)\bigg] 
\\ &- \left[ (D-2)-\left( 1-\frac{\sigma}{L^2} \right)^{-1}\frac{\sigma}{L^2} \right]^{-1} 
 \bigg[ \left( 1-\frac{\sigma}{L^2} \right)^{-1}(\epsilon \partial^2_t \bar\nabla^2 \delta r)-\left( 1-\frac{\sigma}{L^2} \right)^{-1}\frac{\sigma}{L^2}\epsilon \bar\nabla^2 \delta r
\\ & -\epsilon \bar\nabla^2 \bar\nabla^2 \delta r - (D-2) \epsilon \bar\nabla^2 \delta r \bigg] +\frac{2}{(D-2)}\left( 1-\frac{\sigma}{L^2} \right)^{-1}\left[ \epsilon(D-3+\hat\nabla^2) \partial_t\delta r \right] 
\\ & +\left[ \left( 1-\frac{\sigma}{L^2} \right)^{-1} (-\epsilon \partial_t \bar\nabla^2 \delta r) -(D-2)\epsilon \partial_t\delta r -(D-2)\left( 1-\frac{\sigma}{L^2} \right)^{-1}\frac{\sigma}{L^2} \partial_t\delta r \right] 
\\ & -\left[ -(D-2)\left( 1-\frac{\sigma}{L^2} \right)^{-1}(\epsilon \partial^2_t \delta r)-\left( 1-\frac{\sigma}{L^2} \right)^{-1}\frac{\sigma}{L^2}(\epsilon \bar\nabla^2 \delta r) \right] 
\\ &-\frac{1}{D-2} \bigg[\left( 1-\frac{\sigma}{L^2} \right)^{-2}\frac{\sigma}{L^2}\epsilon \partial_t \bar\nabla^2 \delta r - \left( 1-\frac{\sigma}{L^2} \right)^{-1}\epsilon \partial_t \bar\nabla^2 \delta r 
-(D-2) \epsilon \partial_t\delta r 
\\&+(D-2) \left(\frac{\sigma}{L^2}\right)^2\left( 1-\frac{\sigma}{L^2} \right)^{-2}\partial_t\delta r \bigg] 
\\ & - \frac{1}{(D-2)^2}\left( 1-\frac{\sigma}{L^2} \right)^{-2}\epsilon\left[(D-3+\bar\nabla^2)(D-3+\bar\nabla^2) \partial_t\delta r\right] 
\\ & - \frac{1}{(D-2)^3}\left( 1-\frac{\sigma}{L^2} \right)^{-2} \left[ - \hat{\nabla}^2\hat{\nabla}^2\hat{\nabla}^2 \delta r - (D-2) \hat{\nabla}^2\hat{\nabla}^2 \delta r \right] 
 \\ &- \frac{1}{(D-2)^2}\left( 1-\frac{\sigma}{L^2} \right)^{-1} \left[ 3 \epsilon\left( \hat{\nabla}^2\hat{\nabla}^2 \delta r+(D-2)\hat{\nabla}^2\delta r \right) \right] 
\\ &+\frac{3}{D-2} \left( 1-\frac{\sigma}{L^2} \right)^{-2} \left[ -(D-2) \frac{\sigma}{L^2} \epsilon \partial^2_t \delta r - \left(\frac{\sigma}{L^2}\right)^2 \epsilon \hat{\nabla}^2 \delta r \right] 
\end{split}
\end{equation*}
\begin{equation}\label{eqdiv}
\begin{split}
 &- \frac{3}{D-2} \bigg[ -\left( 1-\frac{\sigma}{L^2} \right)^{-2}\left(\frac{\sigma}{L^2}\right) \epsilon \partial_t \hat{\nabla}^2 \delta r -(D-2) \frac{\sigma}{L^2}\left( 1-\frac{\sigma}{L^2} \right)^{-1} \epsilon \partial_t \delta r 
\\ &-(D-2) \left(\frac{\sigma}{L^2}\right)^2\left( 1-\frac{\sigma}{L^2} \right)^{-2}\partial_t\delta r  \bigg] 
+ \frac{3}{D-2} \left( 1-\frac{\sigma}{L^2} \right)^{-1}\left[ -(D-2)(\epsilon \partial^2_t \delta r)-\frac{\sigma}{L^2}(\epsilon \bar\nabla^2 \delta r) \right] 
\\ &-\frac{3}{D-2} \left[ \left( 1-\frac{\sigma}{L^2} \right)^{-1} (-\epsilon \partial_t \bar\nabla^2 \delta r) -(D-2)(\epsilon \partial_t\delta r) -(D-2) \frac{\sigma}{L^2}\left( 1-\frac{\sigma}{L^2} \right)^{-1}\partial_t\delta r \right] 
\\ &-\frac{2}{D-2}\frac{\sigma}{L^2}\left( 1-\frac{\sigma}{L^2} \right)^{-2} \left[ -(D-2)(\epsilon \partial^2_t \delta r)-\frac{\sigma}{L^2}(\epsilon \bar\nabla^2 \delta r) \right] \\ & +\frac{2}{D-2} \frac{\sigma}{L^2}\left( 1-\frac{\sigma}{L^2} \right)^{-2} \bigg[  (-\epsilon \partial_t \bar\nabla^2 \delta r) -(D-2)\left( 1-\frac{\sigma}{L^2} \right)^{1}(\epsilon \partial_t\delta r)-(D-2) \frac{\sigma}{L^2}\partial_t\delta r \bigg] 
\\ & -\frac{1}{D-2} \frac{\sigma}{L^2}\left( 1-\frac{\sigma}{L^2} \right)^{-2} \bigg[ \frac{1}{D-2} \bigg((-\epsilon \bar\nabla^2 \bar\nabla^2 \delta r) - (D-2) (\epsilon \bar\nabla^2 \delta r)\bigg)  \bigg]
\end{split}
\end{equation}
Now we reinstate the factors of $r_H$. \footnote{We use the dimensional analysis to replace $L$ by $\frac{L}{r_H}$ and replace $\omega^s$ by $\omega^s r_H$. Where, $r_H$ is defined in terms of $r_0$ in \eqref{euc}.} We expand the shape fluctuations
\begin{equation}\label{rsph}
\delta r= \sum_{l,m}a_{lm} Y_{lm} e^{-i \omega^s_l r_H t}
\end{equation} 
where, $Y_{lm}$ are the scalar spherical harmonics on $S^{D-2}$ for which 
\begin{equation}
\bar\nabla^2 Y_{lm} = -l(D+l-3) Y_{lm}.
\end{equation} 
Now, we substitute \eqref{rsph} in \eqref{eqdiv} and solve for the scalar QNM frequencies
\begin{equation}\label{sqf}
\begin{split}
\omega^s r_0 &= \pm \sqrt{l\left(1-\frac{\sigma r_0^2}{L^2}\right)-1}\Bigg[ 1 + \frac{1}{2D}\frac{l-1}{l-\left(1-\frac{\sigma r_0^2}{L^2}\right)^{-1}} \Bigg(\left(2\left(1-\frac{\sigma r_0^2}{L^2}\right)^{-1}+1\right)l \\ &-4\left(1-\frac{\sigma r_0^2}{L^2}\right)^{-1}+2\left(1-\frac{\sigma r_0^2}{L^2}\right)^{-1}\ln\left(1-\frac{\sigma r_0^2}{L^2}\right)\Bigg)  \Bigg] \\ &- i(l-1)\left[1+\frac{1}{D}\left( l-2+\ln\left(1-\frac{\sigma r_0^2}{L^2}\right) \right)\right]
\end{split}
\end{equation}
Upto the required order, the answer \eqref{sqf} agrees with the respective answer given in equations (D.3),(D.4) of \cite{QNM:Emparan}.

Similarly we now calculate the vector QNM frequencies. Note that we have solved \eqref{eqdiv}. So, the $\delta r$ and $\Phi$ terms in \eqref{lineq} will drop out and we have
\begin{equation}\label{eqvel}
\begin{split}
& E_a^{tot} \equiv  \left[ (D-2)\left( 1-\frac{\sigma}{L^2} \right)^{\frac{1}{2}}-\left( 1-\frac{\sigma}{L^2} \right)^{\frac{-1}{2}}\frac{\sigma}{L^2} \right]^{-1} \left[-\left( 1-\frac{\sigma}{L^2} \right)^{-1}(\epsilon \partial_t^2 \delta v_a)  + \epsilon \bar\nabla^2 \delta v_a \right]  
\\ & +\left[  \left( 1-\frac{\sigma}{L^2} \right)^{\frac{1}{2}}(\epsilon \delta v_a) +\left( 1-\frac{\sigma}{L^2} \right)^{-\frac{1}{2}}\frac{\sigma}{L^2} \delta v_a \right] 
 -\left[ \left( 1-\frac{\sigma}{L^2} \right)^{-\frac{1}{2}}(\epsilon \partial_t \delta v_a) \right] 
\\ &-\frac{1}{D-2} \left( 1-\frac{\sigma}{L^2} \right)^{-\frac{1}{2}} \bigg[ \left( 1-\frac{\sigma}{L^2} \right) \epsilon \delta v_a - \left(\frac{\sigma}{L^2}\right)^2\left( 1-\frac{\sigma}{L^2} \right)^{-1}\delta v_a \bigg] 
\\ &+ \frac{1}{(D-2)^3}\left( 1-\frac{\sigma}{L^2} \right)^{-\frac{3}{2}}\left[\bar\nabla^2\bar\nabla^2 \delta v_a\right] -\frac{2}{(D-2)^2}\left( 1-\frac{\sigma}{L^2} \right)^{-\frac{1}{2}}\left[ \epsilon\hat\nabla^2 \delta v_a \right]  
\\ &+\frac{3}{D-2} \left( 1-\frac{\sigma}{L^2} \right)^{-\frac{3}{2}} \left[ \frac{\sigma}{L^2} \epsilon \partial_t \delta v_a \right] + \frac{3}{D-2} \left[ \left( 1-\frac{\sigma}{L^2} \right)^{-\frac{1}{2}}(\epsilon \partial_t \delta v_a)\right]
\\ &- \frac{3}{D-2}\left( 1-\frac{\sigma}{L^2} \right)^{-\frac{1}{2}} \left[ \frac{\sigma}{L^2} \epsilon \delta v_a + \left(\frac{\sigma}{L^2}\right)^2\left( 1-\frac{\sigma}{L^2} \right)^{-1}\delta v_a  \right]  
\\ &-\frac{3}{D-2} \left[  \left( 1-\frac{\sigma}{L^2} \right)^{\frac{1}{2}}(\epsilon \delta v_a) + \frac{\sigma}{L^2}\left( 1-\frac{\sigma}{L^2} \right)^{-\frac{1}{2}}\delta v_a \right] 
\\ &-\frac{2}{D-2}\frac{\sigma}{L^2} \left[ \left( 1-\frac{\sigma}{L^2} \right)^{-\frac{3}{2}}(\epsilon \partial_t \delta v_a) \right] 
\\ & +\frac{2}{D-2} \frac{\sigma}{L^2} \left[ \left( 1-\frac{\sigma}{L^2} \right)^{-\frac{1}{2}}(\epsilon \delta v_a)+ \frac{\sigma}{L^2}\left( 1-\frac{\sigma}{L^2} \right)^{-\frac{3}{2}}\delta v_a \right] 
\end{split}
\end{equation}
We expand the $\delta v_a$ fluctuations as
\begin{equation}\label{usph}
\delta v_a = \sum_{l,m} b_{lm} Y_a^{lm} e^{-i \omega^v_l r_H t}
\end{equation}
where, $Y_a^{lm}$ are the vector spherical harmonics on $S^{D-2}$ for which 
\begin{equation}
\bar\nabla^2 Y_a^{lm}=- [(D+l-3)l-1]Y_a^{lm}
\end{equation}
We Substitute \eqref{usph} in \eqref{eqvel} and solve for vector QNM frequencies
\begin{equation}\label{vqf}
 \omega^v r_0 = -i(l-1)\left[ 1+\frac{1}{D}\left( l-1+\ln\left(1-\frac{\sigma r_0^2}{L^2}\right) \right) \right] 
\end{equation}
Upto the required order, the answer \eqref{vqf} agrees with the respective answer given in equation (D.2) of \cite{QNM:Emparan}.

\subsection{{Quasinormal Modes for AdS Schwarzschild black brane}}

 Now we shall repeat the above analysis for the case of uniform planar membrane in AdS. This membrane corresponds to AdS Schwarzschild black brane with horizon topology of $R^{D-2}\times R$ in Poincare patch metric. Here we consider membrane fluctuations in time and all the $D-2$ spatial brane directions.
 
 The background metric in Poincare patch coordinates is
 \begin{equation}\label{ppold}
  ds^2 = -\hat{r}^2 d\hat{t}^2 + \frac{d\hat{r}^2}{\hat{r}^2} + \hat{r}^2 d\hat{x}^ad\hat{x}_a
 \end{equation}
 Where we have set AdS radius $L=1$, i.e. $\Lambda=(D-1)(D-2)$.
 For our convenience we use the following notation for this section
 \begin{equation}
 n \equiv D-1
 \end{equation}
 We consider a uniform planar membrane located at the location $\hat{r}=r_0$. We find it convenient to perform the following rescaling
 \begin{equation}
  \hat{r} = r_0 r, \quad \hat{t} = \frac{t}{r_0}, \quad  \hat{x}^a = \frac{x^a}{r_0}
 \end{equation}
With this rescaling, the background metric \eqref{ppold} becomes 
 \begin{equation}\label{ppmet}
  ds^2_{(bgd)} = g_{AB} dX^AdX^B = -r^2 dt^2 + \frac{dr^2}{r^2} + r^2 dx^adx_a
 \end{equation}
 Where now $r=1$ is the location of the uniform membrane. We will consider the time dependence of the shape and velocity fluctuations of the form $$ e^{-i\hat{\omega}\hat{t}}=e^{-i\omega t} , \quad \text{where}~~\hat{\omega}= \omega r_0 $$
 This choice means that the new coordinates in \eqref{ppmet} are all dimensionless.
 
We consider the fluctuations around the uniform planar membrane as
 \begin{equation}\label{bbflu}
  \begin{split}
   r &= 1 + \epsilon \delta r(t,a) \\
   u &= u_0 dt + \epsilon \delta u_t(t,a) dt + \epsilon \delta u_b(t,a) dx^{b} 
  \end{split}
 \end{equation}
 Where $\epsilon$ is the amplitude of fluctuations and we work upto linear order in $\epsilon$.\\
 Upto linear order, the induced metric on the membrane worldvolume becomes
 \begin{equation}\label{ppind}
  ds^2 = g_{\mu\nu}^{(ind)} dy^\mu dy^\nu = -(1 + 2 \epsilon \delta r) dt^2 + (1 + 2 \epsilon \delta r) dx^adx_a 
 \end{equation}
 Upto linear order, $u_\mu g^{\mu\nu}_{(ind)} u_\nu=-1$ implies 
 \begin{equation}\label{eq:utexp1}
  u_t = u_0 +\epsilon \delta u_t = -(1+\epsilon\delta r)
 \end{equation}

The covariant derivative with respect to induced metric \eqref{ppind} is denoted by $\hat{\nabla}$ and of the background metric \eqref{ppmet} is denoted by $\nabla$. Also ${\cal K}_{\mu\nu}$ and ${\cal K}$ are defined in the same way as the previous subsection.
So we now again consider the membrane equations \eqref{membrane} and \eqref{2ndorderfinal}.\\
Substituting the equations \eqref{bbflu} and \eqref{eq:utexp1} in the  LHS \eqref{2ndorderfinal} (see appendix (\ref{app:QNMPoincare}) for details) we find that $ E_t^{tot}$ is of order $ {\cal O}(\epsilon^2)$, and the `$a$' components of the equation becomes
\begin{equation}\label{eq:fluceqcomp}
\begin{split}
& E_a^{tot} =\epsilon\bigg[ -\partial_t \delta u_a -\frac{\partial_t^2 \delta u_a}{n} + \frac{\partial^2 \delta u_a}{n} +\frac{\partial_t^4 \delta u_a}{n^3}-2 \frac{\partial_t^2\partial^2 \delta u_a}{n^3}+ \frac{\partial^2\partial^2 \delta u_a}{n^3}+2\frac{\partial_t^2 \delta u_a}{n^2}\\&-2\frac{\partial^2 \delta u_a}{n^2}+2\frac{\partial_t \delta u_a}{n} \bigg] +\bigg[ -\partial_a \delta r - \partial_t\partial_a\delta r -\frac{\partial_a\partial_t^2 \delta r}{n}  + \frac{\partial^2\partial_a \delta r}{n} + 2\frac{\partial_a\partial_t \delta r}{n} \\&+ \frac{\partial_a\partial^4_t \delta r-2\partial_t^2\partial_a\partial^ 2\delta r+\partial_a\partial^2\partial^2\delta r }{n^3} + 3\frac{\partial_a\partial_t^2\delta r-\partial_a\partial^2\delta r}{n^2} + \frac{\partial_a\partial_t^2 \delta r}{n^2}-\frac{\partial_a\partial^2 \delta r}{n^2}\\&+2\frac{\partial_a\partial_t \delta r}{n}+ 2\frac{\partial_a \delta r}{n} \bigg]=0
 \end{split}
 \end{equation}
Similarly the expansion of equation \eqref{membrane} to linear order in fluctuation leads to the following equation
\begin{equation}\label{delu}
      \hat{\nabla}.u=0= \epsilon \partial^a \delta u_a + \epsilon (n-1) \partial_t \delta r 
     \end{equation}\\
  Now to find the scalar QNM frequencies, the relevant equations are \eqref{delu} and $\partial^a E_a^{tot}$. Finding $\partial^aE_a^{tot}$ and substituting \eqref{delu} we get
\begin{equation}\label{diveq}
  \begin{split}
 &-(n-1)\epsilon\bigg[ -\partial^2_t \delta r -\frac{\partial_t^3 \delta r}{n} + \frac{\partial_t\partial^2 \delta r}{n} +\frac{\partial_t^5 \delta r}{n^3}-2 \frac{\partial_t^3\partial^2 \delta r}{n^3}+ \frac{\partial_t\partial^2\partial^2 \delta r}{n^3}+2\frac{\partial_t^3 \delta r}{n^2}\\&-2\frac{\partial_t\partial^2 \delta r}{n^2}+2\frac{\partial_t^2 \delta r}{n} \bigg]+\bigg[ -\partial^2 \delta r - \partial_t\partial^2\delta r -\frac{\partial^2\partial_t^2 \delta r}{n}  + \frac{\partial^2\partial^2 \delta r}{n} + 2\frac{\partial^2\partial_t \delta r}{n} \\&+ \frac{\partial^2\partial^4_t \delta r-2\partial_t^2\partial^2\partial^ 2\delta r+\partial^2\partial^2\partial^2\delta r }{n^3} + 3\frac{\partial^2\partial_t^2\delta r-\partial^2\partial^2\delta r}{n^2} + \frac{\partial^2\partial_t^2 \delta r}{n^2}-\frac{\partial^2\partial^2 \delta r}{n^2}\\&+2\frac{\partial^2\partial_t \delta r}{n}+ 2\frac{\partial^2 \delta r}{n} \bigg]= 0
  \end{split}
 \end{equation}
 We consider the plane wave expansion of the shape fluctuations
\begin{equation}\label{rfl}
 \delta r = \delta r^0 e^{-i\omega t} e^{ik_ax^a}
\end{equation}
 We then substitute \eqref{rfl} into \eqref{diveq} and solve for scalar QNM frequencies (where we take $k\sim{\cal O}(\sqrt{n})$)\footnote{
  It turns out, as in 1st order, that the orders of temporal and spatial frequencies are related by factor of $\left(\frac{1}{\sqrt n}\right)$. 
  This can be seen from the equation \eqref{delu}, where there is a relative factor of $(n-1)$ between the divergence of velocity fluctuations and the shape fluctuations. So we cannot have both the temporal and spacial frequencies of the same order.  \\
Here we demanded that the temporal frequency is of order ${\cal O}(1)$, but no restriction was put on the spatial frequencies. Such scaling is consistent with the present $\left(\frac{1}{D}\right)$ expansion. See \cite{arbBack}  and \cite{yogesh2}  for detailed explanation.}
\begin{equation}
 \omega_s = \pm \frac{k}{\sqrt{n}}\left(1+\frac{1+2k^2/n}{2n}\right)-\frac{ik^2}{n}\left(1-\frac{1}{n}\right),~~~~\text{where}~~k^2=k_ak^a~~\text{and} ~~k = \sqrt{k^2}
\end{equation}
 Hence we can write the most general solution of \eqref{diveq}
\begin{equation}
 \delta r = \delta r^0_1 e^{-i\omega_1 t} e^{ik_ax^a} + \delta r^0_2 e^{-i\omega_2 t} e^{ik_ax^a}
\end{equation}
where,
\begin{equation}
\begin{split}
&\omega_1 = \frac{k}{\sqrt{n}}\left(1+\frac{1+2k^2/n}{2n}\right)-\frac{ik^2}{n}\left(1-\frac{1}{n}\right),\\&\omega_2=-\frac{k}{\sqrt{n}}\left(1+\frac{1+2k^2/n}{2n}\right)-\frac{ik^2}{n}\left(1-\frac{1}{n}\right)
\end{split}
\end{equation}
Similarly, we can write the form of the most general solution of \eqref{delu} and \eqref{eq:fluceqcomp} (Note that there is only one vector QNM frequency)
\begin{equation}\label{gsu}
  \delta u_a = \delta r^0_1 V_a^1 e^{-i\omega_1 t} e^{ik_ax^a} + \delta r^0_2 V_a^2 e^{-i\omega_2 t} e^{ik_ax^a}+ v_a e^{-i\omega_v t} e^{ik_ax^a}
\end{equation}
where $V_a^1$ and $V_a^2$ are vectors along $k_a$, and $v_a$ is any vector which satisfies $v_ak^a=0$.\\
Substituting \eqref{gsu} into \eqref{delu} and \eqref{eq:fluceqcomp} and solving we find
\begin{equation}
\begin{split}
&\omega_v = -\frac{ik^2}{n} \left(1+{\cal O}(n^{-2})\right), \quad V_a^1 = \left[ -i\left(1-\frac{1}{n}\right)+\frac{\sqrt{n}}{k}\left(1+\frac{1+2k^2/n}{2n}\right) \right]k_a, \\ & V_a^2 = \left[ -i\left(1-\frac{1}{n}\right)-\frac{\sqrt{n}}{k}\left(1+\frac{1+2k^2/n}{2n}\right) \right]k_a
\end{split}
\end{equation}
Thus, we see that there is no subleading correction to $\omega_v$. Collecting the results for light QNM frequencies 
\begin{equation}\label{bbqnm}
\begin{split}
\omega_s &= \pm \frac{k}{\sqrt{n}}\left(1+\frac{1+2k^2/n}{2n}\right)-\frac{ik^2}{n}\left(1-\frac{1}{n}\right)\\  \omega_v &= -\frac{ik^2}{n} \left(1+{\cal O}(n^{-2})\right)
\end{split}
\end{equation}
Upto the required order, the answers \eqref{bbqnm} agree with the respective answers given in equations (4.23),(4.24),(4.25) of \cite{QNM:Emparan}.

\section{Future directions}\label{sec:future}
In this note we have found new dynamical `black-hole' type solutions of the Einstein equations in presence of cosmological constant in an expansion in the inverse powers of dimension. We have done the calculation upto second subleading order. The space-time, determined here,  will necessarily possess an event horizon. The dynamics of the horizon could be mapped to the dynamics of a velocity field on a dynamical membrane, embedded in the asymptotic background. We have determined  the equation for this dual dynamics  of the membrane and the velocity field also in an expansion in $\left(\frac{1}{D}\right)$ .

There are several directions along which we could proceed from here.\\
As we have mentioned in the introduction, one of our key motivation for this second subleading calculation is to have some insight in entropy production, which is expected to take place only at this order. Calculation of this entropy production along with the effective stress tensor for the membrane (see \cite{radiation} for the stress tensor at first order)  could be one immediate project.
 
 As a check we have matched the spectrum of the Quasi-Normal modes. This gives a check on the equation of motion for the membrane. Another important check would be to match the metric with the large dimension limit of known  black hole solutions. Apart from just a check  on our results, this exercise could also give hints to some exact but non-trivial solutions of our membrane equations. This might lead to some techniques to solve the membrane equation analytically.
 
 It would also be interesting to see how these solutions compare with another perturbative techniques to solve Einstein equations, namely derivative expansion and the correspondence with fluid dynamics
(Along these lines, see \cite{Dandekar:2017aiv} for a detailed study of the comparison between the Improved large D membrane formalism and the Fluid Gravity).

\section*{Acknowledgment}

It is a great pleasure to thank Shiraz Minwalla for initiating discussions on this topic and for his numerous suggestions throughout the course of this work.  We would also like to thank Bidisha Chakrabarty and Arunabha Saha for collaboration at the initial stage. We would  like to thank Suman Kundu, Poulomi Nandi for illuminating discussions. P.B. would like to acknowledge the hospitality of ICTS, HRI and SINP while this work was in progress.
Y.D. would like to acknowledge the hospitality of ICTS, IIT-Kanpur and IISER-Pune while this work was in progress.
The work of Y.D. was supported by the Infosys Endowment for the study of the Quantum Structure of Spacetime, as well as an Indo Israel (UGC/ISF) grant.  We  would  also like to acknowledge our debt to the people of India for their steady and generous support to research in the basic sciences. 

\appendix
\section{Calculation of the sources - $S_{AB}$}\label{app:calsource}

In this section we shall give  the details of the calculation of $S_{AB}$. As mentioned before,  the source will be given by $E_{AB}$ calculated on $\left(G^{(0)}_{AB}\right)=g_{AB}+\psi^{-D} O_A O_B$.\\

We shall follow Appendix(B) of \cite{arbBack} for computation.
The first step would be to decompose the source in the following way.
\begin{equation}\label{source:dec2}
\begin{split}
&S_{AB}\equiv E_{AB}|_{G^{(0)}_{AB}}\\
=~&R_{AB}|_{G^{(0)}_{AB}}-(D-1)\lambda G^{(0)}_{AB}\\
=~&-(D-1)\lambda ~\psi^{-D} O_A O_B+\underbrace{\nabla_C\left[\delta\Gamma^C_{AB}|_{\text{lin}}\right] }_{\delta R_{AB}|_\text{lin}}
+\underbrace{\nabla_C\left[\delta\Gamma^C_{AB}|_{\text{non-lin}}\right]}_{\delta R^{(1)}_{AB}|{\text{non-lin}}}\underbrace{- \left[\delta\Gamma^C_{BE}\right]\left[\delta\Gamma^E_{AC}\right]}_{\delta R^{(2)}_{AB}|{\text{non-lin}}}\\
\end{split}
\end{equation}
where
\begin{equation}\label{eq:gammadecomp2}
\begin{split}
&\delta\Gamma^A_{BC}|_{\text{lin.}}=\frac{1}{2}\left\{\nabla_B(\psi^{-D}O_C O^A)+\nabla_C(\psi^{-D}O_B O^A)-\nabla^A(\psi^{-D}O_B O_C)\right\}\\
&\delta\Gamma^A_{BC}|_{\text{non-lin}}=\frac{1}{2}\psi^{-D}O^A(O\cdot \nabla)(\psi^{-D} O_B O_C)\\
&\delta\Gamma^A_{BC}=\delta\Gamma^A_{BC}|_{\text{lin.}}+\delta\Gamma^A_{BC}|_{\text{non-lin.}}
\end{split}
\end{equation}
At first we present the calculation of $\delta R^{(2)}_{AB}|_{\text{non-linear}}$
\begin{equation}
\begin{split}
\delta R^{(2)}_{AB}|_{\text{non-lin.}}&=\underbrace{-\left[\delta\Gamma^C_{BE}|_{\text{lin.}}\right]\left[\delta\Gamma^E_{AC}|_{\text{lin.}}\right]}_{\text{Term-1}}\underbrace{-\left[\delta\Gamma^C_{BE}|_{\text{lin.}}\right]\left[\delta\Gamma^E_{AC}|_{\text{non-lin.}}\right]}_{\text{Term-2}}\\
&\underbrace{-\left[\delta\Gamma^C_{BE}|_{\text{non-lin.}}\right]\left[\delta\Gamma^E_{AC}|_{\text{lin.}}\right]}_{\text{Term-3}}\underbrace{-\left[\delta\Gamma^C_{BE}|_{\text{non-lin.}}\right]\left[\delta\Gamma^E_{AC}|_{\text{non-lin.}}\right]}_{\text{Term-4}}
\end{split}
\end{equation}
As previously, in this case also, Term-2=Term-3=Term-4=0;\\ \\
Now we need to calculate Term-1.
\begin{equation}\label{RABnonlinear2}
\begin{split}
\delta R^{(2)}_{AB}|_{\text{non-lin.}}&=-\left[\delta\Gamma^C_{BE}|_{\text{lin.}}\right]\left[\delta\Gamma^E_{AC}|_{\text{lin.}}\right]\\
&=\frac{1}{2}\psi^{-2D}(\nabla_E O^C)(\nabla^E O_C)O_B O_A-\frac{1}{2}\nabla_E(\psi^{-D}O_B O^C)\nabla_C(\psi^{-D}O_A O^E)\\
&=-\frac{1}{2}[(O\cdot\nabla)(\psi^{-D} O_B)][(O\cdot\nabla)(\psi^{-D}O_A)]+\psi^{-2D}\left(\frac{DN}{\psi}\right)~Q~O_A O_B\\
&+\frac{\psi^{-2D}}{2}(\nabla_E O_C)(\nabla^E O^C-\nabla^C O^E)O_B O_A-\psi^{-2D}~Q^2~ O_B O_A\\
\end{split}
\end{equation}
$$ \text{Where, }Q\equiv u^E(O\cdot\nabla)n_E$$

\begin{equation}\label{finalR2}
\begin{split}
\delta R^{(2)}_{AB}|_{\text{non-lin.}}&=-\frac{1}{2}[(O\cdot\nabla)(\psi^{-D} O_B)][(O\cdot\nabla)(\psi^{-D}O_A)]+\psi^{-2D} K~Q~O_A O_B\\
&+\frac{\psi^{-2D}}{2}\left[(\nabla_E O_C)(\nabla^E O^C-\nabla^C O^E)-2~Q^2+2~Q\frac{(n\cdot\nabla)K}{K}\right]O_B O_A
\end{split}
\end{equation}
In deriving \eqref{finalR2} we have used,
\begin{equation}
\begin{split}
\frac{DN}{\psi}=K+\frac{(n\cdot\nabla) K}{K}
\end{split}
\end{equation}

Now we proceed to the calculation of $\delta R^{(1)}_{AB}|_{\text{non-lin.}}$

\begin{equation}
\begin{split}
&~~~~\delta R^{(1)}_{AB}|_{\text{non-lin.}}\\
&=\nabla_C\left[\frac{1}{2}\psi^{-D} O^C(O\cdot\nabla)\left(\psi^{-D}O_A O_B\right)\right]\\
&=\left(\frac{\psi^{-D}}{2}\right) \bigg[(\nabla\cdot O)~ (O\cdot\nabla)\left(\psi^{-D} O_BO_A\right)+O_A (O\cdot\nabla)\left[(O\cdot\nabla)(\psi^{-D}O_B)\right]\bigg] \\
&~+ \frac{1}{2}\left[(O\cdot\nabla)\left(\psi^{-D} O_A\right)\right]~\left[(O\cdot\nabla)\left(\psi^{-D} O_B\right)\right]
+ \frac{1}{2}(O\cdot\nabla)\left[\psi^{-2D}O_B(O\cdot\nabla)O_A\right]\\ \\
&=\frac{1}{2}[(O\cdot\nabla)(\psi^{-D}O_A)][(O\cdot\nabla)(\psi^{-D}O_B)]-\frac{\psi^{-2D}}{2}(O\cdot\nabla)[K~ O_A O_B]\\
&+\frac{\psi^{-2D}}{2}\left(\frac{DN}{\psi}-\nabla\cdot O\right)\left(\frac{DN}{\psi}-2~Q\right)O_A O_B\\
&+\frac{\psi^{-2D}}{2}\left[3~Q^2+2~(O\cdot\nabla)Q-(O\cdot\nabla)\left(\frac{(n\cdot\nabla)K}{K}\right)-\frac{(n\cdot\nabla)K}{K}2~Q\right]O_A O_B
\end{split}
\end{equation}
Now,
\begin{equation}\label{DNDN}
\begin{split}
&\left(\frac{DN}{\psi}-\nabla\cdot O\right)\left(\frac{D~N}{\psi}-2~Q\right)\\
&=\bigg[\frac{(n\cdot\nabla)K}{K}+\frac{(n\cdot\nabla)^2K}{K^2}-2~\frac{[(n\cdot\nabla)K]^2}{K^3}+\tilde{\nabla}\cdot u-\frac{1}{K}(u\cdot\nabla)\left(\frac{(n\cdot\nabla)K}{K}\right)\\
&~~~~~~~~~~~~-\frac{(u\cdot \nabla)K}{K}+\frac{1}{K}\frac{(n\cdot\nabla)K}{K}\frac{(u\cdot\nabla)K}{K}\bigg]\left[K+\frac{(n\cdot\nabla)K}{K}-2~Q\right]\\
&=K\left(\tilde{\nabla}\cdot u\right)+(O\cdot \nabla)K+\frac{(n\cdot\nabla)^2K}{K}-2\left[\frac{(n\cdot\nabla)K}{K}\right]^2+\frac{(O\cdot\nabla)K}{K}\frac{(n\cdot\nabla)K}{K}\\
&-2Q\frac{(O\cdot\nabla)K}{K}-(u\cdot\nabla)\left(\frac{(n\cdot\nabla)K}{K}\right)+\frac{(n\cdot\nabla)K}{K}\frac{(u\cdot\nabla)K}{K}+(\tilde{\nabla}\cdot u)\frac{(n\cdot\nabla)K}{K}-2Q(\tilde{\nabla}\cdot u)
\end{split}
\end{equation}
Where, $\tilde{\nabla}$ is defined in \eqref{tildedef}\\
In deriving \eqref{DNDN} we have used (see \ref{derivdelusub} for derivation),
\begin{equation}\label{eq:delusub}
\nabla\cdot u=\tilde{\nabla}\cdot u-\frac{(u\cdot\nabla) K}{K}-\frac{1}{K}(u\cdot\nabla)\left(\frac{(n\cdot\nabla)K}{K}\right)+\frac{1}{K}\frac{(n\cdot\nabla)K}{K}\frac{(u\cdot\nabla)K}{K}
\end{equation}

Using, \eqref{DNDN} we get the final expression of $\delta R^{(1)}_{AB}|_{\text{non-lin.}}$,
\begin{equation}\label{RABnonlinear1}
\begin{split}
&~~~~\delta R^{(1)}_{AB}|_{\text{non-lin.}}\\
&=\frac{\psi^{-2D}}{2} K(\tilde{\nabla}\cdot u)~O_A O_B-\psi^{-2D}K~Q~O_A O_B+\frac{1}{2}\left[(O\cdot\nabla)\left(\psi^{-D}O_A\right)\right]\left[(O\cdot\nabla)(\psi^{-D}O_B)\right]\\
&+\frac{\psi^{-2D}}{2}\bigg[3~Q^2+2(O\cdot\nabla)Q-2~Q\left(\frac{(n\cdot\nabla)K}{K}+\frac{(O\cdot\nabla)K}{K}\right)+(\tilde{\nabla}\cdot u)\left(\frac{(n\cdot\nabla)K}{K}-2Q\right)\bigg]O_A O_B
\end{split}
\end{equation}
Adding \eqref{finalR2} and \eqref{RABnonlinear1} we get
\begin{equation}\label{RABnonlinear}
\begin{split}
&~~~~\delta R_{AB}|_{\text{non-lin.}}\\
&\equiv\delta R^{(1)}_{AB}|_{\text{non-lin.}}+\delta R^{(2)}_{AB}|_{\text{non-lin.}}\\
&=\frac{1}{2}\psi^{-2D}~K(\tilde{\nabla}\cdot u)~O_A O_B+\frac{1}{2}\psi^{-2D}\bigg[(\nabla_E O^C)(\nabla^E O_C-\nabla_C O^E)+Q^2+2(O\cdot\nabla)Q\\
&~~~~~~~~~~~~~~~~~~~~~~-2Q\frac{(O\cdot\nabla)K}{K}+(\tilde{\nabla}\cdot u)\left(\frac{(n\cdot\nabla)K}{K}-2Q\right)\bigg]O_A O_B\\
\end{split}
\end{equation}
Let us note the presence of  `$K(\tilde\nabla\cdot u)$ ' term in $\delta R_{AB}|_{\text{non-lin.}}$. From the membrane equation at first subleading order, it follows that this term is of order ${\cal O}(1)$ on $\psi=1$ hypersurface. This is sort of `anomalous',  since naive order counting suggests that this term should be or order ${\cal O}(D^2)$ and this may not be the case once we are away from the membrane.\\
Now for any generic term, which is of order ${\cal O}(1)$  when evaluated on  $(\psi =1)$ hypersurface, will have corrections of order ${\cal O}\left(1\over D\right)$ (or further suppressed) as one goes away from $\psi=1$.  While integrating the ODEs, this is the reason we could ignore all the implicit $\psi$ dependence  in the source. However from the above discussion we could see that such reasoning does not work for  `$K(\tilde\nabla\cdot u)$ ' (or in fact any such `anomalous' term). Below we shall examine this term in more detail.\\
We can expand $(\tilde{\nabla}\cdot u)$ in $\left[\psi -1 = {R\over D}\right]$ as follows
\begin{equation}\label{nabuoff1}
\begin{split}
\tilde{\nabla}\cdot u =~&(\tilde{\nabla}\cdot u)\bigg{|}_{\psi=1}+\frac{\psi-1}{N}(n\cdot\nabla)(\tilde{\nabla}\cdot u)\bigg{|}_{\psi=1}+~\frac{(\psi-1)^2}{2N^2}\left[\frac{(n\cdot\nabla)N}{N}\right]\bigg{|}_{\psi=1}\left[(n\cdot\nabla)(\tilde{\nabla}\cdot u)\right]\bigg{|}_{\psi=1}\\
&+\frac{(\psi-1)^2}{2N^2}\left[(n\cdot\nabla)(n\cdot\nabla)\left(\tilde{\nabla}\cdot u\right)\right]\bigg{|}_{\psi=1} + {\cal O}(\psi-1)^3\\
=~&\left(\tilde{\nabla}\cdot u\right)\bigg{|}_{R=0}-~R\bigg[{\tilde{\nabla}\cdot E\over K}\bigg]_{R=0}-\frac{R^2}{2}\bigg[\left(\frac{(n\cdot\nabla)K}{K^3}\right)\left(\tilde{\nabla}\cdot E\right)\bigg]_{R=0}\\
&+R^2\bigg[\left(\frac{D^2}{K^3}\right)~\mathfrak{s}_2\bigg]_{R=0} + {\cal O}\left(1\over D\right)^2\\
=~&\left(\tilde{\nabla}\cdot u\right)\bigg{|}_{R=0}-~R\bigg[{\tilde{\nabla}\cdot E\over K}\bigg]_{R=0}+R^2\bigg[\left(\frac{D^2}{K^3}\right)~\mathfrak{s}_2\bigg]_{R=0} + {\cal O}\left(1\over D\right)^2\\
\end{split}
\end{equation}
Where $E_A$ is  given in  equation \eqref{eq:constraint2}.\\
In the second line we have used the following two identities (to prove  them we have used {\it Mathematica Version-11}), 
\begin{equation}\label{eq:mathematica2}
\begin{split}
(n\cdot\nabla)(\tilde{\nabla}\cdot u)\bigg{|}_{R=0}&=-(\tilde{\nabla}\cdot E)\bigg{|}_{R=0}  + {\cal O}\left(1\over D\right)\\
(n\cdot \nabla)(n\cdot\nabla)(\tilde{\nabla}\cdot u)\bigg{|}_{R=0}&=2~D^2\left(\frac{\mathfrak{s}_2}{K}\right)\bigg{|}_{R=0}+{\cal O}(1)\\
\end{split}
\end{equation}
Clearly  the second and the third term in the last line of equation \eqref{nabuoff1} (which encode the value of $(\tilde\nabla\cdot u)$ off the membrane) could contribute in $\delta R_{AB}|_{\text{non-lin.}}$ at order ${\cal O}(1)$.\\
Substituting \eqref{nabuoff1} in equation \eqref{RABnonlinear} we find
\begin{equation}
\begin{split}
&~~~~\delta R_{AB}|_{\text{non-lin.}}\\
&=\psi^{-2D}\left(\frac{K}{2}\right)\left[\left(\tilde{\nabla}\cdot u\right)_{\psi=1}-R\bigg(\frac{\tilde{\nabla}\cdot E}{K}\bigg)_{\psi=1} -\frac{1}{2K}\left[{\nabla}_{(E}u_{F)}{\nabla}_{(C}u_{D)}P^{FC} P^{ED}\right]\right]O_A O_B\\
&~~~~+\frac{\psi^{-2D}}{2}R^2\left(\frac{D^2}{K^2}\right)\left(\mathfrak{s}_2\right)O_A O_B\\
&-\psi^{-2D}\Bigg[2u^A K^C_A\frac{\nabla_C K}{K}-(\nabla_C u_A)(\nabla^C u^A)-(u\cdot K\cdot K\cdot u)+3\left(\frac{(u\cdot\nabla)K}{K}\right)^2\\
&-\frac{K}{D}\left(\frac{(u\cdot\nabla)K}{K}\right)+\frac{K}{D}(u\cdot K\cdot u)-2~\frac{(u\cdot\nabla)K}{K}(u\cdot K\cdot u)- u^E u^F{\bar R}_{EDFC}O^C O^D\Bigg]O_A O_B\\
&=\psi^{-2D}\left(\frac{K}{2}\right)\left[\left(\tilde{\nabla}\cdot u\right)_{\psi=1}-R\bigg(\frac{\tilde{\nabla}\cdot E}{K}\bigg)_{\psi=1} -\frac{1}{2K}\left[{\nabla}_{(E}u_{F)}{\nabla}_{(C}u_{D)}P^{FC} P^{ED}\right]\right]O_A O_B\\
&+\frac{\psi^{-2D}}{2}R^2\left(\frac{D^2}{K^2}\right)\left(\mathfrak{s}_2\right)O_A O_B~ -\psi^{-2D}\Bigg[\left(\frac{u\cdot{\nabla}K}{K}\right)^2+4~u^A {K}^B_A\frac{{\nabla}_B {K}}{K}-(\tilde{\nabla}_A u_B)(\tilde{\nabla}^A u^B)\\
&-(u\cdot {K}\cdot u)^2-2~\frac{\tilde{{\nabla}}^A {K}}{K}\left[(u\cdot{\nabla})u_A\right]-\left[(u\cdot{\tilde{\nabla}})u_A\right]\left[(u\cdot\tilde{\nabla})u^A\right]+2\left[(u\cdot{\nabla})u^A\right](u^B {K}_{BA})\\
&-3(u\cdot {K}\cdot {K}\cdot u)-\frac{\tilde{{\nabla}}_A {K}}{K}\frac{\tilde{{\nabla}}^A {K}}{K}-\frac{K}{D}\left(\frac{u\cdot{\nabla}{K}}{K}-u\cdot {K}\cdot u\right)+u^E u^F n^D n^C\bar{R}_{CEFD}\Bigg]O_A O_B\\ \\
&=e^{-2R}\left(\frac{K}{2}\right)\left[\left(\tilde{\nabla}\cdot u\right)_{R=0} -\frac{1}{2K}\left({\nabla}_{E}u_{F}+{\nabla}_{F}u_{E}\right)\left({\nabla}_{C}u_{D} + \nabla_D u_C\right)P^{FC} P^{ED}\right]O_A~O_B\\
&~~~~+\left(\frac{e^{-2R}}{2}\right)\Bigg[-R\left({\tilde{\nabla}\cdot E}\right)_{R=0}+R^2\left(\frac{D^2}{K^2}~\mathfrak{s}_2\right)_{R=0}\Bigg]O_A O_B-e^{-2R} \left(\mathfrak{s}_1\right)O_A O_B
\end{split}
\end{equation}
where
\begin{equation}
\begin{split}
\mathfrak{s}_1&=u^E u^F n^D n^C\bar{R}_{CEFD}+\left(\frac{u\cdot{\nabla}K}{K}\right)^2+\frac{\tilde{\nabla}_A {K}}{K}\left[4~u^B {K}^A_B-2\left[(u\cdot{\nabla})u_A\right]-\frac{\tilde{\nabla}^A {K}}{K}\right]\\
&-(\tilde{\nabla}_A u_B)(\tilde{\nabla}^A u^B)-(u\cdot {K}\cdot u)^2-\left[(u\cdot\tilde{\nabla})u_A\right][(u\cdot\tilde{\nabla})u^A]+2\left[(u\cdot{\nabla})u^A\right](u^B {K}_{BA})\\
&~~~~~~~~~~~~-3(u\cdot {K}\cdot {K}\cdot u)-\frac{K}{D}\left(\frac{u\cdot{\nabla}{K}}{K}-u\cdot {K}\cdot u\right)\\ \\
&\mathfrak{s}_2=\frac{K^2}{D^2}\Bigg[-\frac{{K}}{D}\left(\frac{u\cdot{\nabla}{K}}{K}-u\cdot {K}\cdot u\right)- 2~\lambda- (u\cdot {K}\cdot {K}\cdot u)+2 \left(\frac{{\nabla}_A{K}}{K}\right)u^B {K}^A_B-\left(\frac{u\cdot{\nabla}{K}}{K}\right)^2\\
&+2\left(\frac{u\cdot{\nabla}{K}}{K}\right)(u\cdot {K}\cdot u)-\left(\frac{\tilde{\nabla}^D K}{K}\right)\left(\frac{\tilde{\nabla}_D K}{K}\right)-(u\cdot {K}\cdot u)^2+n^B n^D u^E u^F\bar{R}_{FBDE}\Bigg]
\end{split}
\end{equation}

Now we shall calculate those terms in Ricci tensor that are linear in $\psi^{-D}$
\begin{equation}
\begin{split}
\delta R_{AB}|_{\text{lin.}}&=\nabla_C\left[\delta\Gamma^C_{BA}|_{\text{lin.}}\right]\\
&=\underbrace{\frac{1}{2}\nabla_C\left\{\nabla_B\left(\psi^{-D}O_A O^C\right)\right\}}_{T_1}+\underbrace{\frac{1}{2}\nabla_C\left\{\nabla_A\left(\psi^{-D}O_B O^C\right)\right\}}_{T_2}\underbrace{-\frac{1}{2}\nabla_C\left\{\nabla^C\left(\psi^{-D}O_A O_B\right)\right\}}_{T_3}
\end{split}
\end{equation}
\begin{equation}\label{T_1}
\begin{split}
T_1&=\frac{1}{2}\nabla_C\left\{\nabla_B\left(\psi^{-D}O_A O^C\right)\right\}\\
&=\frac{1}{2}[\nabla_C,\nabla_B]\left(\psi^{-D}O_A O^C\right) +\frac{1}{2}\nabla_B\nabla_C \left(\psi^{-D}O_A O^C\right)\\
&=\frac{\psi^{-D}}{2}\left(\bar R_{BD} O^D O_A+\bar{R}_{CBAD}O^D O^C\right)- \frac{1}{2}\nabla_B \left[\psi^{-D}\left\{\left(\frac{DN}{\psi}-\nabla\cdot O\right)O_A - Q~O_A\right\}\right]\\
&=\frac{\psi^{-D}}{2}\left(\bar R_{BD} O^D O_A+\bar{R}_{CBAD}O^D O^C\right) +\left(\frac{DN}{2\psi}\right)\psi^{-D}\left[\frac{DN}{\psi} - \nabla\cdot O- Q\right]n_B O_A\\
&~~~~~~~~~~~-\frac{1}{2}\psi^{-D}\nabla_B\left\{\left(\frac{DN}{\psi}-\nabla\cdot O-Q\right)O_A\right\}\\
&=\frac{\psi^{-D}}{2}\left(\bar R_{BD} O^D O_A+\bar{R}_{CBAD}O^D O^C\right) +\frac{\psi^{-D}}{2}\bigg[(n\cdot\nabla)K+K(\nabla\cdot u-Q)\bigg]n_B O_A\\
&~~+\frac{\psi^{-D}}{2}\left[\frac{(n\cdot\nabla)^2 K}{K}-2\left(\frac{(n\cdot\nabla)K}{K}\right)^2-\frac{K}{D}\left(\frac{(n\cdot\nabla)K}{K}\right)\right]n_B O_A+\frac{\psi^{-D}}{2}\left(\frac{K}{D}\right)\left(\nabla_B O_A\right)\\
&~~~~~~-\frac{\psi^{-D}}{2}O_A \nabla_B\left[\frac{(n\cdot\nabla)K}{K}-2~\frac{(u\cdot\nabla)K}{K}+u\cdot K\cdot u+\tilde{\nabla}\cdot u\right]
\end{split}
\end{equation}

Similarly, we will get $T_2$ by interchanging $A$ and  $B$  indices

\begin{equation}\label{T_3}
\begin{split}
T_3&=-\frac{1}{2}\nabla_C\nabla^C(\psi^{-D}O_B O_A)\\
&=-\frac{1}{2}\left(\nabla^2\psi^{_D}\right)O_AO_B -\left(\nabla_C \psi^{-D}\right) \left(\nabla^CO_A O_B\right) - \frac{\psi^{-D}}{2}\nabla^2(O_A O_B)\\
&=\psi^{-D}\bigg[\left(\frac{DN}{\psi}\right)(n\cdot\nabla)\left( O_A O_B\right) - \frac{1}{2}\nabla^2(O_A O_B)\bigg]
\end{split}
\end{equation}

Adding $T_1,T_2,T_3$ we get the expression for $\delta R_{AB}|_{\text{lin.}}$
\begin{equation}\label{delrlin}
\begin{split}
&~~~\delta R_{AB}|_{\text{lin}}\\
&=\psi^{-D}~(D-1)~\lambda~O_A O_B+\psi^{-D}\bar{R}_{CABD} O^D O^C+\psi^{-D} K~(n\cdot\nabla)(O_A O_B)\\
&+\frac{\psi^{-D}}{2}(n_B O_A+n_A O_B)[(n\cdot\nabla)K+K(\nabla\cdot u-Q)]-\frac{\psi^{-D}}{2}\left(O_A\nabla^2 O_B+O_B\nabla^2 O_A\right)\\ \\
&+\frac{\psi^{-D}}{2}\bigg\{\frac{(n\cdot\nabla)^2 K}{K}-2~\left[\frac{(n\cdot\nabla)K}{K}\right]^2-\frac{K}{D}~\frac{(n\cdot\nabla)K}{K}\bigg\}(n_B O_A+O_B n_A)\\
&+\psi^{-D}\bigg\{\left[\frac{(n\cdot\nabla) K}{K}\right](n\cdot\nabla)(O_A O_B)-(\nabla_C O_A)(\nabla^C O_B)\bigg\}+\frac{\psi^{-D}}{2}\frac{K}{D}\left[\nabla_B O_A+\nabla_A O_B\right]\\
&-\frac{\psi^{-D}}{2}(O_A\nabla_B+O_B\nabla_A)\left[\frac{(n\cdot\nabla)K}{K}-2~\frac{(u\cdot\nabla)K}{K}+u\cdot K\cdot u+\tilde{\nabla}\cdot u\right]
\end{split}
\end{equation}

Now, we shall decompose the source in the way as mentioned in \eqref{eq: decompsource}. Note that the decomposition of a general 2-index symmetric tensor ($C_{AB}$) is the following
\begin{equation}\label{simpdecomp}
\begin{split}
C_{AB}&=P_A^D P_B^E C_{DE}+(P^E_A O_B+P^E_B O_A) C_{ED} u^D+(P^E_A n_B+P_B^E n_A) C_{ED} O^D\\
&+(n_A O_B+n_B O_A)(O^E C_{ED} u^D)+O_A O_B(u^E C_{ED}u^D)+n_A n_B(O^E C_{ED}O^D)
\end{split}
\end{equation}
Using \eqref{simpdecomp} we shall first decompose each of the tensor structure appearing in \eqref{delrlin}
\begin{equation}\label{idten1}
\begin{split}
(n\cdot\nabla) (O_A O_B) = ~&2 \left[u^C (n\cdot \nabla) n_C\right]O_A O_B + (O_A P^C_B + O_B P^C_A) (n\cdot\nabla) O_C\\
= ~&2 \left[u^C (n\cdot \nabla) n_C\right]O_A O_B +(O_A P^C_B + O_B P^C_A) (u\cdot\nabla) O_C
\end{split}
\end{equation}

\begin{equation}\label{idten2}
\begin{split}
&O_B \nabla^2 O_A+O_A\nabla^2 O_B\\
&=2\left[K[u^D(n\cdot\nabla)n_D]+(u\cdot\nabla)K-u^D K^C_D\left(\frac{\nabla_C K}{K}\right)+u^D(n\cdot\nabla)^2n_D+(\nabla_C u_D)(\nabla^C u^D)\right]O_A O_B\\
&~~~~~~~~~-[\left(\nabla^C O_D\right)(\nabla_C O^D)][n_A O_B+n_B O_A ] + (O_B P^C_A+O_A P^C_B )\nabla^2O_C
\end{split}
\end{equation}

\begin{equation}\label{idten3}
\begin{split}
(\nabla_C  O_A) (\nabla^C O_B)&=(u^D \nabla_C n_D)(u^E\nabla^C n_E)O_A O_B +( \nabla_D O_C)(\nabla^D O_{C'})P^C_A P^{C'}_B\\
&~~~~~~+(O_B P^C_A+O_A P^C_B)[(\nabla_F O_C)(u^D\nabla^F n_D)]
\end{split}
\end{equation}

\begin{equation}\label{idten4}
\begin{split}
\nabla_B  O_A+\nabla_A O_B&=2~(u\cdot K\cdot u)O_A O_B+~Q~(n_A O_B+n_B O_A)+P^C_A P^{C'}_B(\nabla_C  O_{C'}+\nabla_{C'} O_C)\\
&~~~~~~+(O_B P_A^C+O_A P^C_B)[(u\cdot\nabla)O_C+u^D K_{CD}]
\end{split}
\end{equation}

\begin{equation}\label{idten5}
\begin{split}
&(O_A\nabla_B+O_B\nabla_A)\left[\frac{(n\cdot\nabla)K}{K}-2~\frac{(u\cdot\nabla)K}{K}+u\cdot K\cdot u+\tilde{\nabla}\cdot u\right]\\
&=-2~\frac{(u\cdot\nabla)K}{D}~O_A O_B-\left(O_A P^C_B+O_B P^C_A\right)\frac{\nabla_C K}{D}\\
&+(O_A n_B+O_B n_A)(O\cdot\nabla)\left[\frac{(n\cdot\nabla)K}{K}-2~\frac{(u\cdot\nabla)K}{K}+u\cdot K\cdot u+\tilde{\nabla}\cdot u\right]\\
\end{split}
\end{equation}
\begin{equation}
\begin{split}
\bar{R}_{CABD} O^D O^C&=P^E_A P^F_B~ \bar{R}_{CEFD}O^D O^C+O_A O_B ~u^E u^F\bar{R}_{CEFD} O^D O^C\\
&+(P^E_A O_B+P^E_B O_A)\bar{R}_{CEFD}O^D O^C u^F
\end{split}
\end{equation}\\
Using \eqref{idten1}, \eqref{idten2}, \eqref{idten3}, \eqref{idten4}, \eqref{idten5} we can decompose $\delta R_{AB}|_{\text{lin}}$ in the following way
\begin{equation}
\begin{split}
\delta R_{AB}|_{\text{lin}}&=\delta R^{(S_1)}_{\text{lin}} O_A O_B+\delta R^{(S_2)}_{\text{lin}}(n_A O_B+n_B O_A)+\delta R^{(S_3)}_{\text{lin}} n_A n_B+\delta R^{(tr)}_{\text{lin}}P_{AB}\\
&+(O_A P^C_B+O_B P^C_A)\left[\delta R^{(V_1)}_{\text{lin}}\right]_C+(n_A P^C_B+n_B P^C_A)\left[\delta R^{(V_2)}_{\text{lin}}\right]_C+\left[\delta R^{(T)}_{\text{lin}}\right]_{AB}
\end{split}
\end{equation}
Where
\begin{equation}
\begin{split}
\delta R^{(S_1)}&=\psi^{-D}(D-1)~\lambda+\psi^{-D}\bigg[u^E u^F\bar{R}_{CEFD} n^D n^C-(u\cdot\nabla)\left(\frac{(n\cdot\nabla)K}{K}\right)+u^A K^C_A\frac{\nabla_C K}{K}\\
&~~-\frac{(n\cdot\nabla)K}{K}\frac{(u\cdot\nabla)K}{K}+2\frac{(n\cdot\nabla)K}{K}[u^C(n\cdot\nabla)n_C]-(u^D\nabla_C n_D)(u^E\nabla^C n_E)\\
&~~-u^A(n\cdot\nabla)^2n_A-(\nabla_C u_A)(\nabla^C u^A)+\frac{K}{D}(u\cdot K\cdot u)+\frac{K}{D}\frac{(u\cdot\nabla)K}{K}\bigg]\\
&=\psi^{-D}(D-1)~\lambda+\psi^{-D}\bigg[2u^A K^C_A\frac{\nabla_C K}{K}-(\nabla_C u_A)(\nabla^C u^A)-(u\cdot K\cdot K\cdot u)-\frac{K}{D}\frac{(u\cdot\nabla)K}{K}\\
&~~+3\left(\frac{(u\cdot\nabla)K}{K}\right)^2+\frac{K}{D}(u\cdot K\cdot u)-2~\frac{(u\cdot\nabla)K}{K}(u\cdot K\cdot u)+u^E u^F\bar{R}_{CEFD} n^D n^C\bigg]\\ \\
&=\psi^{-D}(D-1)~\lambda+\psi^{-D} ~\mathfrak{s}_1
\end{split}
\end{equation}
Where,
\begin{equation}
\begin{split}
\mathfrak{s}_1&=\left(\frac{u\cdot{\nabla}K}{K}\right)^2+\frac{\tilde{\nabla}_A {K}}{K}\left[4~u^B {K}^A_B-2\left[(u\cdot{\nabla})u_A\right]-\frac{\tilde{\nabla}^A {K}}{K}\right]-(\tilde{\nabla}_A u_B)(\tilde{\nabla}^A u^B)\\
&-(u\cdot {K}\cdot u)^2-\left[(u\cdot\tilde{\nabla})u_A\right][(u\cdot\tilde{\nabla})u^A]+2\left[(u\cdot{\nabla})u^A\right](u^B {K}_{BA})-3~(u\cdot {K}\cdot {K}\cdot u)\\
&-\frac{K}{D}\left(\frac{u\cdot{\nabla}{K}}{K}-u\cdot {K}\cdot u\right)+u^E u^F\bar{R}_{CEFD}~ n^D n^C
\end{split}
\end{equation}

\begin{equation}\label{deltars2}
\begin{split}
\delta R^{(S_2)}&=\frac{\psi^{-D}}{2}\Bigg[K\left\{\tilde{\nabla}\cdot u-\frac{(u\cdot\nabla)K}{K}-\frac{1}{K}(u\cdot\nabla)\left(\frac{(n\cdot\nabla)K}{K}\right)+\frac{1}{K}\frac{(n\cdot\nabla)K}{K}\frac{(u\cdot\nabla)K}{K}\right\}\\
&+(n\cdot\nabla)K-K~Q+\frac{(n\cdot\nabla)^2 K}{K}-2~\left(\frac{(n\cdot\nabla)K}{K}\right)^2-\frac{K}{D}\frac{(n\cdot\nabla)K}{K}+\frac{K}{D}~Q\\
&+(\nabla^C O_A)(\nabla_C O^A)-(O\cdot\nabla)\left(\frac{(n\cdot\nabla)K}{K}-2~\frac{(u\cdot\nabla)K}{K}+u\cdot K\cdot u+\tilde{\nabla}\cdot u\right)\Bigg]
\end{split}
\end{equation}
We shall massage the above expression for $\delta R^{(S_2)}$ a little more.\\
Let us note the presence of  `$K(\tilde\nabla\cdot u)$ ' term in $\delta R^{(S_2)}$. From the discussion just below the equation \eqref{RABnonlinear} it is clear that we need to take the expansion of $\tilde{\nabla}\cdot u$ in $\psi-1$. The $\psi-1$ expansion of $(\tilde{\nabla}\cdot u)$ is given by \eqref{nabuoff1}
\begin{equation}\label{nabuoff}
\begin{split}
\tilde{\nabla}\cdot u=~&\left(\tilde{\nabla}\cdot u\right)_{R=0}-~R\left[{\tilde{\nabla}\cdot E\over K}\right]_{R=0}+R^2\bigg[\left(\frac{D^2}{K^3}\right)~\mathfrak{s}_2\bigg]_{R=0} + {\cal O}\left(1\over D\right)^2
\end{split}
\end{equation}
Substituting equation \eqref{nabuoff} in equation \eqref{deltars2} we find
\begin{equation}\label{deltars2a}
\begin{split}
\delta R^{(S_2)}&=\frac{\psi^{-D}}{2}\Bigg[K\left(\tilde{\nabla}\cdot u\right)_{R=0}-~R\left({\tilde{\nabla}\cdot E}\right)_{R=0}+R^2\bigg[\left(\frac{D^2}{K^2}\right)\mathfrak{s}_2\bigg]_{R=0}\Bigg]\\
&+\frac{\psi^{-D}}{2}\Bigg[-K\Bigg\{\frac{(u\cdot\nabla)K}{K}+\frac{1}{K}(u\cdot\nabla)\left(\frac{(n\cdot\nabla)K}{K}\right)-\frac{1}{K}\frac{(n\cdot\nabla)K}{K}\frac{(u\cdot\nabla)K}{K}\Bigg\}\\
&+(n\cdot\nabla)K-K~Q+\frac{(n\cdot\nabla)^2 K}{K}-2~\left(\frac{(n\cdot\nabla)K}{K}\right)^2-\frac{K}{D}\frac{(n\cdot\nabla)K}{K}+\frac{K}{D}~Q\\
&+(\nabla^C O_A)(\nabla_C O^A)-(O\cdot\nabla)\left(\frac{(n\cdot\nabla)K}{K}-2~\frac{(u\cdot\nabla)K}{K}+u\cdot K\cdot u+\tilde{\nabla}\cdot u\right)\Bigg]
\end{split}
\end{equation}
Now it turns out that it is possible to rewrite  the last three lines of equation \eqref{deltars2a} in terms of the already defined scalar structures $~{\mathfrak s}_1~$   plus few extra terms which could be expressed  as functions of membrane equation.\\
 We have used {\it Mathematica Version 11} for this  purpose\footnote{More precisely {\it Mathematica} has been used to rearrange $\delta R^{(S_2)}$ on $R=0$ hypersurface . Away from the membrane the calculation is relatively less tedious and could be done by hand.
 On $\psi=1$ i.e., on R=0, $\delta R^{(S_2)}$ becomes
\begin{equation}\label{eq:mathematica1}
\delta R^{(S_2)}\bigg{|}_{R=0}=e^{-R}\left[-\mathfrak{s}_1+\frac{K}{2}\left((\tilde{\nabla}\cdot u)-\frac{1}{2K}\nabla_{(A}u_{B)}\nabla_{(C}u_{D)}P^{AC} P^{BD}\right)\right]\Bigg{|}_{R=0}
\end{equation}
$$\text{Where,    ~~~~~~  }\nabla_{(A}u_{B)}=\nabla_A u_B+\nabla_B u_A$$
 For {\it Mathematica} computation we do have to choose a specific background and coordinate system.  Since we have an independent proof that the final answer is `background-covariant', such a choice does not imply any loss of generality. However, we need to do an appropriate `geometrization' of the answer that we get from {\it Mathematica}, so that we could write it in a `background covariant form' as desired. See \cite{Chmembrane}, \cite{yogesh1} for details of this procedure.}
\begin{equation}\label{finrs2}
\begin{split}
\delta R^{(S_2)}&=e^{-R}\left[-\mathfrak{s}_1+\frac{K}{2}\left((\tilde{\nabla}\cdot u)-\frac{1}{2K}\nabla_{(A}u_{B)}\nabla_{(C}u_{D)}P^{AC} P^{BD}\right)\right]\Bigg{|}_{R=0}\\
&+\frac{e^{-R}}{2}\Bigg[-R\left({\tilde{\nabla}\cdot E}\right)_{R=0}+R^2\bigg[\left(\frac{D^2}{K^2}\right)\mathfrak{s}_2\bigg]_{R=0}\Bigg] + {\cal O}\left(1\over D\right)^2
\end{split}
\end{equation}
This type of rewriting helps to see the consistency of the set of coupled ODEs manifestly (see section - \ref{check1}). 

Let us continue with derivation for the rest of the components of the source.
\begin{equation}
\begin{split}
\delta R^{(S_3)}&=0
\end{split}
\end{equation}

\begin{equation}
\begin{split}
\delta R^{(tr)}&=\frac{\psi^{-D}}{2}\frac{P^{CC'}}{D-2}\left[-2(\nabla_D O_C)(\nabla^D O_{C'})+\frac{K}{D}(\nabla_C O_{C'}+\nabla_{C'}O_C)\right]\\
&=\frac{\psi^{-D}}{2}\frac{1}{D-2}\left[-2~P^{CC'}(\nabla_D n_C)(\nabla^D n_{C'})+\frac{K}{D}P^{CC'}(\nabla_C n_{C'}+\nabla_{C'}n_C)\right]+{\cal{O}}\left(\frac{1}{D}\right)\\
&=\frac{\psi^{-D}}{2}\frac{1}{D-2}\left(-2\frac{K^2}{D}+2\frac{K^2}{D}\right)+{\cal{O}}\left(\frac{1}{D}\right)\\
&=0
\end{split}
\end{equation}

\begin{equation}\label{deltarv2}
\begin{split}
&~~~~\left[\delta R^{(V_1)}_{\text{lin}}\right]_A\\
&=\frac{\psi^{-D}}{2}P^C_A\left[2~K(u\cdot\nabla)O_C-\nabla^2 O_C\right]+\frac{\psi^{-D}}{2}P^C_A\Bigg[2~\bar{R}_{ECFD}~O^D O^E u^F\\
&+2\frac{(n\cdot\nabla)K}{K}[(u\cdot\nabla)O_C]+\frac{\nabla_C K}{D}\-2(\nabla_F O_C)(u^D\nabla^F n_D)+\frac{K}{D}(u\cdot\nabla)O_C+\frac{K}{D}(u^D K_{CD})\Bigg]\\ \\
&=\frac{e^{-R}}{2}P^C_A\left[2~K(u\cdot\nabla)O_C-\nabla^2 O_C\right]\bigg{|}_{\psi=1}+\frac{e^{-R}}{2}\left(\frac{\psi-1}{N}\right)(n\cdot\nabla)\left[P^C_A\left(2K(u\cdot\nabla)O_C-\nabla^2 O_C\right)\right]\bigg{|}_{\psi=1}\\
&~~~+\frac{e^{-R}}{2}P^C_A\Bigg[2~\bar{R}_{ECFD}~O^D O^E u^F+2\frac{(n\cdot\nabla)K}{K}[(u\cdot\nabla)O_C]+\frac{\nabla_C K}{D}\\
&~~~~~~~~~~~~~-2(\nabla_F O_C)(u^D\nabla^F n_D)+\frac{K}{D}(u\cdot\nabla)O_C+\frac{K}{D}u^D K_{CD}\Bigg]_{\psi=1}\\
&=\left(\frac{e^{-R}}{2}\right)\left[K~E_A^{\text{vector}}-2R\left(\frac{D}{K}\right)\mathfrak{v}_A\right]
\end{split}
\end{equation}
In the last line we have used the following two identities (see appendix \ref{vecder} and \ref{vecmemder} for derivation)
\begin{equation}\label{vectoriden}
\begin{split}
&~~~~(n\cdot\nabla)\left[P^C_A\left(2~K(u\cdot\nabla)O_C-\nabla^2 O_C\right)\right]_{R=0}=-2D~\mathfrak{v}_A
\end{split}
\end{equation}\\
\begin{equation}\label{vecmemder1}
\begin{split}
&P^C_A\bigg[2~K(u\cdot\nabla)O_C-\nabla^2 O_C+2~\bar{R}_{ECFD}~O^D O^E u^F+2\frac{(n\cdot\nabla)K}{K}[(u\cdot\nabla)O_C]+\frac{\nabla_C K}{D}\\
&~~~~~~~~~~~~~-2(\nabla_F O_C)(u^D\nabla^F n_D)+\frac{K}{D}(u\cdot\nabla)O_C+\frac{K}{D}u^D K_{CD}\bigg]_{\psi=1}=K~E_A^{\text{vector}}
\end{split}
\end{equation}
Where $E_A^{\text{vector}}$ is the subleading (see equation \eqref{eq:constraint2} ) membrane equation, and ${\mathfrak v}_A$ is given by
\begin{equation}
\begin{split}
\mathfrak{v}_{A}&={P}^B_{A}\Bigg[\frac{K}{D}\left(n^D u^E O^F \bar{R}_{FBDE}\right)+\frac{K^2}{2D^2}\left(\frac{{\nabla}_B {K}}{K}+(u\cdot{\nabla})u_B-2~u^D {K}_{D B}\right)\\
&~~~~~~~~~~~~~~-{P}^{F D} \left(\frac{{\nabla}_F {K}}{D}-\frac{K}{D} (u^E {K}_{E F})\right)\left({K}_{D B}-\nabla_D u_B\right)\Bigg]\\
\end{split}
\end{equation}

Note that  the simplification  of $\left[\delta R^{(V_1)}_{\text{lin}}\right]$ involves the same issues as in  $\delta R^{(S_2)}$.  The first line of the RHS of equation \eqref{deltarv2} is  of order ${\cal O}(D)$  by naive order counting. However,  because of the membrane equation at first subleading order, this is of ${\cal O}(1)$ on  $\psi = 1$ hypersurface. Away  from the hypersurface this may not be the case and we have to expand the first line around $\psi=1$ and take into account at least the first term in the expansion.  This is what has been done in the second line of equation \eqref{deltarv2}. In the final step we have re-written $\left[\delta R^{(V_1)}_{\text{lin}}\right]$ in terms of already-defined vector structure ${\mathfrak v}_A$ plus terms proportional to membrane equation.

The rest of the components of $S_{AB}$ are easy to compute without any further subtlety.
\begin{equation}
\begin{split}
\left[\delta R^{(V_2)}_{\text{lin}}\right]_C&=0
\end{split}
\end{equation}

\begin{equation}\label{tensorsource}
\begin{split}
&~~~~\left[\delta R^{(T)}_{\text{lin}}\right]_{AB}\\
&=\frac{\psi^{-D}}{2}P^C_A P^{C'}_B\left[2~ \bar{R}_{FCC'D}O^D O^F-2(\nabla_D O_C)(\nabla^D O_{C'})+\frac{K}{D}(\nabla_C O_{C'}+\nabla_{C'}O_C)\right]\\
&~~~~-\frac{\psi^{-D}}{2}\frac{P_{AB}}{D-2}P^{CC'}\left[-2(\nabla_D O_C)(\nabla^D O_{C'})+\frac{K}{D}(\nabla_C O_{C'}+\nabla_{C'}O_C)\right]\\
&=\frac{\psi^{-D}}{2}P^C_A P^{C'}_B\left[2~ \bar{R}_{FCC'D}O^D O^F-2(\nabla_D O_C)(\nabla^D O_{C'})+\frac{K}{D}(\nabla_C O_{C'}+\nabla_{C'}O_C)\right]\\
&=\psi^{-D}P^C_A P^{C'}_B\left[\frac{K}{D}\left(K_{CC'}-\frac{\nabla_C u_{C'}+\nabla_{C'}u_C}{2}\right)- P^E_F(K_{EC}-\nabla_E u_C)(K^F_{~~C'}-\nabla^F u_{C'})\right]\\
&~~~~~~~~~~~~~~~~+~\psi^{-D}P^C_A P^{C'}_B~ \bar{R}_{FCC'D}O^D O^F\\
&=\psi^{-D}~\mathfrak{t}_{AB}
\end{split}
\end{equation}
Where,
\begin{equation}
\begin{split}
\mathfrak{t}_{AB}&=P^C_A P^{D}_B\Bigg[+ \bar{R}_{FCDE}O^E O^F+\frac{K}{D}\left(K_{CD}-\frac{\nabla_C u_{D}+\nabla_{D}u_C}{2}\right)\\
&~~~~~~~~~~~~~~- P^{EF}(K_{EC}-\nabla_E u_C)(K_{FD}-\nabla_F u_{D})\Bigg]
\end{split}
\end{equation}\\
In deriving \eqref{tensorsource} we have used the following identity
\begin{equation}
P^C_A(\nabla_D O_C)=P^E_D P^C_A(\nabla_E O_C)-O_D[P^C_A(u\cdot\nabla)O_C]
\end{equation}
Which follows from the subsidiary condition.

\section{Some identities}\label{app:identity}
 In this appendix we shall prove some of the identities that we have used to compute the metric correction.
\subsection{The derivation of the Identity \eqref{eq:divtensor2}}\label{divderivation}
\begin{equation}
[\mathfrak{t}_1]_{CC'}=P^A_C P^B_{C'}\left[\frac{K}{D}\left(K_{AB}-\frac{{\nabla}_A u_B+{\nabla}_{B}u_A}{2}\right)- P^D_E(K_{D A}-{\nabla}_D u_A)(K^E_{~~B}-{\nabla}^E u_B)\right]
\end{equation}

\begin{equation}
\begin{split}
&~~~~\nabla^C [\mathfrak{t}_1]_{CC'}\\
&=\underbrace{\frac{K}{D}\nabla^C\left(P^A_C P^B_{C'}K_{AB}\right)}_{\text{Term-1}}-\underbrace{\left[\nabla^C\left\{P^A_CP^D_F(K_{DA}-\nabla_D u_A)\right\}\right]\left[P^B_{C'}P^{EF}(K_{EB}-\nabla_E u_B)\right]}_{\text{Term-2}}\\
&-\underbrace{\frac{K}{D}\nabla^C\left(P^A_C P^B_{C'}\frac{{\nabla}_A u_B+{\nabla}_{B}u_A}{2}\right)}_{Term-3}-\underbrace{\left[P^A_CP^D_F(K_{DA}-\nabla_D u_A)\right]\left[\nabla^C\left\{P^B_{C'}P^{EF}(K_{EB}-\nabla_E u_B)\right\}\right]}_{Term-4}
\end{split}
\end{equation}
After a bit of straight forward calculation the each of the above terms become
\begin{equation}\label{tr-1}
\begin{split}
\text{Term-1}\equiv\frac{K}{D}P^E_{C'}\nabla_E K
\end{split}
\end{equation}
\begin{equation}\label{tr-2}
\text{Term-2}\equiv P^{EA}P^B_{C'}[\nabla_E K-K (u^D K_{DE})](K_{AB}-\nabla_A u_B)
\end{equation}
\begin{equation}\label{tr-3}
\text{Term-3}\equiv\frac{K}{2D}P^E_{C'}[\nabla_E K+K(u\cdot\nabla)u_E]
\end{equation}
\begin{equation}\label{tr-4}
\text{Term-4}\equiv\frac{K}{D}P^F_{C'}[K~u^D K_{DF}-K(u\cdot\nabla)u_F]
\end{equation}
Adding \eqref{tr-1}, \eqref{tr-2}, \eqref{tr-3} and \eqref{tr-4} we get 
\begin{equation}
\begin{split}
\nabla^C [\mathfrak{t}_1]_{CC'}&=\frac{K}{2D}{P}^B_{C}\left[{\nabla}_B {K}+{K}(u\cdot{\nabla})u_B-2{K}(u^A {K}_{A B})\right]\\
&~~~~~~~~~~~~~-{P}^{B D} {P}^A_C\left({\nabla}_B {K}-{K} (u^E {K}_{E B})\right)\left[{K}_{D A}-\nabla_D u_A\right]
\end{split}
\end{equation}

\subsection{The derivation of scalar structure $\mathfrak{s}_2$ \eqref{vecdiv2}}\label{divderivation2}
The scalar structure $\mathfrak{s}_2$ is defined as 
\begin{equation}
\mathfrak{s}_2={\nabla\cdot\mathfrak{v}\over D}
\end{equation}
\begin{equation}
\begin{split}
\mathfrak{v}_{A}&={P}^B_{A}\Bigg[\frac{K}{D}\left(n^D u^E O^F \bar{R}_{FBDE}\right)+\frac{K^2}{2D^2}\left(\frac{{\nabla}_B {K}}{K}+(u\cdot{\nabla})u_B-2~u^D {K}_{D B}\right)\\
&~~~~~~~~~~~~~~-{P}^{F D} \left(\frac{{\nabla}_F {K}}{D}-\frac{K}{D} (u^E {K}_{E F})\right)\left({K}_{D B}-\nabla_D u_B\right)\Bigg]\\
\end{split}
\end{equation}
Now,
\begin{equation}
\begin{split}
\nabla^A \mathfrak{v}_A&=-K\Bigg[\frac{K^2}{2D^2}\left(\frac{(n\cdot\nabla)K}{K}+n^B(u\cdot\nabla)u_B\right)-P^{FD}\left(\frac{\nabla_F K}{D}-\frac{K}{D}u^E K_{EF}\right)\left(-n^B\nabla_D u_B\right)\\
&+\frac{K}{D}n^D u^E O^F n^B\bar{R}_{FBDE}\Bigg]+P^B_A\Bigg[\frac{K^2}{2D^2}\left(\frac{\nabla^A \nabla_B K}{K}+\nabla^A[(u\cdot\nabla)u_B]-2~u^D\nabla^A K_{DB}\right)\\
&-\left(\nabla^A P^{FD}\right)\left(\frac{\nabla_F K}{D}\right)K_{DB}-P^{FD}\left(\frac{\nabla^A\nabla_F K}{D}-\frac{K}{D}~u^E\nabla^A(K_{EF})\right)K_{DB}\\
&-P^{FD}\left(\frac{\nabla_F K}{D}-\frac{K}{D}\left(u^E K_{EF}\right)\right)\left(\nabla^A K_{DB}-\nabla^A\nabla_D u_B\right)+\frac{K}{D}\left(K^{AD}\right)u^E O^F \bar{R}_{FBDE}\Bigg]\\ \\
&=\frac{K^2}{D}\Bigg[-\frac{K}{2D}\left(\frac{(n\cdot\nabla)K}{K}-u\cdot K\cdot u\right)+P^{FD}\left(\frac{\nabla_F K}{K}-u^E K_{EF}\right)\left(u^B\nabla_D n_B\right)\\
&+n^D u^E u^F n^B\bar{R}_{FBDE}+\frac{1}{2D}\left(\frac{\nabla^2 K}{K}\right)-\frac{\lambda}{2}-\frac{1}{D}\left(u^D\nabla^A K_{DA}\right)+\frac{K}{D}\left(\frac{(n\cdot\nabla)K}{K}\right)\\
&-P^F_A\frac{1}{D}\left(\frac{\nabla^A \nabla_F K}{K}-u^E\nabla^A K_{EF}\right)-P^{FD}\left(\frac{\nabla_F K}{K}-u^E K_{EF}\right)\left(\frac{\nabla^A K_{DA}}{K}\right)-\lambda~\Bigg]
\end{split}
\end{equation}
Now using
\begin{equation}
\begin{split}
\frac{\nabla^2 K}{K^2}&=\frac{\tilde{\nabla}^2 K}{K^2}+\frac{(n\cdot\nabla)K}{K}+{\cal O}\left(\frac{1}{D}\right)\\
{\text{and,}}~~~~ \frac{\tilde{\nabla}^2 K}{K^2}&=2\left(\frac{u\cdot\nabla K}{K}\right)-u\cdot K\cdot u+\frac{\lambda(D-1)}{K}
\end{split}
\end{equation}\\
We get the final expression
\begin{equation}
\begin{split}
\nabla^A \mathfrak{v}_A&=\frac{K^2}{D}\Bigg[n^B n^D u^E u^F\bar{R}_{FBDE}-\frac{{K}}{D}\left(\frac{u\cdot{\nabla}{K}}{K}-u\cdot {K}\cdot u\right)- 2~\lambda\\
&~~~- (u\cdot {K}\cdot {K}\cdot u)+2 \left(\frac{{\nabla}_A{K}}{K}\right)u^B {K}^A_B-\left(\frac{u\cdot{\nabla}{K}}{K}\right)^2\\
&~~~+2\left(\frac{u\cdot{\nabla}{K}}{K}\right)(u\cdot {K}\cdot u)-\left(\frac{\tilde{\nabla}^D K}{K}\right)\left(\frac{\tilde{\nabla}_D K}{K}\right)-(u\cdot {K}\cdot u)^2\Bigg]\\
&=D~ \mathfrak{s}_2
\end{split}
\end{equation}

\subsection{The derivation of the Identity \eqref{eq:delusub}}\label{derivdelusub}
\begin{equation}
\begin{split}
\nabla\cdot u=\tilde{\nabla}\cdot u-\frac{(u\cdot\nabla) K}{K}-\frac{1}{K}(u\cdot\nabla)\left(\frac{(n\cdot\nabla)K}{K}\right)+\frac{1}{K}\frac{(n\cdot\nabla)K}{K}\frac{(u\cdot\nabla)K}{K}
\end{split}
\end{equation}

\begin{equation}
\begin{split}
\nabla\cdot u&=\tilde{\nabla}\cdot u+n_B(n\cdot\nabla)u^B\\
&=\tilde{\nabla}\cdot u-u^B\left[\psi K+\psi \frac{(n\cdot\nabla)N}{N}-N\right]^{-1}\tilde{\nabla}_B\left[\psi K+\psi \frac{(n\cdot\nabla)N}{N}-N\right]
\end{split}
\end{equation}
In the last line we have used the following relation
\begin{equation}
ND=\psi K+\psi\frac{(n\cdot\nabla)N}{N}-N
\end{equation}

\begin{equation}
\begin{split}
\nabla\cdot u&=\tilde{\nabla}\cdot u-u^B\left[\psi K+\psi \frac{(n\cdot\nabla)N}{N}-N\right]^{-1}\tilde{\nabla}_B\left[\psi K+\psi \frac{(n\cdot\nabla)N}{N}-N\right]\\
&=\tilde{\nabla}\cdot u-\left[1-\frac{(n\cdot\nabla)N}{NK}+\frac{N}{\psi K}\right]\left[\frac{(u\cdot\nabla)K}{K}+\frac{1}{K}(u\cdot\nabla)\left\{\frac{(n\cdot\nabla)N}{N}-\frac{N}{\psi}\right\}\right]\\
&=\tilde{\nabla}\cdot u-\frac{(u\cdot\nabla)K}{K}-\frac{1}{K}(u\cdot\nabla)\left\{\frac{(n\cdot\nabla)N}{N}-\frac{N}{\psi}\right\}+\left[\frac{(n\cdot\nabla)N}{NK}-\frac{N}{\psi K}\right]\frac{(u\cdot\nabla)K}{K}\\
&=\tilde{\nabla}\cdot u-\frac{(u\cdot\nabla)K}{K}-\frac{1}{K}(u\cdot\nabla)\left[\frac{(n\cdot\nabla)K}{K}\right]+\frac{1}{K}\left(\frac{(n\cdot\nabla)K}{K}\right)\left(\frac{(u\cdot\nabla)K}{K}\right)\\
\end{split}
\end{equation}

In the last line we have used
\begin{equation}
\frac{(n\cdot\nabla)N}{N}=\frac{(n\cdot\nabla)K}{K}+\frac{K}{D}
\end{equation}\\

\subsection{The derivation of the identity \eqref{vectoriden}}\label{vecder}

\begin{equation}
\begin{split}
&~~~~(n\cdot\nabla)\left[P^C_D\left\{2~K(u\cdot\nabla)O_C-\nabla^2 O_C\right\}\right]\\
&=(n\cdot\nabla)\left[P^C_D\left\{-2~K(n\cdot\nabla)u_C+\nabla^2 u_C\right\}\right]\\
&=\underbrace{\left[(n\cdot\nabla)P^C_D\right]\left[-2~K(n\cdot\nabla)u_C+\nabla^2 u_C\right]}_{\text{1 st Term}}+\underbrace{P^C_D(n\cdot\nabla)\left[-2~K(n\cdot\nabla)u_C+\nabla^2 u_C\right]}_{\text{2 nd Term}}
\end{split}
\end{equation}

\begin{equation}\label{1st term}
\begin{split}
&\textbf{1 st Term}\equiv\left[(n\cdot\nabla)P^C_D\right]\left[-2~K(n\cdot\nabla)u_C+\nabla^2 u_C\right]\\
&~~~=-n_D[(n\cdot\nabla)n^C]\left[-2~K(n\cdot\nabla)u_C+\nabla^2 u_C\right]+u_D[(n\cdot\nabla)u^C]\left[-2~K(n\cdot\nabla)u_C+\nabla^2 u_C\right]\\
&~~~~~~~~-[(n\cdot\nabla)n_D]\left[-2~K n^C(n\cdot\nabla)u_C+n^C\nabla^2 u_C\right]\\
&~~~=0
\end{split}
\end{equation}
Where, we have used
\begin{equation}
\begin{split}
&~~~~(n\cdot\nabla)n_D=-u_D\left[u^B(n\cdot\nabla)n_B\right]+P_D^B(n\cdot\nabla)n_B\\
&~~~~(n\cdot\nabla)u_D=n_D\left[n^B(n\cdot\nabla)u_B\right]+P_D^B(n\cdot\nabla)u_B\\
\text{And, }&-2~K(n\cdot\nabla)u_C+\nabla^2 u_C=n_C\left[2~K u_D(n\cdot\nabla)n^D-u_D\nabla^2 n^D\right]
\end{split}
\end{equation}\\
The third one follows from the fact that,
\begin{equation}
\begin{split}
&~~~~P^C_B\left[-2~K(n\cdot\nabla)u_C+\nabla^2 u_C\right]\\
&=P^C_B\left[\tilde{\nabla}^2 u_C-~K(n\cdot\nabla)u_C\right]\\
&=P^C_B\left[\tilde{\nabla}^2 u_C-\tilde{\nabla}_C K-K(u\cdot\nabla)u_C+K u^D K_{DC}\right]\\
&=0
\end{split}
\end{equation}
Where, $\left[E_1\right]_B^{\text{vector}}$ is the leading order membrane equation.

\begin{equation}\label{ter2}
\begin{split}
\textbf{2 nd Term}&\equiv P^C_D(n\cdot\nabla)\left[-2~K(n\cdot\nabla)u_C+\nabla^2 u_C\right]\\
&=P^C_D\left\{-2[(n\cdot\nabla)K][(n\cdot\nabla)u_C]-2~K~(n\cdot\nabla)[(n\cdot\nabla)u_C]+(n\cdot\nabla)(\nabla^2 u_C)\right\}
\end{split}
\end{equation}
Now,
\begin{equation}\label{ndeldot2}
\begin{split}
&~~~~P^C_D(n\cdot\nabla)(\nabla^2 u_C)\\
&=P^C_D~ n^E\nabla_E\nabla_F \nabla^F u_C\\
&=P^C_D~ n^E[\nabla_E,\nabla_F] \nabla^F u_C+P^C_D~ n^E\nabla_F\nabla_E \nabla^F u_C\\
&=P^C_D\left[-\lambda~(D-1)(n\cdot\nabla)u_C+n^E\bar{R}_{EFCB}\left(\nabla^F u^B\right)+ n^E \nabla^F[\nabla_E,\nabla_F]u_C+ n^E \nabla^F \nabla_F \nabla_E u_C\right]\\
&=P^C_D\bigg[-\lambda~(D-1)(n\cdot\nabla)u_C+n^E\bar{R}_{EFCB}\left(\nabla^F u^B\right)+n^E u^B\left(\nabla^F\bar{R}_{EFCB}\right)+n^E\bar{R}_{EFCB}\left(\nabla^F u^B\right)\\
&~~~+\tilde{\nabla}^2[(n\cdot\nabla)u_C]-(\nabla^2 n^E)(\nabla_E u_C)-2~(\nabla_F n^E)(\nabla^F\nabla_E u_C)+K~(n\cdot\nabla)[(n\cdot\nabla)u_C]\bigg]\\
&=P^C_D\bigg[\tilde{\nabla}^2[(n\cdot\nabla)u_C]-(\nabla^2 n^E)(\nabla_E u_C)-2~(\nabla_F n^E)(\nabla^F\nabla_E u_C)+K~(n\cdot\nabla)[(n\cdot\nabla)u_C]\\
&~~~~-~\lambda~(D-1)(n\cdot\nabla)u_C\bigg]\\
&=P^C_D\bigg[\frac{\tilde{\nabla}^2\tilde{\nabla}^2 u_C}{K}-\frac{1}{K^2}\left(\tilde{\nabla}^2 K\right)\tilde{\nabla}^2 u_C-\left(\tilde{\nabla}^2 n_C\right)\frac{u\cdot{\nabla}K}{K}-(\nabla^2 n^E)(\nabla_E u_C)\\
&~~~~-2~(\nabla_F n^E)(\nabla^F\nabla_E u_C)+K~(n\cdot\nabla)[(n\cdot\nabla)u_C]-\lambda~(D-1)(n\cdot\nabla)u_C\bigg]\\
\end{split}
\end{equation}
In the last line we have used,
\begin{equation}
\begin{split}
P^C_D\tilde{\nabla}^2[(n\cdot\nabla)u_C]&=P^C_D\tilde{\nabla}^2\left[P^E_C\frac{\tilde{\nabla}^2 u_E}{K}-n_C\frac{u\cdot{\nabla}K}{K}\right]\\
&=P^C_D\left[\frac{\tilde{\nabla}^2\tilde{\nabla}^2 u_C}{K}-\frac{1}{K^2}\left(\tilde{\nabla}^2 K\right)\tilde{\nabla}^2 u_C-\left(\tilde{\nabla}^2 n_C\right)\frac{u\cdot{\nabla}K}{K}\right]
\end{split}
\end{equation}
Using \eqref{ndeldot2} in \eqref{ter2} we get,
\begin{equation}
\begin{split}
&~~\textbf{2-nd Term}\\
&=-P^C_D\lambda(D-1)(n\cdot\nabla)u_C+P^C_D\bigg[-2[(n\cdot\nabla)K][(n\cdot\nabla)u_C]-K~(n\cdot\nabla)[(n\cdot\nabla)u_C]+\frac{\tilde{\nabla}^2\tilde{\nabla}^2 u_C}{K}\\
&~~~-\frac{1}{K^2}\left(\tilde{\nabla}^2 K\right)\tilde{\nabla}^2 u_C-\left(\tilde{\nabla}^2 n_C\right)\frac{u\cdot{\nabla}K}{K}-(\nabla^2 n^E)(\nabla_E u_C)-2~(\nabla_D n^E)(\nabla^D\nabla_E u_C)\bigg]
\end{split}
\end{equation}\\
Using the following identity whose derivation is a bit lengthy, and we are skipping the derivation
\begin{equation}\label{difficult}
\begin{split}
&~~~~P^C_B(n\cdot\nabla)[(n\cdot\nabla)u_C]\\
&=P^C_B\Bigg[-4~\frac{u\cdot{\nabla}K}{K}\left[(u\cdot{\nabla})u_C\right]+\left[(u\cdot{\nabla})u_C\right](u\cdot K\cdot u)-7~\frac{u\cdot{\nabla}K}{K}\frac{{\nabla}_C K}{K}+\frac{\tilde{\nabla}^2\tilde{\nabla}^2 u_C}{K^2}\\
&+3~(u\cdot K\cdot u)\frac{{\nabla}_C K}{K}-\frac{K}{D}u^DK_{DC}+4\left(u^DK_{DC}\right)\frac{u\cdot{\nabla}K}{K}-u^DK_{DC}(u\cdot K\cdot u)-2K^D_C\frac{\nabla_D K}{K}\\
&-2(u_E K^{ED})({\nabla}_D u_C)+2~K^{AF}K_{AC} u_F-2\frac{\lambda(D-1)}{K}\frac{\tilde{\nabla}^2 u_C}{K}-2~u^F n^E O^A\bar{R}_{EFCA}\Bigg]
\end{split}
\end{equation}
Now,
\begin{equation}
\begin{split}
&~~\textbf{2-nd Term}\\
&=P^C_B\left[-\frac{K^2}{D}\left((u\cdot{\nabla})u_C-u^D K_{DC}+\frac{{\nabla}_C K}{K}\right)\right]+P^C_B K\bigg[2~u^F n^E O^A\bar{R}_{EFCA}\\
&+2~K^D_C\frac{\nabla_D K}{K}+2(u_E K^{ED})({\nabla}_D u_C)-2~K^{AF} K_{AC} u_F-2~\frac{\tilde{\nabla}^E K}{K}(\nabla_E u_C)\\
&-2\frac{u\cdot{\nabla}K}{K}(u\cdot{\nabla})u_C+2\frac{u\cdot{\nabla}K}{K} u^D K_{DC}+2(u\cdot K\cdot u)[(u\cdot{\nabla})u_C]-2(u\cdot K\cdot u)(u^D K_{DC})\bigg]\\
&=-2~D~\mathfrak{v}_B
\end{split}
\end{equation}

Finally, we get
\begin{equation}
\begin{split}
&~~~~(n\cdot\nabla)\left[P^C_D\left\{2~K(u\cdot\nabla)O_C-\nabla^2 O_C\right\}\right]=-2D~\mathfrak{v}_D
\end{split}
\end{equation}

\subsection{The derivation of the identity \eqref{vecmemder1}}\label{vecmemder}
We can divide the L.H.S. of \eqref{vecmemder1} as follows
\begin{equation}\label{b26}
\begin{split}
&P^C_B\bigg[2~K(u\cdot\nabla)O_C-\nabla^2 O_C+2~n^D O^E u^F\bar{R}_{ECFD}~+2\frac{(n\cdot\nabla)K}{K}[(u\cdot\nabla)O_C]+\frac{\nabla_C K}{D}\\
&~~~-2(\nabla_F O_C)(u^D\nabla^F n_D)+\frac{K}{D}(u\cdot\nabla)O_C+\frac{K}{D}u^D K_{CD}\bigg]\equiv P^C_A ~\nabla^2 u_C-P^C_A \nabla^2 n_C+W
\end{split}
\end{equation}
where $W$ is what we get by subtracting off $ P^C_A ~\nabla^2 u_C-P^C_A \nabla^2 n_C$ from the LHS of equation \eqref{b26}.\\
First we shall simpify $W$
\begin{equation}\label{temr3}
\begin{split}
W&=P^C_B\bigg[2~K(u\cdot\nabla)O_C+2~n^D O^E u^F\bar{R}_{ECFD}~+2\frac{(n\cdot\nabla)K}{K}[(u\cdot\nabla)O_C]+\frac{\nabla_C K}{D}\\
&~~~~-2(\nabla_F O_C)(u^D\nabla^F n_D)+\frac{K}{D}(u\cdot\nabla)O_C+\frac{K}{D}u^D K_{CD}\bigg]\\
&=P^C_B\bigg[2K \left(u^D K_{DC}\right)-2K(u\cdot\nabla)u_C+2~u^D K_{DC}\left(\frac{(u\cdot\nabla)K}{K}-u\cdot K\cdot u\right)\\
&~~~~-2[(u\cdot\nabla)u_C]\left(\frac{(u\cdot\nabla)K}{K}-u\cdot K\cdot u\right)+\frac{\nabla_C K}{D}-2~u_D K_{FC}  K^{FD}\\
&~~~~+2(\nabla_F u_C)\left(u_D K^{FD}\right)+\frac{K}{D}[(u\cdot\nabla)u_C]+2~n^DO^E u^F\bar{R}_{ECFD}\bigg]
\end{split}
\end{equation}  
Now, we shall simplify $P^C_A~\nabla^2 n_C$
\begin{equation}
\begin{split}
P^C_B\nabla^2 n_C&=P^C_B \nabla^D\left(\nabla_D n_C\right)\\
&=P^C_B\nabla^D\left[K_{DC}+n_D(n\cdot\nabla)n_C\right]\\
&=\underbrace{P^C_B\nabla^D K_{DC}}_{\text{$T_1$}}+\underbrace{P^C_B K(n\cdot\nabla)n_C}_{\text{$T_2$}}+\underbrace{P^C_B(n\cdot\nabla)\left[(n\cdot\nabla)n_C\right]}_{\text{$T_3$}}
\end{split}
\end{equation}
\begin{equation}\label{term1}
\begin{split}
\text{$T_1$}&\equiv P^C_B\nabla^D K_{DC}\\
&=P^C_B \nabla^D K_{CD}\\
&=P^C_B \nabla^D\left(\Pi^E_C\nabla_E n_D\right)\\
&=P^C_B \left[(\nabla^D \Pi^E_C)(\nabla_E n_D)+\Pi^E_C\left(\nabla^D\nabla_E n_D\right)\right]\\
&=P^C_B \left\{-(\nabla^D n_C)[(n\cdot\nabla) n_D]+\Pi^E_C \nabla_E\nabla^D n_D\right\}+P^E_B[\nabla_D,\nabla_E]n^D\\
&=-P^C_B K^D_C\left(\frac{{\nabla}_D K}{K}\right)+P^C_B {\nabla}_C K-P^E_B \bar{R}_{DEC}^{~~~~~~D}n^C\\
&=-P^C_B K^D_C\left(\frac{{\nabla}_D K}{K}\right)+P^C_B~{\nabla}_C K
\end{split}
\end{equation}
\begin{equation}\label{term2}
\begin{split}
\text{$T_2$}&\equiv P^C_B K[(n\cdot \nabla)n_C]\\
&=P^C_B ~K~ \frac{{\nabla}_C(N D)}{N D}\\
&=P^C_B K\frac{1}{\psi K+\psi\frac{(n\cdot\nabla)N}{N}-N}{\nabla}_C\left(\psi K+\psi\frac{(n\cdot\nabla)N}{N}-N\right)\\
&=P^C_B \left(1-\frac{(n\cdot\nabla)N}{N K}+\frac{N}{\psi K}\right){\nabla}_C\left( K+\frac{(n\cdot\nabla)N}{N}-\frac{N}{\psi}\right)\\
&=P^C_B~{\nabla}_C\left( K+\frac{(n\cdot\nabla)N}{N}-\frac{N}{\psi}\right)+P^C_B \left(-\frac{(n\cdot\nabla)N}{N K}+\frac{N}{\psi K}\right){\nabla}_C K\\
&=P^C_B{\nabla}_C K+P^C_B{\nabla}_C\left(\frac{(n\cdot\nabla)K}{K}\right)-P^C_B\left(\frac{(n\cdot\nabla)K}{K}\right)\left(\frac{{\nabla}_C K}{K}\right)
\end{split}
\end{equation}
In the first line we have used
\begin{equation}
ND=\psi K+\psi\frac{(n\cdot \nabla)N}{N}-N
\end{equation}
And, in the last line we have used
\begin{equation}
\frac{(n\cdot\nabla)N}{N}=\frac{(n\cdot\nabla)K}{K}+\frac{K}{D}
\end{equation}
\begin{equation}\label{term3}
\begin{split}
\text{$T_3$}&\equiv P^C_B(n\cdot\nabla)[(n\cdot\nabla)n_C]\\
&=P^C_B\left[(n\cdot\nabla)\Pi^D_C\right]\left(\frac{\nabla_D N}{N}\right)+P^C_B(n\cdot\nabla)\left(\frac{\nabla_C N}{N}\right)\\
&=-P^C_B\left[(n\cdot\nabla)n_C\right]\left(\frac{(n\cdot\nabla) N}{N}\right)-P^C_B\frac{1}{N^2}[(n\cdot\nabla)N](\nabla_C N)+P^C_B\frac{1}{N}[(n\cdot\nabla)(\nabla_C N)]\\
&=-P^C_B\left(\frac{{\nabla}_C K}{K}\right)\left(\frac{(n\cdot\nabla) N}{N}\right)-P^C_B\left(\frac{(n\cdot\nabla)N }{N}\right)\left(\frac{{\nabla}_C K}{K}\right)+P^C_B\frac{1}{N}n^D\nabla_C\nabla_D N\\
&=-2P^C_B\left(\frac{{\nabla}_C K}{K}\right)\left(\frac{(n\cdot\nabla) N}{N}\right)+P^C_B\frac{1}{N}\nabla_C[(n\cdot\nabla) N]-P^C_B\frac{1}{N}\left(\nabla_C n^D\right)(\nabla_D N)\\
&=-2P^C_B\left(\frac{{\nabla}_C K}{K}\right)\left(\frac{(n\cdot\nabla) N}{N}\right)+P^C_B\nabla_C\left(\frac{(n\cdot\nabla) N}{N}\right)+\frac{1}{N^2}P^C_B(\nabla_C N)[(n\cdot\nabla)N]\\
&~~~~~~~~~~~~~~~~~~~~~~~~~~~~~~-P^C_B\frac{1}{N}\left(\nabla_C n^D\right)(\nabla_D N)\\
&=-2P^C_B\left(\frac{{\nabla}_C K}{K}\right)\left(\frac{(n\cdot\nabla) K}{K}+\frac{K}{D}\right)+P^C_B\nabla_C\left(\frac{(n\cdot\nabla) K}{K}+\frac{K}{D}\right)\\
&~~~~~~~+P^C_B\left(\frac{{\nabla}_C K}{K}\right)\left(\frac{(n\cdot\nabla) K}{K}+\frac{K}{D}\right)-P^C_B K^D_C\left(\frac{{\nabla}_D K}{K}\right)\\
&=-2P^C_B\left(\frac{{\nabla}_C K}{K}\right)\left(2\frac{(u\cdot\nabla) K}{K}-u\cdot K\cdot u\right)+P^C_B\nabla_C\left(\frac{\tilde{\nabla}^2 K}{K^2}\right)+P^C_B\frac{{\nabla}_C K}{K}\frac{\lambda(D-1)}{K}\\
&~~~~~~~+P^C_B\left(\frac{{\nabla}_C K}{K}\right)\left(2\frac{(u\cdot\nabla) K}{K}-u\cdot K\cdot u\right)-P^C_B K^D_C\frac{{\nabla}_D K}{K}
\end{split}
\end{equation}
In the last line we have used 
\begin{equation}
\frac{(n\cdot\nabla)K}{K}=\frac{\tilde{\nabla}^2 K}{K^2}-\frac{(D-1)\lambda}{K}-\frac{K}{D}
\end{equation}
And, divergence of leading order vector membrane equation
\begin{equation}
\frac{\tilde{\nabla}^2 K}{K^2}=2\frac{u\cdot{\nabla}K}{K}-u\cdot K\cdot u+\frac{\lambda(D-1)}{K}
\end{equation}

Adding \eqref{term1} \eqref{term2} and \eqref{term3} we get 
\begin{equation}\label{del2nc}
\begin{split}
P^C_B\nabla^2 n_C
&=P^C_B\bigg[2{\nabla}_C K-2 K^D_C\left(\frac{{\nabla}_D K}{K}\right)+\frac{2}{K^2}{\nabla}_C\left(\tilde{\nabla}^2 K\right)-2~\frac{{\nabla}_C K}{K}\frac{\lambda(D-1)}{K}\\
&~~~~~~~~~~~~~~~~~~~-6\left(\frac{{\nabla}_C K}{K}\right)\left(2\frac{(u\cdot{\nabla}) K}{K}-u\cdot K\cdot u\right)\bigg]
\end{split}
\end{equation}\\
Now, we shall simplify $P^C_B{\nabla}^2 u_C$
\begin{equation}\label{del2u1}
\begin{split}
&~~~~P^C_B\tilde{\nabla}^2 u_C\\
&=P^C_B\tilde{\nabla}^E\left(\Pi^F_E\Pi^D_C\nabla_F u_D\right)\\
&=P^C_B\Pi^E_M\nabla^M\left(\Pi^F_E\Pi^D_C\nabla_F u_D\right)\\
&=P^C_B\Pi^N_M\left(\nabla^M\Pi^F_N\right)(\nabla_F u_C)+P^C_B\Pi^F_M\left(\nabla^M\Pi^D_C\right)(\nabla_F u_D)+P^C_B\Pi^F_M\nabla^M\nabla_F u_C\\
&=P^C_B\bigg[- \Pi^N_M~ n^F\left(\nabla^M n_N\right)(\nabla_F u_C)-\Pi^F_M n^D\left(\nabla^M n_C\right)(\nabla_F u_D)+\nabla^2 u_C- n^F n_M \nabla^M \nabla_F u_C\bigg]\\
&=P^C_B\bigg[-n^F~K(\nabla_F u_C)-n^D\left(\nabla^M n_C\right)(\nabla_M u_D)+n^D[(n\cdot\nabla)n_C][(n\cdot\nabla)u_D]\\
&~~~~~~~~~~~~+\nabla^2 u_C-n_M\nabla^M\left(n^F\nabla_F u_C\right)+n_M\left(\nabla^M n^F\right)(\nabla_F u_C)\bigg]\\
&=P^C_B\bigg[-K[(n\cdot\nabla)u_C]-\left(\nabla^M n_C\right)(n^D\nabla_M u_D)+[(n\cdot\nabla)n_C][n^D(n\cdot\nabla)u_D]\\
&~~~~~~~~~~~~+\nabla^2 u_C-(n\cdot\nabla)[(n\cdot\nabla)u_C]+\left[(n\cdot\nabla)n^F\right](\nabla_F u_C)\bigg]\\
&\Rightarrow P^C_B\nabla^2 u_C=P^C_B\bigg[\tilde{\nabla}^2 u_C+K[(n\cdot\nabla)u_C]+\left(\nabla^M n_C\right)(n^D\nabla_M u_D)\\
&~~~~~~~~~~-[(n\cdot\nabla)n_C][n^D(n\cdot\nabla)u_D]+(n\cdot\nabla)[(n\cdot\nabla)u_C]-\left[(n\cdot\nabla)n^F\right](\nabla_F u_C)\bigg]\\
\end{split}
\end{equation}
\begin{equation}\label{del2u2}
\begin{split}
\text{Now, } &~~~~P^C_B\left(\nabla^M n_C\right)\left(n^D\nabla_M u_D\right)\\
&=-P^C_B\left[K^M_C+n^M(n\cdot\nabla)n_C\right]\left[u_DK_M^D+u_D n_M(n\cdot\nabla)n^D\right]\\
&=-P^C_B K^M_C K^D_M u_D-P^C_B[(n\cdot\nabla)n_C]\left[u^D(n\cdot\nabla)n_D\right]\\
&=-P^C_B K^M_C K^D_M u_D-P^C_B\frac{{\nabla}_C K}{K}\left(\frac{u\cdot\hat{\nabla}K}{K}\right)
\end{split}
\end{equation}

Putting \eqref{del2u2} in \eqref{del2u1} we get
\begin{equation}\label{del2u3}
\begin{split}
P^C_B\nabla^2 u_C&=P^C_B\tilde{\nabla}^2 u_C+P^C_B~K[(n\cdot\nabla)u_C]-P^C_BK^M_C K^D_M u_D\cancel{-P^C_B\frac{{\nabla}_C K}{K}\left(\frac{u\cdot{\nabla}K}{K}\right)}\\
&\cancel{+P^C_B\frac{{\nabla}_C K}{K}\left(\frac{u\cdot{\nabla}K}{K}\right)}+P^C_B~(n\cdot\nabla)[(n\cdot\nabla)u_C]-P^C_B\frac{\tilde{\nabla}^F K}{K}(\nabla_F u_C)\\
\Rightarrow P^C_B\nabla^2 u_C &=P^C_B\tilde{\nabla}^2 u_C+P^C_B~ K[(n\cdot\nabla)u_C]-P^C_BK^M_C K^D_M u_D+P^C_B~(n\cdot\nabla)[(n\cdot\nabla)u_C]\\
&~~~~~~~~~~~~~~~~~-P^C_B\frac{\tilde{\nabla}^F K}{K}(\nabla_F u_C)\\
\end{split}
\end{equation}
As we have mentioned before derivation of $P^C_B(n\cdot\nabla)[(n\cdot\nabla)u_C]$ is lengthy, we shall use the result mentioned in eq\eqref{difficult}\\
Using \eqref{difficult} for $P^C_B(n\cdot\nabla)[(n\cdot\nabla)u_C]$ we get the final expression for $P^C_B\nabla^2 u_C$
\begin{equation}\label{del2ucf}
\begin{split}
&P^C_B\nabla^2 u_C
=P^C_B\Bigg[\tilde{\nabla}^2 u_C+K[(n\cdot\nabla)u_C]-4\frac{(u\cdot{\nabla})K}{K}\left[(u\cdot{\nabla})u_C\right]+\left[(u\cdot{\nabla})u_C\right](u\cdot K\cdot u)\\
&~~~-7\left(\frac{u\cdot{\nabla}K}{K}\right)\frac{{\nabla}_C K}{K}-\frac{\tilde{\nabla}_D K}{K}\left({\nabla}^D u_C\right)+3~(u\cdot K\cdot u)\frac{{\nabla}_C K}{K}+\frac{\tilde{\nabla}^2\tilde{\nabla}^2 u_C}{K^2}-\frac{K}{D}u^DK_{DC}\\
&~~~+4\left(u^DK_{DC}\right)\frac{u\cdot{\nabla}K}{K}-u^DK_{DC}(u\cdot K\cdot u)-2~K^D_C\frac{\nabla_D K}{K}-2(u_E K^{ED})({\nabla}_D u_C)\\
&~~~+K^{AF}K_{AC} u_F-2\frac{(D-1)\lambda}{K}\left(\frac{{\nabla}_C K}{K}-u^E K_{EC}+(u\cdot{\nabla})u_C\right)-2~n^E u^F O^A \bar{R}_{EFCA}\Bigg]\\
\end{split}
\end{equation}
Adding \eqref{temr3} \eqref{del2nc} and \eqref{del2ucf} we get the final expression
\begin{equation}
\begin{split}
&\frac{1}{K}\left(P^C_B ~\nabla^2 u_C-P^C_B \nabla^2 n_C+W\right)\\
&=\bigg[\frac{\tilde{\nabla}^2 u_C}{{K}}-\frac{\tilde{\nabla}_C{{K}}}{{K}}+u^E {{K}}_{E C}-u\cdot\tilde{\nabla} u_C\bigg]{ P}^C_B+\bigg[\frac{\tilde{\nabla}^2\tilde{\nabla}^2 u_C}{{{K}}^3}-\frac{u^E {{K}}_{E D} {{K}}^D_C}{{K}}-\frac{(\tilde{\nabla}_C{{K}})(u\cdot\tilde{\nabla}{{K}})}{{{K}}^3}\\
&-\frac{(\tilde{\nabla}_E{{K}})(\tilde{\nabla}^E u_C)}{{{K}}^2}-\frac{2{{K}}^{D E}\tilde{\nabla}_D\tilde{\nabla}_E u_C}{K^2}-\frac{\tilde{\nabla}_C\tilde{\nabla}^2{{K}}}{{{K}}^3}+\frac{\tilde{\nabla}_C({{K}}_{ED} {{K}}^{ED} {{K}})}{K^3}+3\frac{(u\cdot {{K}}\cdot u)(u\cdot\tilde{\nabla} u_C)}{{{K}}}\\
&-3\frac{(u\cdot {{K}}\cdot u)(u^E {{K}}_{E C})}{{{K}}}-6\frac{(u\cdot\tilde{\nabla}{{K}})(u\cdot\tilde{\nabla} u_C)}{{{K}}^2}+6\frac{(u\cdot\tilde{\nabla}{{K}})(u^E {{K}}_{E C})}{{{K}}^2}+3\frac{u\cdot\tilde{\nabla} u_C}{D-3}\\
&-3\frac{u^E {{K}}_{E C}}{D-3}-\frac{(D-1)\lambda}{{{K}}^2}\bigg(\frac{\tilde{\nabla}_C {{K}}}{{K}}-2u^D {{K}}_{D C}+2(u\cdot\tilde{\nabla})u_C\bigg)\bigg]{ P}^C_B\\
&\equiv E_{B}^{\text{vector}}
\end{split}
\end{equation}
\\
Where, in the last step we have used the following identity
\begin{equation}
\begin{split}
P^C_B(n\cdot\nabla)u_C=&P^C_B \Bigg[\frac{\nabla_C K}{K}+\frac{1}{K}\nabla_C\left(\frac{\tilde{\nabla}^2 K}{K^2}-\frac{(D-1)\lambda}{K}-\frac{K}{D}\right)-u^D K_{DC}+(u\cdot\nabla)u_C\\
&~~~~~~~-\frac{1}{K}\left(\frac{\nabla_C K}{K}\right)\left(2\frac{(u\cdot\nabla)K}{K}-u\cdot K\cdot u-\frac{K}{D}\right)\Bigg]
\end{split}
\end{equation}

\section{QNM for AdS/dS  Schwarzschild Black hole:\\ Details of the calculation}\label{app:QNMGlobal}
In this subsection we shall present several computational details. We shall follow \cite{Chmembrane} and \cite{arbBack}. Steps are tedious but a straightforward extension of what has been done in \cite{arbBack}.

The answers for non-zero components of Christoffel symbols for metric \eqref{globmet} are  (denoting the metric on unit sphere by $\bar g_{ab}$, its Christoffel symbols by $\bar\Gamma^a_{bc}$ and the covariant derivatives with respect to $\bar g_{ab}$ by $\bar\nabla_a$)
\begin{equation}
 \begin{split}
  \Gamma^r_{ab} &= -r \left( 1-\frac{\sigma r^2}{L^2} \right)\bar{g}_{ab},~~~
  \Gamma^a_{rb} = \frac{1}{r} \delta^a_b,~~~
  \Gamma^r_{tt} = -r \left( 1-\frac{\sigma r^2}{L^2} \right)\frac{\sigma}{L^2} \\
    \Gamma^t_{rt} &= -r  \left( 1-\frac{\sigma r^2}{L^2} \right)^{-1} \frac{\sigma}{L^2},~~~
  \Gamma^r_{rr} = r  \left( 1-\frac{\sigma r^2}{L^2} \right)^{-1} \frac{\sigma}{L^2},~~~\Gamma^a_{bc}=\bar\Gamma^a_{bc} \\   
 \end{split}
\end{equation}
 The normal to membrane evaluates to
 \begin{equation}
  \begin{split}
   n_r &=  \left( 1-\frac{\sigma r^2}{L^2} \right)^{-\frac{1}{2}},~~
   n_t =  \left( 1-\frac{\sigma r^2}{L^2} \right)^{-\frac{1}{2}} (-\epsilon \partial_t \delta r) ,~~
   n_a =  \left( 1-\frac{\sigma r^2}{L^2} \right)^{-\frac{1}{2}} (-\epsilon \bar\nabla_a \delta r) 
  \end{split}
 \end{equation}
 $\nabla_An_B$ evaluates to
\begin{equation}
 \begin{split}
  \nabla_rn_r &= 0,~ ~~~
  \nabla_rn_t = \left( 1-\frac{\sigma r^2}{L^2} \right)^{-\frac{3}{2}}\frac{2\sigma r}{L^2}(-\epsilon \partial_t \delta r) \\
  \nabla_tn_r &= \left( 1-\frac{\sigma r^2}{L^2} \right)^{-\frac{3}{2}}\frac{\sigma r}{L^2}(-\epsilon \partial_t \delta r),~~~\nabla_an_t = (-\epsilon \partial_t \bar\nabla_a \delta r)\left( 1-\frac{\sigma r^2}{L^2} \right)^{-\frac{1}{2}}  \\
  \nabla_tn_t &= \left( 1-\frac{\sigma r^2}{L^2} \right)^{-\frac{1}{2}}(-\epsilon \partial^2_t \delta r) + \left( 1-\frac{\sigma r^2}{L^2} \right)^{\frac{1}{2}}\frac{\sigma r}{L^2} \\
  \nabla_rn_a &= (-\epsilon \bar\nabla_a \delta r)\left[ \frac{\sigma r}{L^2}\left( 1-\frac{\sigma r^2}{L^2} \right)^{-\frac{3}{2}}-\frac{1}{r}\left( 1-\frac{\sigma r^2}{L^2} \right)^{-\frac{1}{2}} \right] \\
  \nabla_an_r &= (\epsilon \bar\nabla_a \delta r)\frac{1}{r}\left( 1-\frac{\sigma r^2}{L^2} \right)^{-\frac{1}{2}},~~~
  \nabla_tn_a = (-\epsilon \partial_t \bar\nabla_a \delta r)\left( 1-\frac{\sigma r^2}{L^2} \right)^{-\frac{1}{2}} \\
\nabla_an_b &= \left( 1-\frac{\sigma r^2}{L^2} \right)^{-\frac{1}{2}}(-\epsilon \bar\nabla_a \bar\nabla_b \delta r) + r\left( 1-\frac{\sigma r^2}{L^2} \right)^{\frac{1}{2}}\bar{g}_{ab}
 \end{split}
\end{equation}
The projector $P_{A}^{B}=\delta^A_B - n^A n_B $ evaluates to
\begin{equation}
\begin{split}
 & P^r_r = 0,\quad P^t_t = 1,\quad P^a_b = \delta^a_b,\quad P^t_a = 0,\quad P^a_t = 0,\\
 & P^r_t = \epsilon \partial_t \delta r,\quad P^t_r = \left( 1-\frac{\sigma r^2}{L^2} \right)^{-2} (-\epsilon \partial_t \delta r), \\
 & P^r_a = \epsilon \bar\nabla_a \delta r,\quad P^a_r = \frac{1}{r^2}\left( 1-\frac{\sigma r^2}{L^2} \right)^{-1} (\epsilon \bar\nabla^a \delta r)
\end{split}
\end{equation}
The spacetime form of Extrinsic curvature $K_{AB}=\Pi_{A}^{C} {\nabla}_C n_B$ evaluates to 
\begin{equation}\label{Cstb}
 \begin{split}
  K_{rr} &= 0,~~~ 
  K_{rt} = \left( 1-\frac{\sigma r^2}{L^2} \right)^{-\frac{3}{2}} \frac{\sigma r}{L^2} (-\epsilon \partial_t \delta r),~~~
  K_{ra} = \frac{1}{r} \left( 1-\frac{\sigma r^2}{L^2} \right)^{-\frac{1}{2}}(\epsilon \bar\nabla_a \delta r) \\
K_{ta} &= \left( 1-\frac{\sigma r^2}{L^2} \right)^{-\frac{1}{2}}(-\epsilon \partial_t \bar\nabla_a \delta r),~~~
  K_{tt} = \left( 1-\frac{\sigma r^2}{L^2} \right)^{-\frac{1}{2}}(-\epsilon \partial^2_t \delta r)+\left( 1-\frac{\sigma r^2}{L^2} \right)^{\frac{1}{2}}\frac{\sigma r}{L^2} \\ 
  K_{ab} &= \left( 1-\frac{\sigma r^2}{L^2} \right)^{-\frac{1}{2}}(-\epsilon \bar\nabla_a\bar\nabla_b \delta r)+r\left( 1-\frac{\sigma r^2}{L^2} \right)^{\frac{1}{2}}\bar{g}_{ab} 
 \end{split}
\end{equation}
Answers for the nonzero components of Christoffel symbols for metric \eqref{lind} are
\begin{equation}
 \begin{split}
  \Gamma^t_{tt} &= -\left( 1-\frac{\sigma }{L^2} \right)^{-1}\frac{\sigma}{L^2}(\epsilon \partial_t \delta r),~~~
  \Gamma^a_{tt} = -\frac{\sigma}{L^2}(\epsilon \bar\nabla^a \delta r) \\
  \Gamma^t_{at} &= -\left( 1-\frac{\sigma }{L^2} \right)^{-1}\frac{\sigma}{L^2}(\epsilon \bar\nabla_a \delta r),~~~
  \Gamma^t_{ab} = \left( 1-\frac{\sigma }{L^2} \right)^{-1}(\epsilon \partial_t \delta r)\bar{g}_{ab} \\
  \Gamma^a_{tb} &= (\epsilon \partial_t \delta r)\delta^a_b,~~~
  \Gamma^a_{bc} = \bar{\Gamma}^a_{bc} + \epsilon(\bar\nabla_b \delta r \delta^a_c+\bar\nabla_c \delta r \delta^a_b-\bar\nabla^a \delta r \bar{g}_{bc})
 \end{split}
\end{equation}
 $\hat{\nabla}_\mu u_\nu$ evaluates to
\begin{equation}
\begin{split}
\hat{\nabla}_t u_t &= 0,~~~
 \hat{\nabla}_t u_a = \epsilon \partial_t \delta u_a - \left( 1-\frac{\sigma}{L^2} \right)^{-\frac{1}{2}}\left(\frac{\sigma}{L^2}\right) (\epsilon \bar\nabla_a \delta r) \\
 \hat{\nabla}_a u_t &= 0,~~~
 \hat{\nabla}_a u_b = \epsilon \bar\nabla_a \delta u_b + \left( 1-\frac{\sigma}{L^2} \right)^{-\frac{1}{2}} (\epsilon \partial_t \delta r) \bar{g}_{ab}
 \end{split}
\end{equation}
The projector ${\cal P}^\mu_\nu\equiv \delta^\mu_\nu + u^\mu u_\nu$ evaluates to
\begin{equation}
 {\mathcal P}^t_t = 0,\quad {\mathcal P}^a_t = -\left( 1-\frac{\sigma }{L^2} \right)^{\frac{1}{2}}(\epsilon \delta u_a),\quad {\mathcal P}^t_a = \left( 1-\frac{\sigma }{L^2} \right)^{-\frac{1}{2}}(\epsilon \delta u_a),\quad {\mathcal P}^a_b = \delta^a_b
\end{equation}
\subsection{Computation of $ {\cal K}_{\mu\nu}$}

We define ${\cal K}_{\mu\nu}$ as the pullback of Extrinsic curvature $K_{MN}$ (which is a spacetime tensor) on the membrane surface
\begin{equation}\label{eq:pullback}
 {\cal K}_{\mu\nu} = \left( \frac{\partial X^M}{\partial y^\mu} \right) \left( \frac{\partial X^N}{\partial y^\nu} \right) K_{MN} \vert_{r= 1 + \epsilon \delta r}
 \end{equation}
 where we denote the coordinates in spacetime $(r,t,\theta^a)$ by $X^M$ and the coordinates on the membrane worldvolume $(t,\theta^a)$ by $y^\mu$.
The extrinsic curvature $K_{AB}$ is defined as 
\begin{equation}
\begin{split}
K_{AB}&=\Pi_{A}^{C} {\nabla}_C n_B,~~~
\text{where}~~~ \Pi_{AC}=g_{AC} -n_A n_C
\end{split}
\end{equation}
Now equation \eqref{eq:pullback} evaluated upto linear order for the QNM calculation implies that 
\begin{equation}\label{eq:pullapply}
\begin{split}
 {\cal K}_{\mu\nu} =\epsilon (\partial_\mu\delta r ) K_{r\nu} +\epsilon (\partial_\nu\delta r ) K_{r\mu} + K_{\mu\nu} + {\cal O}(\epsilon^2)
 \end{split}
\end{equation}
From \eqref{Cstb} we see that $K_{rN}={\mathcal O}(\epsilon)$. Using this fact along with \eqref{eq:pullapply} gives us
 \begin{equation}\label{ECmb}
  \begin{split}
   {\cal K}_{tt} &=  \left( 1-\frac{\sigma}{L^2} \right)^{-\frac{1}{2}} (-\epsilon \partial^2_t \delta r) + \left( 1-\frac{\sigma}{L^2} \right)^{\frac{1}{2}} \left(\frac{\sigma}{L^2}\right) \left( 1 + \epsilon\delta r - \frac{\sigma \epsilon \delta r}{L^2-\sigma} \right) \\
    {\cal K}_{ta} &=  \left( 1-\frac{\sigma}{L^2} \right)^{-\frac{1}{2}} (-\epsilon \partial_t \bar\nabla_a \delta r) \\
     {\cal K}_{ab} &=  \left( 1-\frac{\sigma}{L^2} \right)^{-\frac{1}{2}} (-\epsilon \bar\nabla_a \bar\nabla_b \delta r) + \left( 1-\frac{\sigma}{L^2} \right)^{\frac{1}{2}}  \left( 1 + \epsilon\delta r - \frac{\sigma \epsilon \delta r}{L^2-\sigma} \right)\hat{g}_{ab} \\
  \end{split}
 \end{equation}
Trace of Extrinsic curvature \eqref{ECmb} evaluates to
\begin{equation}
 \begin{split}
  {\mathcal K} &= \left( 1-\frac{\sigma}{L^2} \right)^{-\frac{3}{2}} (\epsilon \partial^2_t \delta r) - \left( 1-\frac{\sigma}{L^2} \right)^{-\frac{1}{2}} \left(\frac{\sigma}{L^2}\right) \left( 1 + \frac{\epsilon L^2 \delta r}{L^2-\sigma} \right) \\ & + \left( 1-\frac{\sigma}{L^2} \right)^{-\frac{1}{2}} (-\epsilon \bar\nabla^2 \delta r) + \left( 1-\frac{\sigma}{L^2} \right)^{\frac{1}{2}}  \left( 1 - \frac{\epsilon L^2 \delta r}{L^2-\sigma} \right) (D-2)
 \end{split}
\end{equation}

\subsection{Computation of the  terms  relevant for the membrane equation}\label{releglobal}
Here, we report the relevant terms needed to evaluate the membrane equation upto linear order. The relevant terms at leading order evaluate to 
  \begin{equation}\label{eq:list1}
   \begin{split}
    u^\nu {\cal K}_{\nu t} &= \frac{\sigma}{L^2} + {\mathcal O}(\epsilon) \\
    u^\nu {\cal K}_{\nu a} &= \left( 1-\frac{\sigma}{L^2} \right)^{-1} (-\epsilon \partial_t \bar\nabla_a \delta r) + \left( 1-\frac{\sigma}{L^2} \right)^{\frac{1}{2}}(\epsilon \delta u_a) \\
    u^\nu \hat{\nabla}_{\nu}u_{t} &= 0\\
    u^\nu \hat{\nabla}_{\nu}u_{a} &= \left( 1-\frac{\sigma}{L^2} \right)^{-\frac{1}{2}}(\epsilon \partial_t \delta u_a)-\left( 1-\frac{\sigma}{L^2} \right)^{-1}\frac{\sigma}{L^2}(\epsilon \bar\nabla_a \delta r) \\
   \hat{\nabla}_t {\mathcal K} &=  {\mathcal O}(\epsilon) \\
 \hat{\nabla}_a {\mathcal K} &=  \left( 1-\frac{\sigma}{L^2} \right)^{-\frac{3}{2}}(\epsilon \partial^2_t \bar\nabla_a \delta r)-\left( 1-\frac{\sigma}{L^2} \right)^{-\frac{3}{2}}\frac{\sigma}{L^2}(\epsilon \bar\nabla_a \delta r) \\
    &+ \left( 1-\frac{\sigma}{L^2} \right)^{-\frac{1}{2}}(-\epsilon \bar\nabla_a \bar\nabla^2 \delta r) - (D-2) \left( 1-\frac{\sigma}{L^2} \right)^{-\frac{1}{2}}(\epsilon \bar\nabla_a \delta r) \\
   \hat{\nabla}^2u_t &= {\mathcal O}(\epsilon) \\
   \hat{\nabla}^2u_a &= -\left( 1-\frac{\sigma}{L^2} \right)^{-1}(\epsilon \partial_t^2 \delta u_a) + \left( 1-\frac{\sigma}{L^2} \right)^{-\frac{3}{2}}\frac{\sigma}{L^2} (\epsilon \partial_t \bar\nabla_a \delta r) \\
    &+ \epsilon \bar\nabla^2 \delta u_a + \left( 1-\frac{\sigma}{L^2} \right)^{-\frac{1}{2}} (\epsilon \partial_t \bar\nabla_a \delta r)
   \end{split}
  \end{equation}
  The relevant terms at subleading order evaluate to 
 \begin{equation}\label{eq:list2}
  \begin{split}
     u^\nu {\cal{K}}_{\nu\mu}{\cal{K}}^\mu_t &= - \left(\frac{\sigma}{L^2}\right)^2 \left( 1-\frac{\sigma}{L^2} \right)^{-\frac{1}{2}} \\  u^\nu {\cal{K}}_{\nu\mu}{\cal{K}}^\mu_a &= \left( 1-\frac{\sigma}{L^2} \right)^{-\frac{3}{2}}\frac{\sigma}{L^2}\epsilon \partial_t \bar\nabla_a \delta r - \left( 1-\frac{\sigma}{L^2} \right)^{-\frac{1}{2}}\epsilon \partial_t \bar\nabla_a \delta r + \left( 1-\frac{\sigma}{L^2} \right) \epsilon \delta u_a \\ \hat{\nabla}^2\hat{\nabla}^2u_t &= {\mathcal O}(\epsilon) \\ \hat{\nabla}^2\hat{\nabla}^2u_a &= \bar\nabla^2\bar\nabla^2 \delta u_a \\ u.\hat\nabla{\mathcal K} &= {\mathcal O}(\epsilon) \\ \hat\nabla^\nu{\mathcal K}\hat\nabla_\nu u_t &= {\mathcal O}(\epsilon) \\ \hat\nabla^\nu{\mathcal K}\hat\nabla_\nu u_a &= {\mathcal O}(\epsilon)^2 \\ {\cal{K}}^{\mu\nu}\hat\nabla_\mu\hat\nabla_\nu u_t &= {\mathcal O}(\epsilon) \\ {\cal{K}}^{\mu\nu}\hat\nabla_\mu\hat\nabla_\nu u_a &= \left( 1-\frac{\sigma}{L^2} \right)^{\frac{1}{2}}\epsilon\hat\nabla^2 \delta u_a \\ \hat{\nabla}_t\hat{\nabla}^2{\mathcal K} &=  {\mathcal O}(\epsilon) \\ \hat{\nabla}_a\hat{\nabla}^2{\mathcal K} &=  -\left( 1-\frac{\sigma}{L^2} \right)^{-\frac{1}{2}} \hat{\nabla}_a\hat{\nabla}^2\hat{\nabla}^2 \delta r - (D-2)\left( 1-\frac{\sigma}{L^2} \right)^{-\frac{1}{2}} \hat{\nabla}_a\hat{\nabla}^2 \delta r \\ \hat{\nabla}_t({\cal{K}}_{\mu\nu}{\cal{K}}^{\mu\nu}{\mathcal K}) &= {\mathcal O}(\epsilon)\\ \hat{\nabla}_a({\cal{K}}_{\mu\nu}{\cal{K}}^{\mu\nu}{\mathcal K}) &= -3(D-2)\left( 1-\frac{\sigma}{L^2} \right)^{\frac{1}{2}} \epsilon\left( \hat{\nabla}_a\hat{\nabla}^2 \delta r+(D-2)\hat{\nabla}_a\delta r \right) \\ (u\cdot{\cal{K}}\cdot u)u^\nu \hat{\nabla}_{\nu}u_{t} &=   {\mathcal O}(\epsilon)\\ (u\cdot{\cal{K}}\cdot u)u^\nu \hat{\nabla}_{\nu}u_{a} &= \left( 1-\frac{\sigma}{L^2} \right)^{-1} \frac{\sigma}{L^2} \epsilon \partial_t \delta u_a - \left( 1-\frac{\sigma}{L^2} \right)^{-\frac{3}{2}}\left(\frac{\sigma}{L^2}\right)^2 \epsilon \hat{\nabla}_a \delta r \\ (u\cdot{\cal{K}}\cdot u)u^\mu {\cal{K}}_{\mu t} &= \left( 1-\frac{\sigma}{L^2} \right)^{-\frac{1}{2}}\left(\frac{\sigma}{L^2}\right)^2 \\ (u\cdot{\cal{K}}\cdot u)u^\mu {\cal{K}}_{\mu a}
    &= -\left( 1-\frac{\sigma}{L^2} \right)^{-\frac{3}{2}}\left(\frac{\sigma}{L^2}\right) \epsilon \partial_t \hat{\nabla}_a \delta r + \frac{\sigma}{L^2} \epsilon \delta u_a \\ (u\cdot\hat{\nabla}{\mathcal K})u^\mu {\cal{K}}_{\mu t} &= {\mathcal O}(\epsilon) \\ (u\cdot\hat{\nabla}{\mathcal K})u^\mu {\cal{K}}_{\mu a} &= {\mathcal O}(\epsilon)^2\\ (u\cdot\hat{\nabla}{\mathcal K})u^\nu \hat{\nabla}_\nu u_t &= {\mathcal O}(\epsilon) \\ (u\cdot\hat{\nabla}{\mathcal K})u^\nu \hat{\nabla}_\nu u_a &= {\mathcal O}(\epsilon)^2
  \end{split}
 \end{equation}

\subsection{Arguments leading to \eqref{lineq}}

Firstly, for convenience, rewrite the membrane equation \eqref{2ndorderfinal} as
\begin{equation*}
E^{tot}_{\mu} \equiv \mathcal{P}^\nu_\mu E_{\nu},\quad \text{where}\quad E_\mu \equiv \frac{\hat{\nabla}^2u_\mu}{\mathcal{K}}-\frac{\hat{\nabla}_\mu {\mathcal K}}{\mathcal{K}}+u^\nu {\cal K}_{\nu\mu}-u^\nu \hat{\nabla}_{\nu}u_{\mu} + \mathellipsis
\end{equation*}
 So, we get
 \begin{equation}\label{med}
\begin{split}
 E^{tot}_t &= E_t {\mathcal P}^t_t +  E_b {\mathcal P}^b_t \\
 E^{tot}_a &= E_t {\mathcal P}^t_a +  E_b {\mathcal P}^b_a
\end{split}
\end{equation}

 We can see for a uniform membrane configuration with spherical symmetry that $E_a$ would be zero and hence we have $E_a\sim{\cal O}(\epsilon)$ in case of fluctuations. Also we see that ${\mathcal P}^t_t= 0$ and ${\mathcal P}^a_t \sim {\mathcal O}(\epsilon)$. Hence we see from \eqref{med} that $E^{tot}_t$ is identically zero at the linear order. Similarly because ${\mathcal P}^t_a = {\mathcal O}(\epsilon)$, only ${\mathcal O}(\epsilon^0)$ pieces of $ E_t$ are relevant for evaluating $E^{tot}_a$ at linear order. Hence in subsection \ref{releglobal} we evaluated only those terms in $E_\mu$ that are relevant for the linearized analysis.\\
 
Substituting the expressions derived in subsection (\ref{releglobal}) in the linearized vector membrane equation in the angular directions we finally get \eqref{lineq}.

\section{QNM for AdS  Schwarzschild black brane:\\ Details of the calculation}\label{app:QNMPoincare}
Just like previous section, here we shall provide the details of the computation required to determine the QNM frequencies for AdS Schwarzschild black brane.

 The answers for nonzero components of Christoffel symbols for the background metric \eqref{ppmet} are
 \begin{equation}
   \Gamma^r_{rr} = \frac{-1}{r},\quad
   \Gamma^r_{ab} = -r^3 \delta_{ab},\quad
   \Gamma^a_{rb} = \frac{1}{r} \delta^{a}_{b}, \quad 
   \Gamma^r_{tt} = r^3,\quad
   \Gamma^t_{rt} = \frac{1}{r}\quad
 \end{equation}
Normal to the membrane evaluates to
 \begin{equation}
  n_r = \frac{1}{r},\quad n_a = \frac{-\epsilon \partial_a \delta r}{r},\quad n_t = \frac{-\epsilon \partial_t \delta r}{r} 
 \end{equation}
  Non zero components of $\nabla_Mn_N$ evaluate to
\begin{equation}
\begin{split}
  \nabla_rn_r &= 0, \quad  \nabla_rn_t = \frac{2\epsilon \partial_t \delta r}{r^2}, \quad \nabla_tn_r = \frac{\epsilon \partial_t \delta r}{r^2},  \quad \nabla_tn_t = -\frac{\epsilon \partial^2_t \delta r}{r} -r^2, \quad \\
  \nabla_rn_a &= \frac{2\epsilon \partial_a \delta r}{r^2}, \quad \nabla_an_r = \frac{\epsilon \partial_a \delta r}{r^2}, \quad \nabla_tn_a = \frac{-\epsilon \partial_t \partial_a \delta r}{r}, \\
  \nabla_an_t &= \frac{-\epsilon \partial_t \partial_a \delta r}{r}, \quad \nabla_an_b = \frac{-\epsilon \partial_a \partial_b \delta r}{r} + r^2 \delta_{ab}
\end{split}
\end{equation}
The projector $P_A^B=\delta_A^B - n_An^B$ evaluates to
\begin{equation}
 \begin{split}
  P^r_r &= 0,\quad P_t^t = 1,\quad P^a_b = \delta^a_b,\quad P^a_t = 0, \quad P^t_a = 0, \\
  P^r_t &= \epsilon \partial_t \delta r, \quad P^t_r = \frac{-\epsilon \partial_t \delta r}{r^4}, \quad \quad P^r_a = \epsilon \partial_a \delta r, \quad \quad P^a_r = \frac{\epsilon \partial^a \delta r}{r^4}
 \end{split}
\end{equation}
Nonzero components of the spacetime form of Extrinsic curvature $K_{MN}$ evaluate to
\begin{equation}
\begin{split}
 K_{rr} &= 0, \quad K_{rt}  = \frac{\epsilon \partial_t \delta r}{r^2} , \quad K_{ra} = \frac{\epsilon \partial_a \delta r}{r^2} \\ K_{tt} &= \frac{-\epsilon \partial^2_t \delta r}{r} - r^2, \quad K_{ta}  =\frac{-\epsilon \partial_t \partial_a \delta r}{r}, \quad K_{ab} = \frac{-\epsilon \partial_a \partial_b \delta r}{r} + r^2 \delta_{ab}
 \end{split}
\end{equation}
Nonzero components of Christoffel symbols for the induced metric \eqref{ppind} evaluate to
\begin{equation}
 \begin{split}
  \Gamma^t_{tt} &= \epsilon \partial_t \delta r, \quad  \Gamma^a_{tt} = \epsilon \partial^a \delta r, \quad \Gamma^t_{at} = \epsilon \partial_a \delta r, \quad \Gamma^t_{ab} = \epsilon \partial_t \delta r\delta_{ab}\\
  \Gamma^a_{tb} &= \epsilon \partial_t \delta r\delta^a_b, \quad \Gamma^a_{bc} = \epsilon  ( \partial_b \delta r \delta^a_c + \partial_c \delta r \delta^a_b - \partial^a \delta r \delta_{bc})
 \end{split}
\end{equation}
The projector  $\mathcal{P}^\mu_\nu = \delta^\mu_\nu + u^\mu u_\nu$ evaluates to
\begin{equation}
  \mathcal{P}^a_b = \delta^a_b, \quad \mathcal{P}^t_t = 0, \quad \mathcal{P}^t_a = \epsilon \delta u_a, \quad  \mathcal{P}^a_t = -\epsilon \delta u_a, 
\end{equation}
Nonzero components of $\hat{\nabla}_\mu u_\nu$ evaluate to
\begin{equation}
\begin{split}
 \hat{\nabla}_t u_t &= 0, \quad \hat{\nabla}_t u_a = \epsilon \partial_t \delta u_a + \epsilon \partial_a \delta r, \quad \hat{\nabla}_a u_t = 0, \\
 \hat{\nabla}_a u_b &= \epsilon \partial_a \delta u_b + \epsilon \partial_t \delta r \delta_{ab}
\end{split}
\end{equation}

\subsection{Computation of ${\cal K}_{\mu\nu}$}
As done previously, ${\cal K}_{\mu\nu}$ is defined as the pullback of spacetime form of  extrinsic curvature $K_{MN}$ on the membrane worldvolume. Doing this procedure we find that the nonzero components of ${\cal K}_{\mu\nu}$ evaluate to
 \begin{equation}
     {\cal  K}_{tt} = -\epsilon \partial^2_t \delta r - (1 + 2\epsilon \delta r), \quad {\cal  K}_{ta} = -\epsilon \partial_t \partial_a \delta r, \quad{\cal  K}_{ab} = -\epsilon \partial_a \partial_b \delta r + (1 + 2\epsilon \delta r) \delta_{ab}
 \end{equation}
 Trace of Extrinsic curvature ${\cal K}_{\mu\nu}$ evaluates to
 \begin{equation}\label{kap}
  \mathcal{K} = n + \epsilon \partial^2_t \delta r
  - \epsilon \partial_a \partial^a \delta r
 \end{equation}
where we raised the index $a$ in \eqref{kap} with $\delta^{ab}$.

 \subsection{Computation of the terms relevant for membrane equation}\label{relepoincare}
 At leading order the relevant terms evaluate to 
 \begin{equation}\label{eq:lis1}
    \begin{split}
     u^\nu {\cal K}_{\nu t} &= -1 + {\mathcal O}(\epsilon) \\
     u^\nu {\cal K}_{\nu a} &= -\epsilon \partial_t\partial_a\delta r+\epsilon \delta u_a \\
     u^\nu \hat{\nabla}_{\nu}u_{t} &= {\mathcal O}(\epsilon)\\
     u^\nu \hat{\nabla}_{\nu}u_{a} &= \epsilon \partial_t \delta u_a + \epsilon \partial_a \delta r \\
    \hat{\nabla}_t {\mathcal K} &=  {\mathcal O}(\epsilon) \\
  \hat{\nabla}_a {\mathcal K} &=  \epsilon \partial_a\partial_t^2 \delta r - \epsilon\partial_a \partial^2 \delta r \\
    \hat{\nabla}^2u_t &= {\mathcal O}(\epsilon) \\
    \hat{\nabla}^2u_a &= -\epsilon \partial_t^2 \delta u_a  + \epsilon \partial^2 \delta u_a 
    \end{split}
   \end{equation}
   While at subleading order the relevant terms evaluate to
   \begin{equation}\label{eq:lis2}
     \begin{split}
        u^\nu K_{\nu\mu}K^\mu_t &= -1 + {\mathcal O}(\epsilon)\\  u^\nu K_{\nu\mu}K^\mu_a &= -2 \epsilon \partial_t \partial_a \delta r + \epsilon \delta u_a \\ \hat{\nabla}^2\hat{\nabla}^2u_t &= {\mathcal O}(\epsilon) \\ \hat{\nabla}^2\hat{\nabla}^2u_a &= \epsilon\partial_t^4 \delta u_a -2\epsilon\partial_t^2 \partial^2\delta u_a + \epsilon\partial^4 \delta u_a \\ u.\hat\nabla{\mathcal K} &= {\mathcal O}(\epsilon) \\ \hat\nabla^\nu{\mathcal K}\hat\nabla_\nu u_t &= {\mathcal O}(\epsilon) \\ \hat\nabla^\nu{\mathcal K}\hat\nabla_\nu u_a &= {\mathcal O}(\epsilon)^2 \\ K^{\mu\nu}\hat\nabla_\mu\hat\nabla_\nu u_t &= {\mathcal O}(\epsilon) \\ K^{\mu\nu}\hat\nabla_\mu\hat\nabla_\nu u_a &= -\epsilon \partial_t^2 \delta u_a  + \epsilon \partial^2 \delta u_a \\ \hat{\nabla}_t\hat{\nabla}^2{\mathcal K} &=  {\mathcal O}(\epsilon) \\ \hat{\nabla}_a\hat{\nabla}^2{\mathcal K} &= -\epsilon \partial_a \partial_t^4 \delta r + 2\epsilon\partial_a\partial_t^2\partial^2 \delta r - \epsilon \partial_a\partial^2\partial^2 \delta r \\ \hat{\nabla}_t(K_{\mu\nu}K^{\mu\nu}{\mathcal K}) &= {\mathcal O}(\epsilon)\\ \hat{\nabla}_a(K_{\mu\nu}K^{\mu\nu}{\mathcal K}) &= 3\epsilon(\partial_a \partial_t^2 \delta r - \partial_a\partial^2 \delta r ) \\ (u.K.u)u^\nu \hat{\nabla}_{\nu}u_{t} &=   {\mathcal O}(\epsilon)\\ (u.K.u)u^\nu \hat{\nabla}_{\nu}u_{a} &= -(\epsilon\partial_t \delta u_a + \epsilon\partial_a \delta r) \\ (u.K.u)u^\mu K_{\mu t} &= 1 + {\cal O}(\epsilon) \\ (u.K.u)u^\mu K_{\mu a} &= \epsilon \partial_t \partial_a \delta r-\epsilon \delta u_a \\ (u.\hat{\nabla}{\mathcal K})u^\mu K_{\mu t} &= {\mathcal O}(\epsilon) \\ (u.\hat{\nabla}{\mathcal K})u^\mu K_{\mu a} &= {\mathcal O}(\epsilon)^2\\ (u.\hat{\nabla}{\mathcal K})u^\nu \hat{\nabla}_\nu u_t &= {\mathcal O}(\epsilon) \\ (u.\hat{\nabla}{\mathcal K})u^\nu \hat{\nabla}_\nu u_a &= {\mathcal O}(\epsilon)^2
     \end{split}
    \end{equation}
%

\subsection{Arguments leading to \eqref{eq:fluceqcomp}}

Following the same trick as previously done, we denote the vector membrane equation as
 \begin{equation}
E^{tot}_{\mu} \equiv \mathcal{P}^\nu_\mu E_{\nu},\quad \text{where}\quad E_{\mu} \equiv\left[\frac{\hat{\nabla}^2 u_\alpha}{\cal{K}}-\frac{\hat{\nabla}_\alpha{\cal{K}}}{\cal{K}}+u^\beta K_{\beta\alpha}-u\cdot\hat{\nabla} u_\alpha\right]+\mathellipsis 
 \end{equation}
Hence we have
\begin{equation}\label{eomdiv}
\begin{split}
 E^{tot}_t &= E_t {\mathcal P}^t_t +  E_b {\mathcal P}^b_t \\
 E^{tot}_a &= E_t {\mathcal P}^t_a +  E_b {\mathcal P}^b_a 
\end{split}
\end{equation}
For the uniform planar membrane we have translational symmetry along the $x_a$ directions, so we have $E_b\sim{\mathcal O}(\epsilon)$ in the case of fluctuations. Note that ${\mathcal P}^t_t = 0$,  ${\mathcal P}^a_t\sim {\mathcal O}(\epsilon)$ and also $E_b\sim{\mathcal O}(\epsilon)$, hence $E^{tot}_t$ vanishes upto linear order.
Note that ${\mathcal P}^t_a \sim {\mathcal O}(\epsilon)$, hence only ${\mathcal O}(\epsilon^0)$ pieces of $E_t$ contribute when we evaluate $E^{tot}_a$ upto linear order.
 Keeping these facts in mind we calculated only those terms that are relevant in subsection \ref{relepoincare}.
 
 Substituting the expressions derived in subsection (\ref{relepoincare}) in the linearized vector membrane equation in the angular directions we finally get \eqref{eq:fluceqcomp}.\\

\bibliography{larged}
\bibliographystyle{JHEP}

\end{document}